\documentclass[12pt]{article}
\usepackage{graphicx}
\usepackage{amssymb}
\usepackage{capt-of}
\usepackage{color}
\usepackage{lineno}
\usepackage[colorlinks,linkcolor=blue,anchorcolor=blue,citecolor=blue,urlcolor=blue]{hyperref}
\usepackage{lmodern}
\usepackage{subfigure}
\usepackage{multirow}
\usepackage{tikz}
\usepackage{textcomp}
\usepackage{wrapfig}
\usetikzlibrary{positioning}
\usepackage{physics}
\usepackage{hepunits}

\graphicspath{
{./}{1.Introduction/}
{2.1MassOrdering/}
{2.2Precision/}
{2.3Supernova/}
{2.4DSNB/}
{2.5Solar/}
{2.6Atmospheric/}
{2.7Geoneutrino/}
{2.8ProtonDecay/}
{2.9OtherPhys/}
{3.1DetectorOverview/}
{3.2CD/}
{3.3Veto/}
{3.4LS/}
{3.5PMTInstrumentation/}
{3.6Electronics/}
{3.7SPMT/}
{3.8Calibration/}
{3.9DAQDCS/}
{3.10Offline/}
{3.11LowBackground/}
{3.13TAO/}
{4Summary/}
}

\begin{document}

\title{ \vspace{1cm} JUNO Physics and Detector}
\author{JUNO Collaboration~\cite{junocollaboration}}
\maketitle

\begin{abstract}
The Jiangmen Underground Neutrino Observatory (JUNO) is a 20 kton liquid scintillator detector in a laboratory at 700-m underground. An excellent energy resolution and a large fiducial volume offer exciting opportunities for addressing many important topics in neutrino and astro-particle physics.
With six years of data, the neutrino mass ordering can be determined at a 3-4$\sigma$ significance and the neutrino oscillation parameters $\sin^2\theta_{12}$, $\Delta m^2_{21}$, and $|\Delta m^2_{32}|$ can be measured to a precision of 0.6\% or better, by detecting reactor antineutrinos from the Taishan and Yangjiang nuclear power plants.
With ten years of data, neutrinos from all past core-collapse supernovae could be observed at a 3$\sigma$ significance; a lower limit of the proton lifetime, $8.34 \times 10^{33}$ years (90\% C.L.), can be set by searching for $p \to \bar{\nu} K^+$; detection of solar neutrinos would shed new light on the solar metallicity problem and examine the vacuum-matter transition region.
A typical core-collapse supernova at a distance of 10~kpc would lead to $\sim5000$ inverse-beta-decay events and $\sim2000$ (300) all-flavor neutrino-proton (electron) elastic scattering events in JUNO. Geo-neutrinos can be detected with a rate of $\sim 400$ events per year.
Construction of the detector is very challenging. In this review, we summarize the final design of the JUNO detector and the key R\&D achievements, following the Conceptual Design Report in 2015~\cite{Djurcic:2015vqa}.
All 20-inch PMTs have been procured and tested. The average photon detection efficiency is 28.9\% for the 15,000 MCP PMTs and 28.1\% for the 5,000 dynode PMTs, higher than the JUNO requirement of 27\%. Together with the $>20$~m attenuation length of the liquid scintillator achieved in a 20-ton pilot purification test and the $>96\%$ transparency of the acrylic panel, we expect a yield of 1345 photoelectrons per MeV and an effective relative energy resolution of $3.02\%/\sqrt{E{\rm (MeV)}}$ in simulations~\cite{Abusleme:2020lur}.
To maintain the high performance, the underwater electronics is designed to have a loss rate $<0.5\%$ in six years.
With degassing membranes and a micro-bubble system, the radon concentration in the 35 kton water pool could be lowered to $<10$~mBq/m$^3$. Acrylic panels of radiopurity $<0.5$~ppt U/Th for the 35.4-m diameter liquid scintillator vessel are produced with a dedicated production line. The 20 kton liquid scintillator will be purified onsite with Alumina filtration, distillation, water extraction, and gas stripping. Together with other low background handling, singles in the fiducial volume can be controlled to $\sim10$~Hz.
 The JUNO experiment also features a double calorimeter system with 25,600 3-inch PMTs, a liquid scintillator testing facility OSIRIS, and a near detector TAO.

\end{abstract}

\tableofcontents
\clearpage


\section{Introduction}
\label{sec:intro}
\subsection{Overview of the JUNO experiment}
\label{subsec:ExpOverview}

The standard three-flavor neutrino oscillation pattern is well established after the observation of the neutrino oscillation in solar, atmospheric, accelerator and reactor neutrino experiments.
Two independent neutrino mass splittings $|\Delta m^2_{31}|=|m^2_3-m^2_1|$ (or $|\Delta
m^2_{32}|=|m^2_3-m^2_2|$) and $\Delta m^2_{21}=m^2_2 - m^2_1$, and three neutrino mixing angles from the Pontecorvo-Maki-Nakagawa-Sakata (PMNS) parametrization~\cite{Maki:1962mu,Pontecorvo:1967fh} were measured with precisions at a level of few percents. However, several unknowns still exist and will be the focus of future neutrino oscillation experiments. They include
\begin{itemize}
\item the Neutrino Mass Ordering (NMO),
\item the leptonic CP-violating phase $\delta$ in the PMNS matrix,
\item the octant of the mixing angle $\theta_{23}$ (i.e., $\theta_{23}<\pi/4$ or $\theta_{23}>\pi/4$).
\end{itemize}

The Jiangmen Underground Neutrino Observatory (JUNO), a 20 kton multi-purpose underground liquid scintillator detector, was proposed with the determination of the neutrino mass ordering as a primary physics goal~\cite{Zhan:2008id,Zhan:2009rs,Li:2013zyd,yellowbook}. The neutrino mass ordering has only two possibilities: the normal ordering (NO, $m_1<m_3$) and the inverted ordering (IO, $m_1>m_3$). The relatively large value of $\theta_{13}$ has provided excellent opportunities to resolve the NMO in various neutrino oscillation experiments, which include a medium baseline ($\sim$50~km) reactor antineutrino $\bar\nu_{e}\to\bar\nu_{e}$ oscillation experiment (JUNO), long-baseline accelerator (anti-)neutrino $\nu_{\mu}\to\nu_{e}$ oscillation experiments (NO$\nu$A~\cite{Ayres:2004js} and DUNE~\cite{Adams:2013qkq}), and atmospheric (anti-)neutrino oscillation experiments (INO~\cite{Kumar:2017sdq}, PINGU~\cite{Aartsen:2014oha}, ORCA~\cite{VanElewyck:2015una}, DUNE~\cite{Adams:2013qkq} and Hyper-K~\cite{Abe:2011ts}).
The accelerator and atmospheric experiments rely on the matter effect in neutrino oscillations (the charge-current interaction between (anti-)$\nu_e$ and electrons in the matter). JUNO is a unique experiment designed to identify the NMO using the oscillation interplay between $\Delta m^2_{31}$ and $\Delta m^2_{32}$~\cite{Petcov:2001sy}. The NMO sensitivity of JUNO has no dependence on the unknown CP-violating phase and the $\theta_{23}$ octant, playing a key role when combined with other neutrino experiments.

\begin{figure}
\begin{centering}
\includegraphics[width=0.55\textwidth]{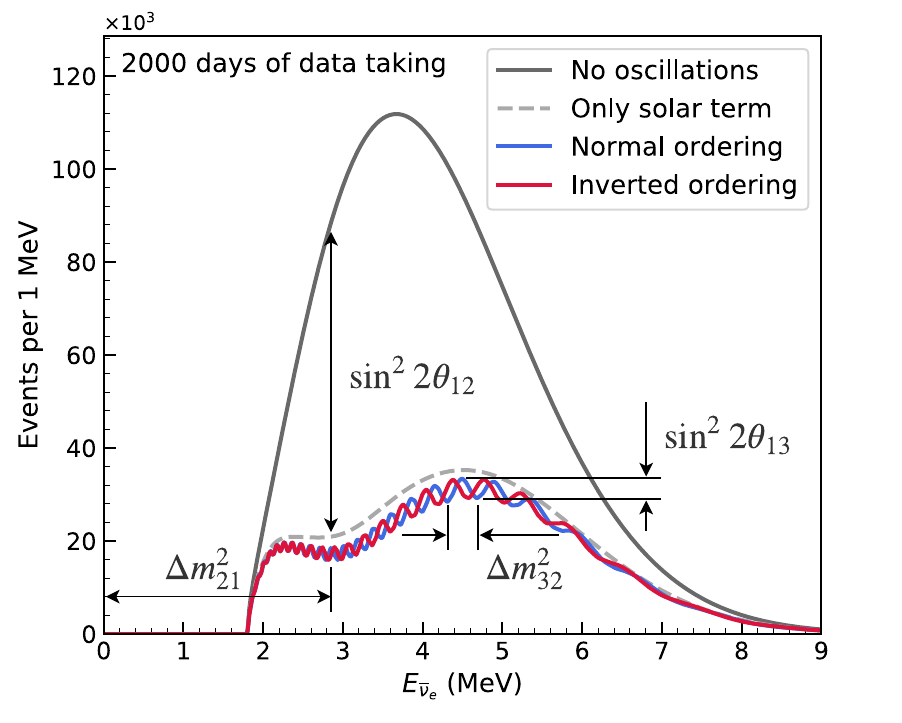}
\par\end{centering}
\caption{\label{fig:oscillation} The expected antineutrino energy spectrum weighted by IBD cross-section with and without oscillation at the JUNO experiment for normal ordering and inverted ordering assuming 2000 days of data-taking. Dependence on the four oscillation parameters is shown.}
\end{figure}

The reactor antineutrino survival probability in vacuum can be written as
\begin{eqnarray}
\label{eq:P_nue2nue}
		P_{\bar{\nu}_e\to\bar{\nu}_e}=1 -\sin^22\theta_{13}(\cos^2\theta_{12}\sin^2\Delta_{31}+\sin^2\theta_{12}\sin^2\Delta_{32})
		 -\cos^4\theta_{13}\sin^22\theta_{12}\sin^2\Delta_{21},
\end{eqnarray}
where $\Delta_{ij}=\Delta m_{ij}^2L/(4E)=(m_i^2-m_j^2)L/(4E)$, in which L is the baseline and E is the antineutrino energy.
At a baseline of 53~km, JUNO will simultaneously measure oscillations driven by small mass splitting ($\Delta m^2_{21}$) and large mass splitting ($\Delta m^2_{31}$ and $\Delta m^2_{32}$) as shown in Fig.~\ref{fig:oscillation}.
The small oscillation peaks in the oscillated antineutrino spectrum contain the NMO information.
Precise measurement of the oscillated antineutrino spectrum is a key for JUNO to determine the NMO.
This requires a 20~kton liquid scintillator detector with an unprecedented relative energy resolution of $\sigma_{\rm E}/E=3\%/\sqrt{E_{\mathrm{vis}}}$, with $E_{\mathrm{vis}}$ being the visible energy in the detector in MeV.

Besides the neutrino mass ordering, the large fiducial volume and the excellent energy resolution of JUNO offer exciting opportunities for addressing many important topics in neutrino and astro-particle physics.

The precision measurement of reactor antineutrino spectrum will also lead to the precise determination of the neutrino oscillation parameters $\sin^2\theta_{12}$, $\Delta m^2_{21}$, and $|\Delta m^2_{32}|$ as illustrated in Fig.~\ref{fig:oscillation}. The expected accuracy of these measurements will be better than 0.6\%, which will play a crucial role in the future unitarity test of the PMNS matrix.

The JUNO detector is not limited to detect antineutrinos from the reactors, but also observe neutrinos/antineutrinos from terrestrial and extra-terrestrial sources, including supernova burst neutrinos, diffuse supernova neutrino background, geoneutrinos, atmospheric neutrinos, and solar neutrinos. For example, a neutrino burst from a typical core-collapse supernova at a distance of 10~kpc (kiloparsec ) would lead to $\sim5000$ inverse-beta-decay events and $\sim2000$ all-flavor neutrino-proton elastic scattering events in JUNO, which are of crucial importance for understanding the mechanism of supernova explosion and for exploring novel phenomena such as collective neutrino oscillations. Detection of 1--2 neutrinos per year from all past core-collapse supernova explosions in the visible universe can further provide valuable information on the cosmic star-formation rate and the average core-collapse neutrino energy spectrum.
Antineutrinos originating from the radioactive decay of uranium and thorium in the Earth can be detected in JUNO with a rate of $\sim 400$ events per year, significantly improving the statistics of existing geoneutrino event samples. Atmospheric neutrino events collected in JUNO can provide independent inputs for determining the mass ordering and the octant of the $\theta_{23}$ mixing angle. Detection of the $^7$Be and $^8$B solar neutrino events at JUNO would shed new light on the solar metallicity problem and examine the spectral transition region between the vacuum and matter-dominated neutrino oscillations.

The JUNO detector provides sensitivity to physics searches beyond the Standard Model. As examples, we highlight the searches for proton decay via the $p \to K^+ + \bar \nu$ decay channel, neutrinos resulting from dark-matter annihilation in the Sun, violation of Lorentz invariance via the sidereal modulation of the reactor neutrino event rate, and the effects of non-standard neutrino interactions.

JUNO was first conceived in 2008~\cite{Zhan:2008id,Zhan:2009rs}. It was approved in 2013 after Daya Bay~\cite{An:2012eh}, Double Chooz~\cite{Abe:2011fz}, and RENO~\cite{Ahn:2012nd} measured an unexpectedly large value of $\theta_{13}$, which meant that the NMO could be determined with current technologies. The civil construction started in 2015. The detector is expected to be ready in 2022, and data-taking is expected in 2023.

In 2018, the Taishan Antineutrino Observatory (TAO, also known as JUNO-TAO) was proposed as a satellite experiment of JUNO to measure the reactor antineutrino spectrum with sub-percent energy resolution~\cite{junotaocdr}. Since Daya Bay~\cite{An:2015nua}, Double Chooz~\cite{Abe:2014bwa}, and RENO~\cite{Bak:2018ydk}, among others, have found that the model prediction on the reactor antineutrino spectrum~\cite{Huber:PhysRevC84,PhysRevC.83.054615} has large discrepancies with data, TAO will provide a reference spectrum for JUNO, and also provide a benchmark measurement to test nuclear databases. TAO will be a ton-level liquid scintillator detector at $\sim30$ meters from a reactor core of the Taishan Nuclear Power Plant (NPP). It is expected to start operation at a similar time scale as JUNO.

The physics potential of the JUNO experiment has been explored in the JUNO Yellow Book~\cite{yellowbook}. In this article, we will review the major goals with updated inputs and improved analyses. Since the release of the Conceptual Design Report of JUNO~\cite{Djurcic:2015vqa}, the detector design has been further optimized. Challenges regarding the detector technologies have been solved with extensive R\&D. The final design of the JUNO detector and key R\&D achievements will be summarized in this article.

\subsection{JUNO experimental site}
\label{subsubsec:site}

The JUNO experiment is located in Jinji town, 43 km to the southwest of Kaiping city, a county-level city in the prefecture-level city Jiangmen in Guangdong province, China. The geographic location is 112$^\circ$31'05"~E and  22$^\circ$07'05"~N. The distances to several megacities, Guangzhou, Shenzhen, and Hong Kong, are all around 200~km.

\begin{figure}[htb]
\centering
\includegraphics[width=0.7\textwidth]{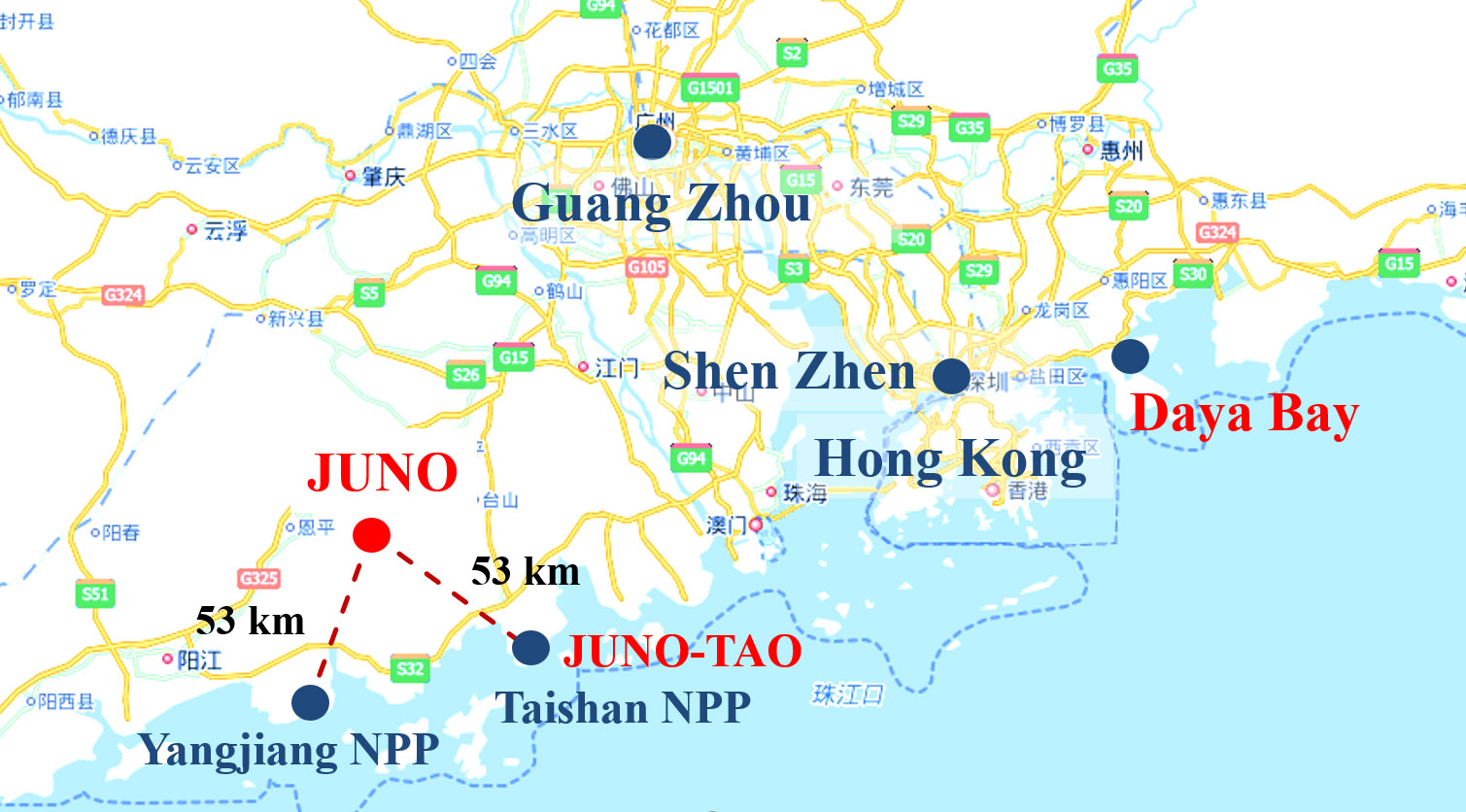}
\caption{JUNO location.
\label{fig:junolocation}}
\end{figure}

As shown in Fig.~\ref{fig:junolocation}, the experimental site is at equal distances of $\sim$ 53~km from the Yangjiang NPP and the Taishan NPP, optimized to have the best sensitivity for determining the mass ordering.

Reactor antineutrino is the primary neutrino source in the JUNO detector. Yangjiang NPP has six reactor cores of 2.9~GW$_{\rm th}$ each (thermal power). All cores are the second-generation pressurized water reactors CPR1000, which is a derivative of Framatome M310. The distances between any two cores of the Yangjiang NPP are between 89~m and 736~m. All six cores are in operation. Taishan NPP has two cores of 4.6~GW$_{\rm th}$ each in operation. Both are third-generation pressurized water reactors EPR. The distance between the two cores is 252.5~m. Possibly another two cores in the Taishan NPP will be built in the future, but the plan is unclear for now. The total thermal power of the Yangjiang and Taishan NPPs is 26.6~GW$_{\rm th}$.

The Daya Bay nuclear complex is 215~km away from the JUNO detector. It includes the Daya Bay NPP, the Ling Ao NPP, and the Ling Ao-II NPP in a spread of 1.1~km, each with 2 cores of 2.9~GW$_{\rm th}$. The Daya Bay and Ling Ao cores are Framatome M310 and the Ling Ao-II cores are CPR1000. They will contribute about 6.4\% of the reactor antineutrino events in the JUNO detector considering oscillation.
Huizhou NPP is under construction with six 2.9~GW$_{\rm th}$ reactor cores and is expected to be ready around 2025.
The plan of Lufeng NPP is unclear now.
The Huizhou site is 265~km away from the JUNO detector and the Lufeng site is more than 300~km away.
There is no other NPP or planned NPP in a radius of 500~km around the JUNO experimental site.

The thermal power of all cores and the baselines (distances to the JUNO detector) are listed in Tab.~\ref{tab:intro:NPP}. The distances from the detector site to the Yangjiang and Taishan cores are surveyed with a Global Positioning System (GPS) to a precision of 1~meter. All these NPPs are constructed and operated by the China General Nuclear Power Group (CGNPG).

\begin{table}[htb]
\centering
\begin{tabular}{|c|c|c|c|c|c|c|c|c|c|c|}\hline
Cores & YJ-1 & YJ-2 & YJ-3 & YJ-4 & YJ-5  & YJ-6 &TS-1 & TS-2 &  DYB  & HZ \\ \hline
Power (GW) & 2.9 & 2.9 & 2.9 & 2.9 & 2.9 & 2.9 & 4.6 & 4.6 & 17.4 & 17.4 \\ \hline
Baseline(km) & 52.74 & 52.82 & 52.41 & 52.49 & 52.11 & 52.19& 52.77 & 52.64  & 215 & 265 \\
\hline
\end{tabular}
\caption{Summary of the thermal power and baseline to the JUNO detector for the
Yangjiang (YJ) and Taishan (TS) reactor cores, as well as the remote reactors of Daya Bay (DYB) and
Huizhou (HZ).\label{tab:intro:NPP}}
\end{table}

Due to the absence of high mountains in the allowed area where the sensitivity to the mass ordering is optimized, the JUNO detector will be deployed in an underground laboratory under the Dashi hill.
Currently, the experiment hall has been excavated. The location has been shifted by $\sim$60~m to the northwest of the originally designed location in Ref.~\cite{yellowbook}.
The elevation of the hill above the detector is 240.6~m above sea level. The dome and the floor of the underground experimental hall are at -403.5~m and -430.5~m, respectively. The detector is located in a cylindrical pit, with the detector center at -452.75~m. Therefore, the vertical overburden for the detector center is 693.35~m (1800~m.w.e). The experimental hall will have two accesses: a 564~m deep vertical shaft and a 1266~m long tunnel with a slope of 42.5\%. The surrounding rock is granite. The average rock density along a 650~m borehole near the experimental hall is measured to be 2.61 g/cm$^3$.
The activities of the $^{238}$U, $^{232}$Th, and $^{40}$K in the rock around the experimental hall are measured to be 120, 106, and 1320 Bq/kg, respectively, with 10\% uncertainties.
The muon rate and average energy in the JUNO detector are 0.004~Hz/m$^2$ and 207 GeV estimated by simulation, taking the surveyed mountain profile into account.

\subsection{JUNO detector}
\label{subsubsec:detector}

\begin{figure}[htb]
    \centering
    \includegraphics[width=0.8\columnwidth]{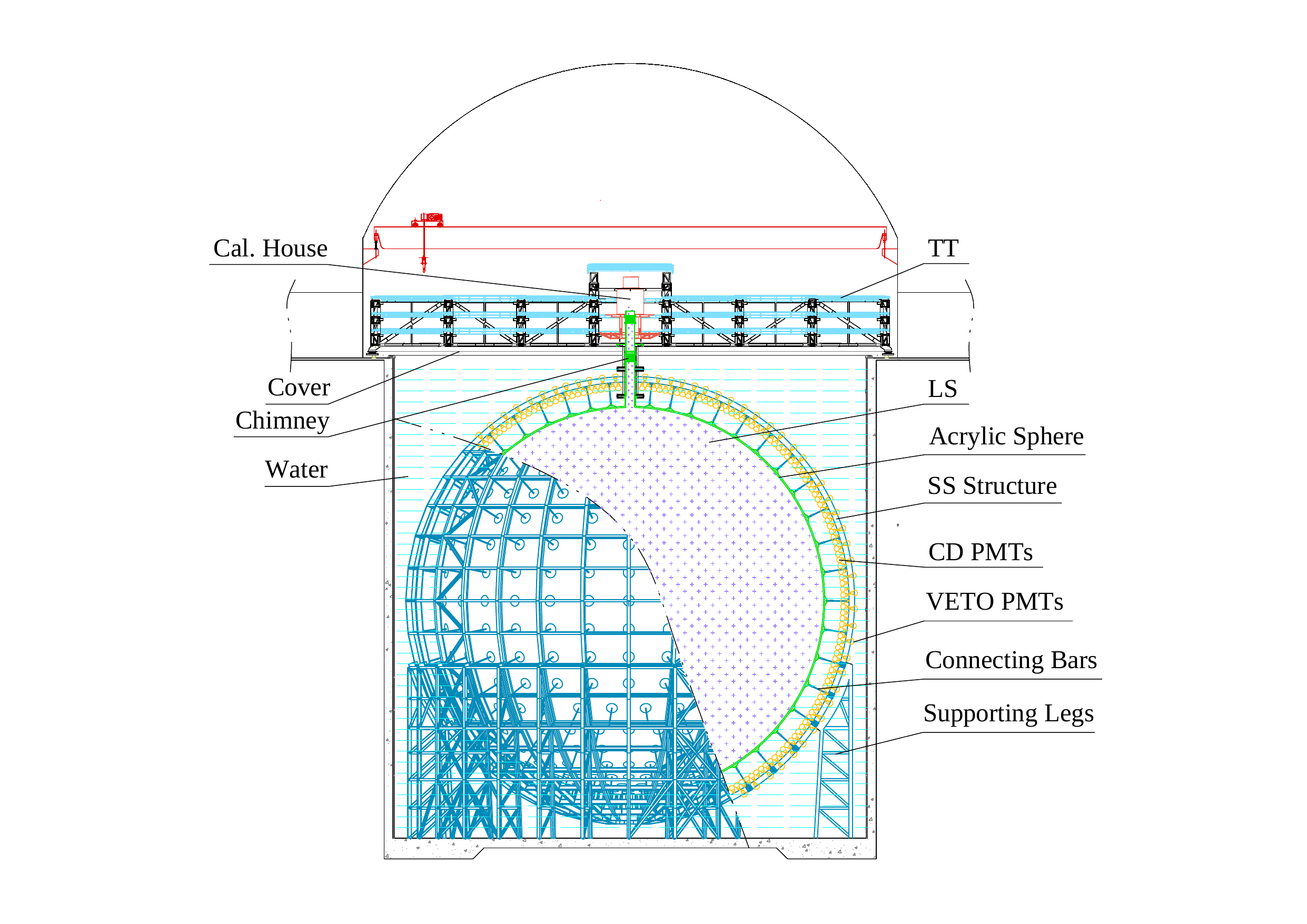}
    \caption{Schematic view of the JUNO detector
    \label{fig:junodetector}}
\end{figure}

The JUNO detector consists of a Central Detector (CD), a water Cherenkov detector and a Top Tracker (TT). A schematic view of the JUNO detector is shown in Fig.~\ref{fig:junodetector}. The CD is a liquid scintillator (LS) detector with a designed effective energy resolution of $\sigma_{\rm E}/E=3\%/\sqrt{E{\rm (MeV)}}$. It contains 20 kton LS in a spherical acrylic vessel, which is submerged in a water pool. The acrylic vessel is supported by a stainless steel (SS) structure via Connecting Bars. The CD Photomultiplier Tubes (PMTs) are installed on the inner surface of the SS structure. The water pool is equipped with PMTs to detect the Cherenkov light from cosmic muons, acting as a veto detector. Compensation coils are mounted on the SS structure to suppress the Earth's magnetic field and minimize its impact on the photoelectron collection efficiency of the PMTs. The CD and the water Cherenkov detector are optically separated. On top of the water pool, there is a plastic scintillator array, i.e. Top Tracker, to accurately measure the muon tracks. A chimney for calibration operations connects the CD to the outside from the top. The calibration systems are operated in the Calibration House, above which a special radioactivity shielding and a muon detector are designed. Detailed description of the detector can be found in Sec.~\ref{sec:detector}.

\subsection{JUNO signal and background}
\label{subsec:Signal}
Reactor antineutrinos constitute the primary signal in the JUNO detector for determining the neutrino mass ordering and for precision measurements of the neutrino oscillation parameters.
All reactors close to JUNO are commercial light-water reactors, where fissions of four isotopes in fuel,  $^{235}$U, $^{238}$U, $^{239}$Pu and $^{241}$Pu, contribute $>99.7\%$ of the antineutrinos.
The antineutrino flux is composed of these four components weighted by the fission rate of the four isotopes, $\phi(E_{\nu}) = \sum_i^4F_iS_i(E_{\nu})$, where $F_i$ is the fission rate and $S_i(E_{\nu})$ is the antineutrino energy spectrum per fission for the $i$-th isotope.
The fission rate can be evaluated based on the reactor running information provided by the NPPs, including the reactor thermal power, the burn-up of the fuel, the fission fractions of four isotopes, and the energy released per fission.
The antineutrino spectrum per fission has been calculated using two methods.
One is based on the summation method~\cite{PhysRevLett.112.202501,PhysRevC.24.1543,PhysRevLett.109.202504}
which sums all the antineutrino energy spectra corresponding to thousands of beta
decay branches for about 1000 isotopes in the fission products, utilizing information in nuclear databases.
The other is the beta conversion method~\cite{Huber:PhysRevC84,PhysRevC.83.054615,PhysRev.113.280, VonFeilitzsch:1982jw,Schreckenbach:1985ep,HAHN1989365}
which converts the measured $\beta$ energy spectra from the individual fission
isotopes $^{235}$U, $^{239}$Pu, and $^{241}$Pu
to the corresponding antineutrino energy spectra. The  $^{238}$U spectrum relies on the summation method and contributes $<10$\% of the total events.
Recent findings of the reactor antineutrino flux and spectrum anomalies have revealed unclear systematic effects in the reactor flux models.
To provide a reliable reference antineutrino spectrum, the JUNO-TAO experiment was proposed as a satellite experiment of JUNO to measure the reactor antineutrino spectrum with sub-percent energy resolution~\cite{junotaocdr}.

JUNO detects electron antineutrinos via inverse beta decay (IBD) interactions, $\bar{\nu}_e+p\to e^{+}+n$.
The $e^+$ quickly deposits its energy and annihilates into two 0.511-MeV photons, which provides a prompt signal. The prompt energy contains both the positron kinetic energy $T_{e^+}$ and the annihilation energy of $2\times 0.511$ MeV. The neutron is mainly captured on protons. After approximately 200 $\mu s$ of scattering in the detector, the capture releases a 2.2-MeV photon, providing a delayed signal.
A set of preliminary antineutrino selection cuts is listed below:
\begin{itemize}
  \item fiducial volume cut $r<17.2$~m;
  \item the prompt energy cut 0.7 MeV $<E_p<$ 12~MeV;
  \item the delayed energy cut 1.9 MeV $<E_d<$ 2.5~MeV;
  \item time interval between the prompt and delayed signal $\Delta T<1.0$~ms;
  \item the prompt-delayed distance cut $R_{p-d}<1.5$~m;
  \item muon veto criteria:
  \begin{itemize}
    \item for muons tagged by the water Cerenkov detector or the Top Tracker, veto the whole LS volume for 1.5 ms;
    \item for well-tracked muons in the Central Detector, veto the detector volume within a cylinder of distance to the muon track $R_{d2\mu}<3$~m and within time to the preceding muon $T_{d2\mu}<1.2$~s;
    \item for tagged, non-trackable muons in the Central Detector, veto the whole LS volume for 1.2~s.
  \end{itemize}
\end{itemize}
The detection efficiency for each cut is shown in Tab.~\ref{tab:mh:sigbkg}.
JUNO will detect 60 IBDs/day with the above selection criteria.

The two major background sources of JUNO are natural radioactivity and the products of cosmic muons.
Natural radioactivity comes from all materials and the environment.
Huge efforts on material screening and a careful arrangement of the experimental apparatus reduce the single event rate to about 10~Hz in the fiducial volume.
The accidental background due to radioactivity and neutrons is further reduced by applying selection cuts using the energy, time and space signatures of both prompt and delayed signals.
Moreover, ${}^{13}$C($\alpha$,n)${}^{16}$O reactions in the liquid scintillator result in correlated background events.
For the reactor antineutrino program, the U/Th concentration is required to be lower than $10^{-15}$~g/g in the liquid scintillator, while the requirement is $10^{-17}$~g/g for the solar neutrino detection.
With 693.35~m (1800~m.w.e) vertical overburden, the muon flux in the detector is 0.004~Hz/m$^2$. The double muon veto systems, formed by a Top Tracker system and a water Cherenkov detector, ensure a high muon tagging efficiency to reject the cosmogenic backgrounds: $^9$Li/$^8$He and fast neutron backgrounds.
An optimized veto strategy has been developed to obtain a $<3\%$ background to signal ratio for $^9$Li/$^8$He and fast neutron backgrounds.
The veto strategy utilizes the correlation time and distance to the parent muon, the muon track and energy deposition, the pulse shapes and the energies of prompt-delayed event pairs.
Geoneutrinos, produced by U/Th in the Earth, have the same signatures as the reactor antineutrinos.
Features of the energy spectrum shapes in the region of $<3$~MeV help to separate the geoneutrino components with a fraction of $\sim 2\%$.
The ratio of the residual backgrounds to the IBD signal, including geoneutrinos, accidentals, $^9$Li/$^8$He, fast neutrons and ${}^{13}$C($\alpha$,n)${}^{16}$O is estimated to be about 6\% using the same method as Ref.~\cite{yellowbook}.

The JUNO detector requires an energy non-linearity uncertainty of better than 1\% and a $3\%/\sqrt{E}$ effective energy resolution to determine the neutrino mass ordering.
The relation between the true energy and the detected energy is non-linear due to quenching effects and Cherenkov light emission. The PMT instrumentation and readout electronics may contribute additional non-linearity for each channel.
Various calibration sources covering most of the IBD energy range will be deployed regularly to calibrate the energy scale to a sub-percent level~\cite{Abusleme:2020lur}.
The novel dual calorimetry with 20-inch and 3-inch PMTs enables a clean determination of the instrumental non-linearity.

As shown in Fig.~\ref{fig:oscillation}, the NMO sensitivity relies on the difference of the multiple small oscillation pattern driven by $\Delta m^2_{31}$ in normal and inverted mass ordering cases.
Precise measurement of the oscillation pattern driven by $\Delta m^2_{31}$ requires unprecedented effective energy resolution.
Otherwise, the small oscillation pattern will be washed out.
The unprecedented energy resolution puts stringent requirements on the transparency of the scintillator and the detection efficiency of PMTs. With the R\&D achievements described in this article, a yield of 1345 p.e./MeV at the detector center is obtained in simulations based on the nominal detector parameters. This contributes a statistical term of 2.73\% in the energy resolution. The p.e.\ yield in the detector is position-dependent. A multi-positional source deployment calibration strategy is devolved to correct the non-uniformity. In general, the fractional energy resolution for a visible energy $E_{\rm vis}$ can be written as an approximate formula
\begin{equation}\label{eqn:approximate-energy-resolution}
\frac{\sigma_{E_{\rm vis}}}{E_{\rm vis}} = \sqrt{ \left( \frac{a}{\sqrt{E_{\rm vis}}} \right)^2 + b^2 +
\left( \frac{c}{E_{\rm vis}} \right)^2 } \,,
\end{equation}
where the $a$ term is the statistical term driven by photostatistics, the $b$ term is dominated by the position non-uniformity, and the $c$ term represents the contribution of background noises. For the neutrino mass ordering determination, it was found that the impact of the $b$ term is 1.6 times larger than that of the $a$ term, and the impact of the $c$ term is 1.6 times smaller than that of the $a$ term~\cite{yellowbook}. Therefore, an effective energy resolution can be defined as  $\sigma_{\rm eff}/E=a_{\rm eff}/\sqrt{E{\rm (MeV)}}$, with
\begin{equation}\label{eqn:effective-energy-resolution}
a_{\rm eff} \equiv \sqrt{a^2 + (1.6\times b)^2 + \left( \frac{c}{1.6} \right)^2} \,.
\end{equation}
Detailed studies on the effective energy resolution that could be achieved with the JUNO detector calibration strategy can be found in Ref.~\cite{Abusleme:2020lur}, taking into account the non-uniformity, vertex smearing, PMT quantum efficiency variation and charge resolution, energy nonlinearity, and PMT noises, etc. For the nominal setup, the $a$ term is 2.61\% (the 2.73\% mentioned above corresponds to the statistics at the detector center while the p.e.\ yield increases with radius of the vertex), the $b$ term is 0.82\%, and the $c$ term is 1.23\%. Thus, an effective energy resolution of $3.02\%/\sqrt{E{\rm (MeV)}}$ is expected for the JUNO MO determination in simulations.

As a multiple-purpose detector, JUNO is sensitive to a range of neutrino sources beyond the reactor antineutrinos. The expected signal and background estimates for various researches and the corresponding physics potentials will be reviewed in Section~\ref{sec:physics}. The expected neutrino signal rates and major background sources are summarized in Tab.~\ref{tab:signalbackground}. Only the MO determination requires a 3\% energy resolution. The $|\Delta m^2_{32}|$ measurement benefits moderately from a high energy resolution. All other studies are not sensitive to the energy resolution. Solar neutrino studies require a U/Th radiopurity of the LS of $1\times10^{-17}$~g/g. Reactor and Geoneutrino studies require $1\times10^{-15}$~g/g and other studies are not sensitive.

\begin{table}[htb]
\centering
\begin{tabular}{|c|c|c|c|}
  \hline
  Research & Expected signal & Energy region & Major backgrounds \\ \hline
  Reactor antineutrino & 60~IBDs/day & 0--12~MeV &  Radioactivity, cosmic muon \\
  Supernova burst & 5000 IBDs at 10~kpc & 0--80~MeV & Negligible \\
  & 2300 elastic scattering & & \\
  DSNB (w/o PSD) & 2--4 IBDs/year & 10--40~MeV & Atmospheric $\nu$ \\
  Solar neutrino & hundreds per year for $^8$B & 0--16~MeV & Radioactivity\\
  Atmospheric neutrino & hundreds per year & 0.1--100~GeV & Negligible\\
  Geoneutrino & $\sim 400$ per year & 0--3~MeV & Reactor $\nu$ \\
  \hline
\end{tabular}
\caption{Summary of detectable neutrino signals in the JUNO experiment and the expected signal rates and major background sources. \label{tab:signalbackground}}
\end{table}

\section{Physics with JUNO}
\label{sec:physics}

\subsection{Neutrino mass ordering}
\label{subsec:nmo}

JUNO determines the NMO using the oscillation interplay between $\Delta m^2_{31}$ and $\Delta m^2_{32}$ at a medium reactor baseline ($\sim$53~km).
The reactor antineutrino survival probability $P_{\bar{\nu}_e\to\bar{\nu}_e}$ is shown in Eq.~\ref{eq:P_nue2nue}.
Although the NMO can be determined as well with long-baseline accelerator or atmospheric neutrino experiments, JUNO is unique since its sensitivity is based on the vacuum oscillations while accelerator and atmospheric experiments rely on the NMO dependence of matter effects.
The JUNO NMO sensitivity has no dependence on the unknown CP-violating phase and the $\theta_{23}$ octant, adding unique information when combined with other neutrino experiments.

At a baseline of 53~km, JUNO will simultaneously measure oscillations driven by the small mass splitting ($\Delta m^2_{21}$) and the large mass splitting ($\Delta m^2_{31}$ and $\Delta m^2_{32}$) as shown in Fig.~\ref{fig:oscillation}. The small oscillation peaks in the oscillated antineutrino spectrum contain the NMO information. Precise measurement of the oscillated antineutrino spectrum is a key for JUNO to determine NMO. This requires a 20~kton liquid scintillator detector with an unprecedented effective energy resolution of 3\%$/\sqrt{E({\mathrm{MeV}})}$.

JUNO detects the reactor antineutrino signal via IBD reaction, $\bar{\nu}_e +p \rightarrow e^+ + n $.
The reactor antineutrino interacts with a proton, creating a positron ($e^+$) and a neutron.
The positron quickly deposits its energy and annihilates into two 0.511-MeV $\gamma$-rays, which gives a prompt signal.
The neutron scatters in the detector until thermalization.
It is then captured on a proton (99\%) or carbon (1\%) within an average time of $\sim\,$200~$\mu$s and produces gammas of 2.2~MeV or 4.95~MeV.
The coincidence of the prompt-delayed signal pair in such a short time significantly reduces backgrounds.
The observable neutrino energy spectrum can be approximately obtained from the prompt signal with a shift of $\sim\,$0.8~MeV.

\def\alphan{\mbox{$^{13}$C($\alpha$, n)$^{16}$O}}
Accidental coincidence background, $^8$He/$^9$Li, fast neutrons and \alphan{} reactions are the major backgrounds for the reactor neutrino program in JUNO.
A fiducial volume cut can significantly reduce the accidental and the \alphan{} backgrounds.
Energy selection, time coincidence, and vertex correlation of the prompt and delayed signals are used in the antineutrino selection to further suppress the accidental background.
To reject the cosmogenic backgrounds such as $^{9}$Li/$^{8}$He and fast neutrons, muon veto cuts need to be optimized to maximize the detector live time and minimize the dead volume losses.
A set of preliminary selection criteria is studied in the JUNO Yellow Book~\cite{yellowbook}.
Tab.~\ref{tab:mh:sigbkg} summarizes the efficiencies of the selection cuts and the corresponding reductions for various backgrounds. JUNO will observe 60 IBD events per day, with about 6\% background contamination.
\begin{table}[htb]
\centering
\begin{tabular}{|c|c|c|c|c|c|c|c|}\hline\hline
Selection & IBD efficiency & IBD & Geo-$\nu$s & Accidental & $^9$Li/$^8$He & Fast $n$ & $(\alpha, n)$ \\ \hline
- & - & 83 & 1.5 & - & 84 & - & - \\ \hline
Fiducial volume & 91.8\% & 76 & 1.4 &  & 77 & 0.1 & 0.05 \\ \cline{1-4}\cline{6-6}
Energy cut & 97.8\% & & & 410 &  &  &  \\ \cline{1-2}
Time cut & 99.1\% & 73 & 1.3 &  & 71 &  &  \\ \cline{1-2}\cline{5-5}
Vertex cut & 98.7\% & & & 1.1 &  &  &  \\ \cline{1-6}
Muon veto & 83\% & 60 & 1.1 & 0.9  & 1.6 &  &  \\ \hline
Combined & 73\% & 60  & \multicolumn{5}{c|}{3.75} \\ \hline
\hline
\end{tabular}
\caption{The efficiencies of antineutrino selection cuts, signal and backgrounds rates, taken from Ref.~\cite{yellowbook}.
\label{tab:mh:sigbkg}}
\end{table}

To obtain the NMO sensitivity, we employ the least-squares method and construct a $\chi^2$ function as
\begin{equation}
\chi^2_{\mathrm{REA}}=\sum^{N_{\mathrm{bin}}}_{i=1}\frac{[M_{i} -
T_{i}(1+\sum_k \alpha_{ik}\epsilon_{k})]^2}{M_{i}} +
\sum_k\frac{\epsilon^2_{k}}{\sigma^2_k}\,,\label{eq:mh:chiREA}
\end{equation}
where $M_{i}$ is the number of measured neutrino events in the $i$-th energy bin, $T_{i}$ is the predicted number of neutrino events with oscillations, $\sigma_k$ is a systematic uncertainty, $\epsilon_{k}$ is the corresponding pull parameter, and $\alpha_{ik}$ is the fraction of neutrino event contribution of the $k$-th pull parameter to the $i$-th energy bin.
The considered systematic uncertainties include the correlated (absolute) reactor uncertainty ($2\%$), the uncorrelated (relative) reactor uncertainty ($0.8\%$), the background uncertainties, the spectral shape uncertainty ($1\%$) and the non-linearity uncertainty.
We fit the spectrum assuming the normal ordering or inverted ordering with the chi-squared method and take the difference of the minima as a measure of the median NMO sensitivity.
The discriminator of the NMO can be defined as
\begin{equation}
\Delta \chi^2_{\mathrm{MO}}=|\chi^2_{\rm min}(\rm NO)-\chi^2_{\rm
min}(\rm IO)|, \label{eq:mh:chisquare}
\end{equation}
where the minimization process is implemented for all the relevant nuisance parameters including oscillation parameters.
With the inputs described above, we obtain $\Delta \chi^2 = 10$, which shows the NMO can be determined with a significance of $3\sigma$ in 6 years of data taking~\cite{yellowbook}.
Several analyses implementing different approaches have yielded consistent results.

Recent results on the measurement of the antineutrino energy spectrum show an anomaly on the spectral shape, known as the 5--MeV bump.
In addition, fine structure~\cite{Sonzogni:2017voo} may exist in the antineutrino energy spectrum.
These findings indicate that the uncertainty in the reactor antineutrino flux model may be underestimated.
The Taishan Antineutrino Observatory (TAO) has been proposed to provide a precise model-independent reference spectrum for JUNO to reduce the flux uncertainty~\cite{junotaocdr}.
The uncertainty of this reference spectrum is estimated to be $\sim1\%$  over most of the energy region, taking into account the statistics with 3 years of data, energy non-linearity effect, energy leakage in a small detector, and fission fraction uncertainties when predicting other reactor cores with the measurement of only one core.

Compared to the assumptions in the Yellow Book, several changes have occurred to the experimental inputs.
\begin{itemize}
\item It has become clear that two of the four planned Taishan reactor cores will not be built in the near term. This will reduce the total reactor power by about 25\%, resulting in a reduction of $\sim\,$2.5 for $\Delta \chi^2$ with 6 years' data.
\item The location of the JUNO experimental hall was shifted by about 60~m to adapt to the underground geological conditions. The cosmic muon flux in the new experimental hall increases by 30\% due to a reduction of the vertical overburden by 58~m. The changes on the baselines and the overburden have negative but negligible impacts on the NMO sensitivity.
\item Recent efforts in optimization of the antineutrino selection show a possibility to minimize the dead time after the muon veto. The live time fraction increases from 83\%  to 93\%, partially compensating for the statistics loss due to less reactor power.
\item All 20-inch PMTs have been procured and tested. The average photon detection efficiency for PMTs in the Central Detector is 29.1\%, comparing to the designed value 27\%  used in Ref.~\cite{yellowbook}. Meanwhile, a more realistic PMT and liquid scintillator optical model has been developed with dedicated studies using a Daya Bay detector~\cite{DYBJUNOLS}, showing a possible photoelectron yield increase and thus an improvement on the energy resolution.
\item Impacts of the energy non-linearity in the measured spectrum have been further studied.
\item Combined analysis of the TAO and JUNO data shows that a moderate improvement could be achieved compared to the simple assumption of 1\% bin-to-bin uncertainty, while the model dependence of the input spectrum~\cite{Forero:2017vrg,Capozzi:2020cxm} can be removed.
\end{itemize}
With all these factors taken into account, the NMO sensitivity is almost the same as that in the Yellow Book.

In addition, as shown in Refs.~\cite{yellowbook,Blennow:2013vta}, electron antineutrino disappearance and muon neutrino (antineutrino) appearance experiments will prefer a different value of $\Delta m_{32}^{2}$ when the wrong NMO hypothesis is tested in a common statistical analysis of the oscillation data.
The corresponding tension provides additional power to exclude the wrong NMO hypothesis and greatly boosts the sensitivity of the NMO measurement.
As a consequence, a combined analysis between JUNO and an accelerator-based or atmospheric-based NMO experiment will reach much better sensitivity than it may be expected from a simple sum of $\Delta \chi^{2}$.
The boosted $\Delta \chi^{2}$ due to the combination effect of different types of experiments relies on the precisions of $\Delta m_{31}^{2}$ (or $\Delta m_{32}^{2}$) from individual experiments.
The determination of $\Delta m_{32}^{2}$ with a high precision of 0.6\% or better (see Section~\ref{subsec:precision}) in JUNO experiment via reactor antineutrinos provides unique information in the NMO combined analysis.
Efforts towards the combined analysis of JUNO+PINGU+ORCA~\cite{Bezerra:2019dao} and JUNO+T2K+NO$\nu$A are also on-going.

\subsection{Precision measurement of neutrino oscillation parameters}
\label{subsec:precision}

As illustrated in Fig.~\ref{fig:oscillation}, JUNO will be the first experiment to observe the effects of the so-called solar and atmospheric oscillations simultaneously. The former, driven by $\Delta m^2_{21}$ and modulated by $\sin^2 2\theta_{12}$, cause the slow (low frequency) oscillation that is responsible for the bulk of the reactor antineutrino disappearance. The latter, driven by $\Delta m^2_{32}$ and modulated by $\sin^2 2\theta_{13}$, cause the higher frequency oscillations riding on top of the slower one. Measuring the oscillated reactor antineutrino spectrum with a resolution of $3\%$ at 1~MeV will enable determining these 4 parameters simultaneously.

An estimate of JUNO's sensitivity to these 4 oscillation parameters was previously reported in Ref.~\cite{yellowbook}. A new study is ongoing that incorporates several important updates, namely:
\begin{itemize}
    \item Updated baselines and cosmogenic backgrounds that reflect the slight adjustments made to the underground cavern's position during construction.
    \item An IBD cross-section that incorporates all radiative corrections and neutron recoil effects.
    \item A reactor antineutrino shape uncertainty consistent with the expectation from the TAO experiment.
    \item A data-driven scintillator non-linearity model inspired from the experience in the Daya Bay experiment~\cite{Adey:2019zfo}.
    \item The updated energy resolution from Ref.~\cite{Abusleme:2020lur}.
    \item The use of JUNO's full Monte Carlo simulation to model detector response effects such as non-uniformities and leakage.
    \item The updated reactor configuration of Tab.~\ref{tab:intro:NPP}.
\end{itemize}

The sensitivity to the oscillation parameters is being independently estimated by several methods relying on different statistical approaches to treat the uncertainties, such as pull parameters and covariance matrices, all of which yield consistent results. The final values of the sensitivities are still being finalized and will be the focus of a separate publication to be released soon. While the precision on $\sin^2 2\theta_{13}$ will not exceed that of existing measurements~\cite{Tanabashi:2018oca}, the $\sin^2 2\theta_{12}$, $\Delta m^2_{21}$, and $|\Delta m^2_{32}|$ parameters will be determined to $0.6\%$ or better after 6 years of exposure, greatly improving over current estimates.

JUNO is currently the only experiment on the horizon that will determine some of these parameters with such high precision. Moreover, its unique approach using medium-baseline reactor antineutrino oscillations will feature very different systematic uncertainties to those of other experiments, most notably accelerator~\cite{Abe:2011ts, Acero:2019ksn, Abe:2018wpn, DUNE_TDR_vol2} and solar~\cite{Aharmim:2011vm,Abe:2016nxk} experiments. As powerful discriminators of neutrino masses and mixing models, constraints to other experiments, and handles to probe the 3-neutrino flavor paradigm well beyond current limits, these estimates will thus provide very valuable input to the community for the foreseeable future.

\subsection{Supernova neutrinos and trigger strategy}
\label{subsec:supernova}

Neutrinos are crucial players during all stages of stellar collapse and explosion. The detection of two dozen neutrino events from Supernova (SN) 1987A in Kamiokande-II~\cite{Hirata:1987hu}, Irvine-Michigan-Brookhaven~\cite{Bionta:1987qt}, and Baksan~\cite{Alekseev:1988gp} experiments, has essentially confirmed the scenario of the delayed neutrino-driven explosion mechanism for core-collapse SNe~\cite{Colgate:1966ax, Bethe:1990mw, Woosley:2002zz, Janka:2006fh, Janka:2017vcp}. However, with the poor statistics from SN 1987A, it is impossible to determine supernova and neutrino parameters, although some hints can be obtained (see, e.g., Ref.~\cite{Vissani:2014doa}, for an overview on the statistical analysis of neutrino data from SN 1987A). Future large water Cherenkov, liquid scintillator, and liquid argon time projection chamber neutrino detectors (e.g., Hyper-Kamiokande, JUNO, and DUNE) are all able to register large statistics of SN neutrinos and have great potential to provide complete flavor information on SN neutrinos.
Meanwhile, numerical models have recently been advanced in better predictions of the measurable SN neutrinos features, see recent reviews Refs.~\cite{Mirizzi:2015eza,Janka:2017vlw}.

Theoretical predictions of the SNe neutrino-emission depict three main phases, namely the shock-breakout burst phase, post-bounce accretion phase and the proto-neutron star cooling phase. It is highly desirable to obtain a clear observation of these discriminative phases that correspond to the dynamical evolution stages of stellar collapse and explosion. The total energy, luminosity evolution, spectral distribution, and the mix of different flavors of the SN neutrinos carry the information of the hydrodynamic conditions and dynamical processes, the characteristics of the progenitor star and its compact remnant. Moreover, before a massive star collapses and forms an SN, i.e., during its silicon burning, a significant number of MeV-energy neutrinos can be produced via thermal processes and nuclear weak interactions. These pre-supernova neutrinos are also useful to investigate late-stage stellar evolution~\cite{Kato:2015faa,Yoshida:2016imf}.

\begin{figure}
  \centering
  \includegraphics[width=10cm]{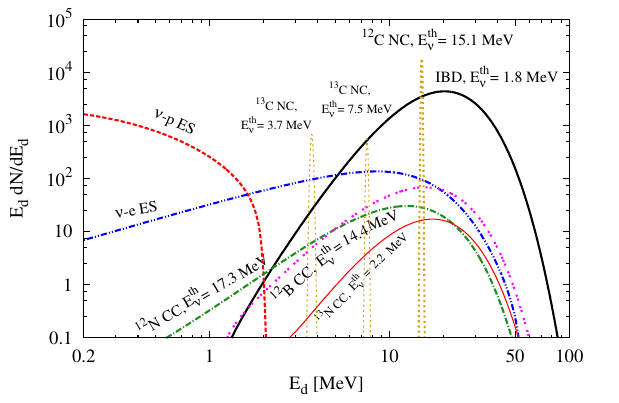}\\
  \caption{The neutrino event spectra with respect to the visible energy $E_d$ in the JUNO detector for a typical SN at 10 kpc. (Image remade from Ref.~\cite{yellowbook})}\label{fig:SNspectra}
\end{figure}

JUNO, with 20 kt LS, has excellent capability of detecting all flavors of the $\mathcal{O}$(10 MeV) postshock neutrinos. The multichannel detection is mainly via the inverse beta decay (IBD), $\overline{\nu}^{}_e + p \to e^+ + n$, the elastic neutrino-electron scattering ($e$ES), $\nu + e^- \to \nu + e^-$, and the elastic neutrino-proton scattering ($p$ES), $\nu + p \to \nu + p$.
For a typical galactic distance of 10 kpc and typical SN parameters, JUNO will register $\sim$5000 IBD events, $\sim$300 $e$ES events and $\sim$2000 $p$ES events. The charged-current (CC) and neutral-current (NC) interactions of neutrinos on $^{\rm 12}$C nuclei are observable as well~\cite{Lu:2016ipr}, and they will register $\sim$200 events and $\sim$300 events, respectively. The $^{\rm 12}$B and $^{\rm 12}$N in the final states of CC interactions on $^{\rm 12}$C are $\beta^-$-emitters with a 20.2 ms and 11 ms half-life, respectively, leading to a prompt-delayed coincident signal. The visible energy spectra caused by the six main reaction channels are shown in Fig.~\ref{fig:SNspectra}.
Note that the available numerical models in the community may predict quite different signal statistics. The models from the Garching group~\cite{garching} and Japan group~\cite{Nakazato:2012qf} predict $\mathcal{O}$(5k) IBD events at JUNO.
The energy spectra of $\overline\nu_e$, $\nu_e$, and $\nu_x$ (i.e., $\nu_\mu$, $\nu_\tau$, and $\overline\nu_\mu$, $\overline\nu_\tau$) are of crucial importance to understand the microscopic physics of SN explosions.
A unique strength of JUNO is the possibility to provide flux and spectral information for the $\nu_x$ flavors based on the NC $^{\rm 12}$C and $p$ES channels, respectively, as first pointed out in Ref.~\cite{Beacom:2002hs}.
In fact, JUNO has the capability to provide flux and spectral information for all flavors, as demonstrated by a model-independent approach~\cite{Li:2017dbg,Li:2019qxi}.

The $\mathcal{O}$(1 MeV) pre-supernova neutrinos could provide a unique and independent early warning for the optical observations of core-collapse SNe. Attributing to the low detection threshold $\mathcal{O}$(100 keV)~\cite{Fang:2019lej}, pre-supernova neutrinos are detectable via the IBD and $e$ES channels. However, the $e$ES channel will suffer from the contaminations by natural radioactivity and cosmogenic products, whereas the IBD channel is less affected. Therefore, the pre-supernova neutrinos can be probed if the accumulated IBD rate significantly exceeds the background level within a certain time window. Limited pointing information becomes available once sufficient IBD events are accumulated~\cite{Li:2020gaz, Mukhopadhyay:2020ubs}.

The JUNO front-end-electronics~\cite{Pedretti:2018xjc} will feature real-time waveform processing by FPGAs (charge reconstruction and timestamp tagging) and a 2 GB DDR3 memory shared by three PMTs. The processed signal will be sent to the data acquisition (DAQ) in triggerless mode, while the raw waveforms will be sent to DAQ once validated by the global trigger electronics. This configuration limits the chance of data loss even in the face of the very high event rate expected for a nearby supernova, because the supernova signals can be temporarily stored in the 2~GB memory. It can completely handle the Supernova explosion beyond 0.5 kpc (more than 2~M events). If a Supernova explosion is even closer, the waveform processing by FPGA may start to have dead time and the memory gets overflow. Furthermore, a dedicated multi-messenger trigger system is under design to achieve an ultra-low detection threshold of $\mathcal{O}$(10 keV). This system will enable JUNO to act as a powerful transient machine for broad-band multi-messenger observation and connect to the global network of multi-messenger observatories. JUNO can be expected to become a major player in the next-generation Supernova Early Warning System (SNEWS2.0)~\cite{Kharusi:2020ovw} for multi-messenger astronomy.

No SN neutrino burst has been observed since SN 1987A, and a recent analysis that combines several independent studies yields an expected rate of $1.63 \pm 0.46$ core-collapse SN per century~\cite{Rozwadowska:2021lll}. The large neutrino detectors like SuperK-Gd/JUNO/Hyper-K/DUNE will be online for several decades, so it seems likely that we will obtain a high-statistics measurement of the neutrino signal from at least one galactic core-collapse SN in the next few decades.
This will not only enable a deep understanding of the explosion mechanism, but also probe the intrinsic properties of the neutrino themselves, e.g., constrain the absolute scale of neutrino masses~\cite{Lu:2014zma}. Supernova neutrino emission is predicted to be very variable. The detection of the integrated signal from all past stellar core-collapse SNe, namely the diffuse SN neutrino background, will improve understanding of the average SN neutrino signal and the underlying cosmology, as discussed in the following section.

\subsection{Diffuse supernova neutrino background}
\label{subsec:dsnb}

While core-collapse Supernovae (SNe) in our own galaxy are rare events, they frequently occur throughout the visible Universe, sending bursts of neutrinos in the direction of the Earth. They all contribute to a low background flux of low-energy neutrinos, the so-called Diffuse Supernova Neutrino Background (DSNB) on the level of $\sim 10 \nu$ cm$^{-2}$s$^{-1}$ (e.g.~\cite{Beacom:2010kk}). Its exact flux and spectrum bear information on the red-shift dependent supernova rate, average SN neutrino energy spectrum and the fraction of black hole formation in core-collapse SNe.

JUNO is in an excellent position to detect the $\bar\nu_e$ component of the DSNB flux. Depending on the DSNB model, we expect about 2$-$4 IBD events per year in the energy range above the reactor $\bar\nu_e$ signal. Given the high light yield, the delayed signal from neutron capture on hydrogen in the LS offers a very efficient tag for background reduction, while pulse-shape discrimination helps to suppress the background from atmospheric-neutrino NC interactions. This will provide JUNO with a DSNB sensitivity competitive to the current Super-Kamiokande+Gadolinium phase.

\subsubsection{DSNB signal}

The energy spectrum of DSNB events in JUNO is given by
\begin{equation}
\dv{N_{\nu}}{E_{\nu}} = N_p \times \sigma_{\nu}(E_{\nu}) \times c \int_0^{\infty} \dv{N(E'_{\nu})}{E'_{\nu}} \times \dv{E'_{\nu}}{E_{\nu}} \times R_\mathrm{SN}(z) \times \left|\dv{t}{z}\right| \dd z,
\label{Eq:DSNBSpectrum} \end{equation}
where $N _p=7.16\times 10^{31}/(10\mbox{ }{\rm kton})$ is the number of protons in the JUNO target, $\sigma_{\nu}(E_{\nu})$ is the energy-dependent cross-section for the IBD reaction \cite{Strumia:2003zx}, and the last term represents the differential DSNB flux. It is computed via a line-of-sight integral of the average SN neutrino spectrum $\dv{ N}{E'_{\nu}}$ (weighted by an initial mass function), multiplied by the core-collapse SN rate $R_\mathrm{SN}(z)$ that evolves over the cosmic history. To take into account the effects of cosmic expansion, $E'_{\nu}$ denotes the neutrino energy at emission, while $E_{\nu}=E'_{\nu}/(1+z)$ is the red-shifted neutrino energy upon detection. The term $|\dv{t}{z}|$ accounts for the expansion history of the Universe and relates $z$ to the cosmic time $t$.

\begin{figure}[htp]
\centering
\includegraphics[width=0.4\textwidth]{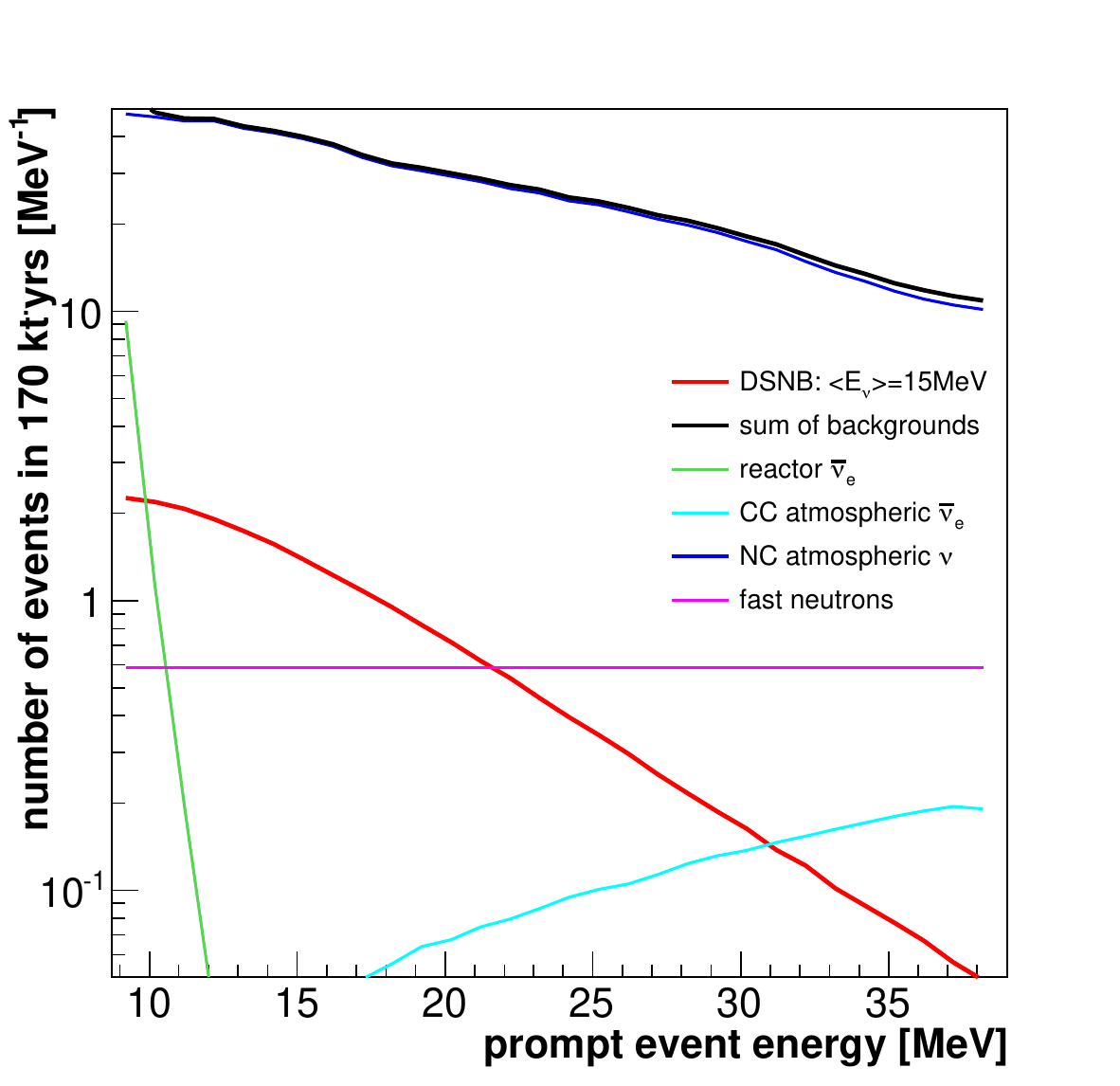}
\hspace{1cm}
\includegraphics[width=0.4\textwidth]{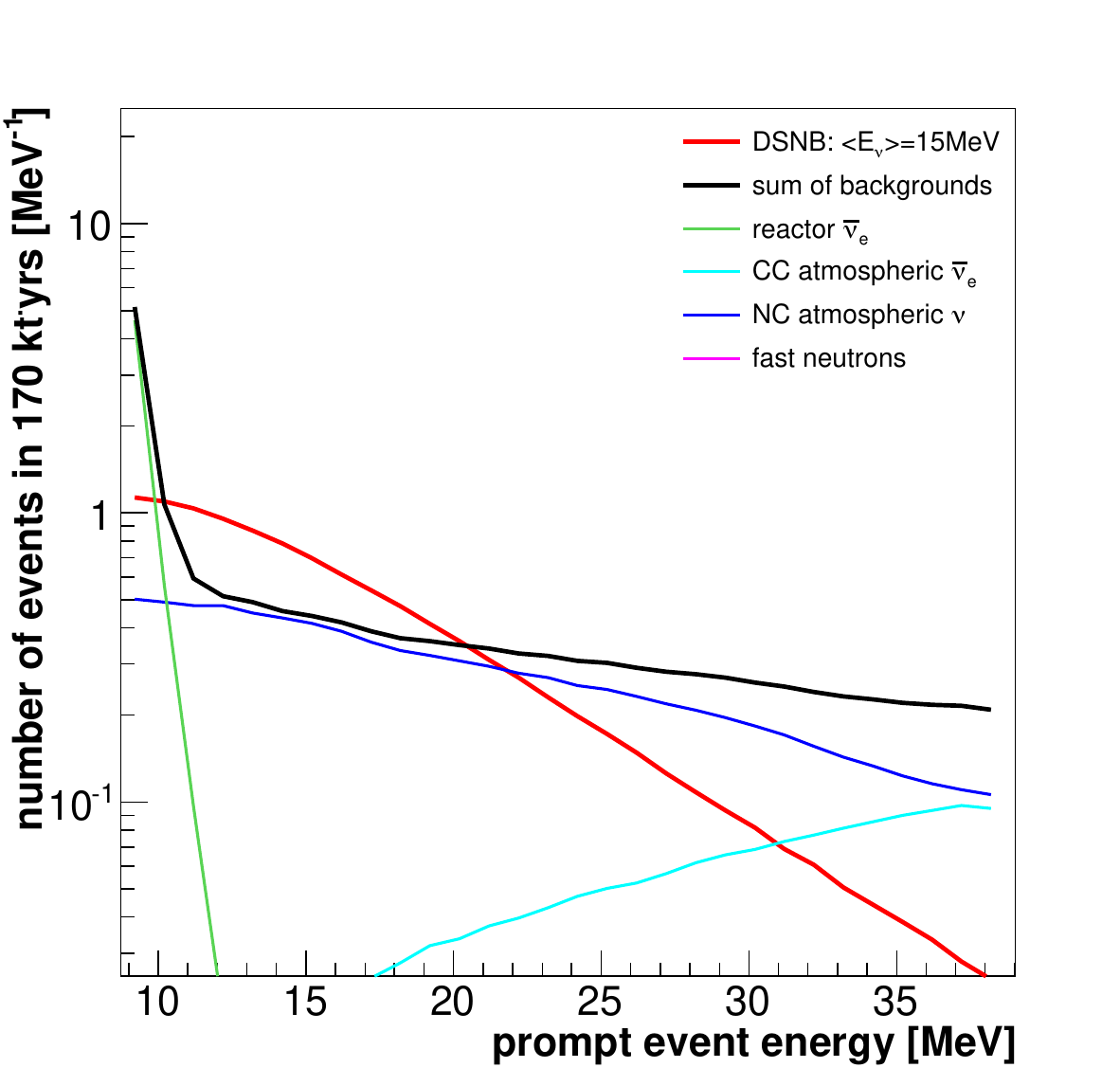}
\caption{Prompt ($e^+$) energy spectra for the DSNB signal ($\langle E_\nu\rangle=15$ MeV) and backgrounds as expected for JUNO. {\it Left:} After basic selection cuts, the background from atmospheric-neutrino NC reactions dominates over the whole range of the observation window from 10 to 30\,MeV.  {\it Right:}  When applying pulse shape discrimination, atmospheric neutrino NC and fast neutron backgrounds are greatly reduced. The DSNB signal dominates in the range from 11 to $\sim$22\,MeV.}
\label{fig:dsnb}
\end{figure}

For the sake of simplicity, we assume the underlying SN neutrino spectra to follow a Maxwell-Boltzmann distribution. The expected DSNB rate and spectra largely depend on the expected mean energy $\langle E_\nu\rangle$. The left panel of Fig.~\ref{fig:dsnb} displays an exemplary event spectrum for $\langle E_\nu\rangle=15$~MeV. Event numbers are plotted as a function of the detected prompt (i.e.~positron) event energy and for 10 years of observation in a 17-kton fiducial volume (see below). Tab.~\ref{tab:dsnb} displays the expected range of the DSNB event rate depending on the average $\langle E_\nu\rangle$.

\subsubsection{Background rejection}

 A variety of backgrounds besets the DSNB signal. The terrestrial flux of $\bar{\nu}_e$'s from reactors and atmospheric neutrinos causes an irreducible background and reduces the DSNB observation window to the range of about 10 to 30\,MeV.

 Secondly, cosmic muons are penetrating the detector. In spallation processes, these muons will generate $\beta/n$-emitting isotopes (mainly $^9$Li) in the target material as well as fast neutrons when passing through the rock surrounding the detector. Both can mimic the IBD signature but are effectively reduced by a coincidence veto in the wake of a through-going muon and a fiducial volume cut, respectively.

 The most critical background for the DSNB search in LS detectors is created by neutral-current (NC) interactions of atmospheric neutrinos with $^{12}$C~\cite{Collaboration:2011jza, Mollenberg:2014pwa}, where the emission of one neutron together with a prompt energy deposit can mimic the signature of IBD signals. To calculate the exclusive final-state distribution of NC events, both the total cross-sections and residual nuclei's deexcitation processes have to be regarded. For this, the neutrino interaction generator GENIE and the nuclear deexcitation tool TALYS have been used~\cite{Cheng:2020aaw}. The corresponding event rates in the observation window are listed in the first column of Tab.~\ref{tab:dsnb}. The NC background rate surpasses the DSNB signal by more than one order of magnitude. This is true for the whole spectral range depicted in Fig.~\ref{fig:dsnb}.

However, the intrinsic pulse-shape discrimination (PSD) capabilities of liquid scintillators can be used to suppress the NC background to an acceptable level. Here, we assume the PSD efficiencies as in the JUNO physics paper~\cite{Sisti:2020jie}, where 50\% of the DSNB signal and 1.1\% of the NC background will survive after the PSD selection. The effect of the PSD on the signal and background rates inside the observation window can be appreciated in the second column of Tab.~\ref{tab:dsnb}. where a signal-to-background ratio $\gtrsim1$ is achieved for all DSNB models considered. The corresponding prompt energy spectra are shown in the right panel of Fig.~\ref{fig:dsnb}. The DSNB signal dominates in the range from 11 to $\sim$22\,MeV. Note that the residual atmospheric NC background rate may be constrained {\it{in situ}} by measuring the radioactive decays of final-state nuclei, bringing down the uncertainty on the background rate to a 10\% level~\cite{Cheng:2020oko}.

\begin{table}
\begin{center}
	\begin{tabular}{ll|cc}
		\hline
		 \multicolumn{2}{l|}{}& \multicolumn{2}{c}{Rate per 170\,kton$\cdot$yrs}\\
		\multicolumn{2}{l|}{Source} & w/o PSD & w/ PSD\\
		\hline
		DSNB signal  &$\langle E_{\bar{\nu}_e} \rangle$ = 12 MeV&  13 & 7 \\
		      &$\langle E_{\bar{\nu}_e} \rangle$ = 15 MeV & 23 & 12 \\
		      &$\langle E_{\bar{\nu}_e} \rangle$ = 18 MeV & 33 & 16\\
		      &$\langle E_{\bar{\nu}_e} \rangle$ = 21 MeV & 39 & 19\\
\hline
		Background & Reactor $\bar{\nu}_e$'s  &0.3  & 0.13 \\
		    & Atmospheric CC &1.3  & 0.7\\
		    & Atmospheric NC &6$\times10^{2}$ & 6.2 \\
		    & Fast neutrons & 11 &0.14 \\
		\hline
	\end{tabular}
	\caption{Signal and background rates in the observation window (10$-$30\,MeV), assuming an exposure of 170\,kton$\cdot$yrs. The first column shows raw event rates, the second the residual rates after background rejection by pulse shape discrimination (PSD). 	\label{tab:dsnb}}
\end{center}
\end{table}

\begin{figure}[htbp]
    \centering
    \includegraphics[width=0.5\textwidth]{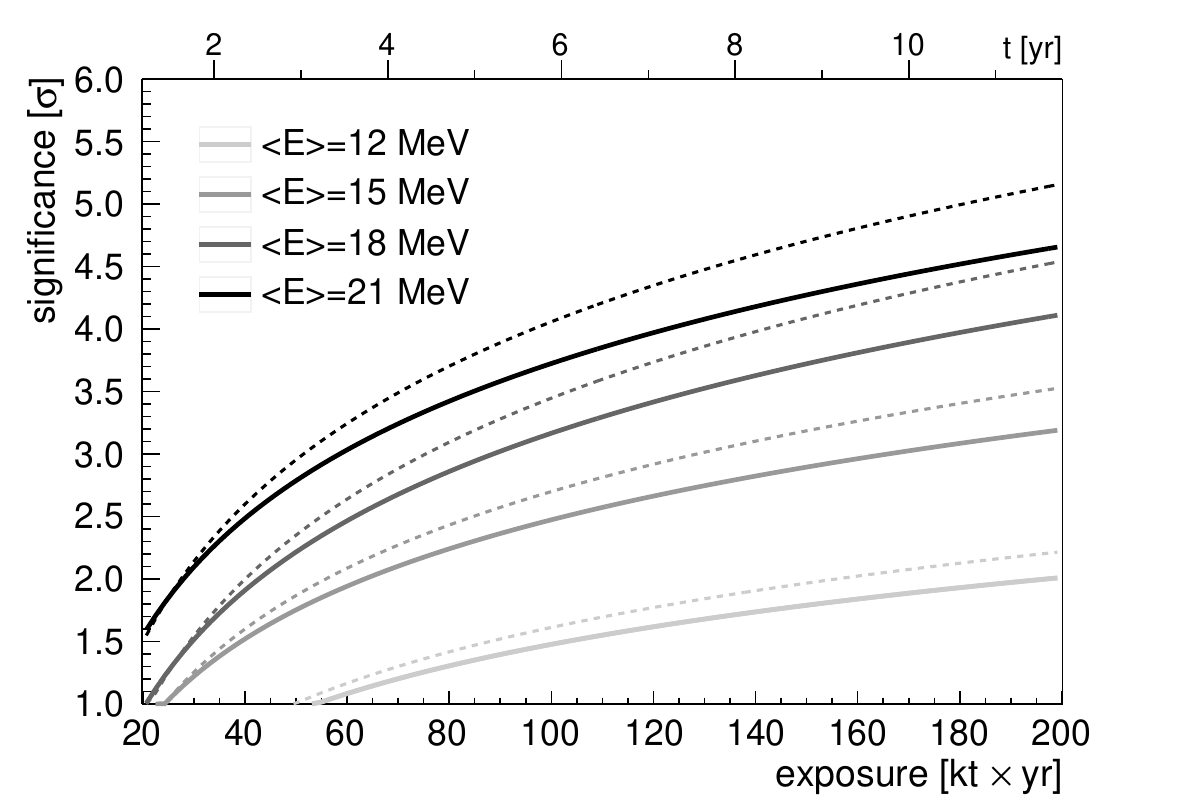}
    \caption{Sensitivity to the DSNB signal vs.~exposure, depending on mean spectral energy and background uncertainty for which 20\% (10\%) is assumed for solid (dashed) lines. }
    \label{fig:dsnb_sensitivity}
\end{figure}

\subsubsection{DSNB sensitivity}

We determine the sensitivity of JUNO for a positive DSNB detection based on an event counting analysis inside the observation window (i.e.~neglecting spectral information). The corresponding event numbers for 10 years of data taking are summarized in the second column of Tab.~\ref{tab:dsnb}. Fig.~\ref{fig:dsnb_sensitivity} illustrates the sensitivity as a function of exposure. The sensitivity depends on the uncertainty of the background rate assumed for the analysis. We expect a level of 10\% based on delayed tagging (see above) but show as well a more conservative 20\%. Sensitivity depends as well greatly on the mean spectral energy $\langle E_{\bar{\nu}_e} \rangle$ and such event rate of the DSNB signal. For $\langle E_{\bar{\nu}_e} \rangle \gtrsim 15 $ MeV, JUNO can be expected to provide 3$\sigma$ evidence of the DSNB signal after 10 years.

\subsection{Solar neutrinos}
\label{subsec:solar}

Solar neutrinos are produced in thermonuclear fusion reactions in the solar core and have played a significant role in the development of neutrino physics and astrophysics.
They represent a natural source of neutrinos with the energy spectrum extending up to about 16\,MeV.
They can reveal direct information about the solar core and can be exploited to study fundamental neutrino properties and neutrino interactions with matter.

In the last decade, substantial experimental progresses have been made by
Borexino and Super-Kamiokande via high precision real-time spectroscopy of solar neutrinos based on the elastic scattering~(ES) off electrons.
The energy dependence of the electron neutrino survival probability $P_{ee}$ measured for $pp$, $^7$Be, $pep$, and $^8$B neutrinos at Borexino~\cite{Agostini:2018uly} is compatible with the transition from the so-called vacuum to the matter-dominated region predicted by the MSW mechanism~\cite{Wolfenstein:1977ue,Mikheev:1986gs}.
Most recently, Borexino observed solar neutrinos from the Carbon-Nitrogen-Oxygen~(CNO) fusion cycle~\cite{Agostini:2020mfq}.
The world's largest sample of $^8$B solar neutrinos collected at Super-K made it possible to study in detail the spectral distortion in the transition region of $P_{ee}$, to constrain the neutrino oscillation parameters such as the mixing angle $\theta_{12}$ and the mass-squared difference $\Delta m^2_{21}$, and to measure the day-night asymmetry of the $^8$B neutrino interaction rate~\cite{Abe:2016nxk,SK2020}.
A mild inconsistency is found between the value of $\Delta m^2_{21}$ reported by solar neutrino experiments and the KamLAND experiment.

There are still several open questions in solar neutrinos.
Firstly, in neutrino oscillation physics, the transition region of $P_{ee}$ is not fully explored since it is difficult to reach an analysis threshold for ES electrons to less than 3~(3.5)~MeV in Borexino~(Super-K).
The exact shape of the transition region is not only sensitive to $\Delta m^2_{21}$ value, but can also be used to test new physics, such as the non-standard interactions, the existence of sterile neutrinos, etc~\cite{Maltoni:2015kca}.
In addition, future high precision measurements of $\sin^2\theta_{12}$ and $\Delta m^2_{21}$ based on solar neutrinos can be used to test the consistency of the standard three neutrino framework and to probe new physics beyond the Standard Model.
On the other hand, solar physics, and especially the question of solar metallicity, will profit from more precise flux measurements of all neutrino components~\cite{yellowbook,Serenelli:2012zw}.
However, any improvement of the flux measurement needs significant efforts in the detector design, construction, and data analysis.

The JUNO experiment, using a Borexino-type liquid scintillator detector but with a much larger mass comparable to the Super-K water Cherenkov detector, has the potential to make significant contributions to the understanding of solar neutrinos.
Initial discussions on $^7$Be and $^8$B neutrinos were reported in the Yellow Book~\cite{yellowbook}.
The feasibility and physics potentials of detecting $^8$B neutrinos have been further explored~\cite{Abusleme:2020zyc}.
In this section, we will summarize the recent $^8$B neutrino studies and discuss briefly the detection of $^7$Be and $pp$ neutrinos.
Sensitivity to the $pep$ and CNO solar neutrinos is hindered by the production of $^{11}$C cosmogenic background, since the JUNO overburden is relatively shallow, for example, comparing to the Gran Sasso underground laboratory where Borexino is located.

\subsubsection{Measurement of \texorpdfstring{$^8$B}{8B} solar neutrinos}
\begin{figure}[htb]
\begin{centering}
\includegraphics[width=.48\textwidth]{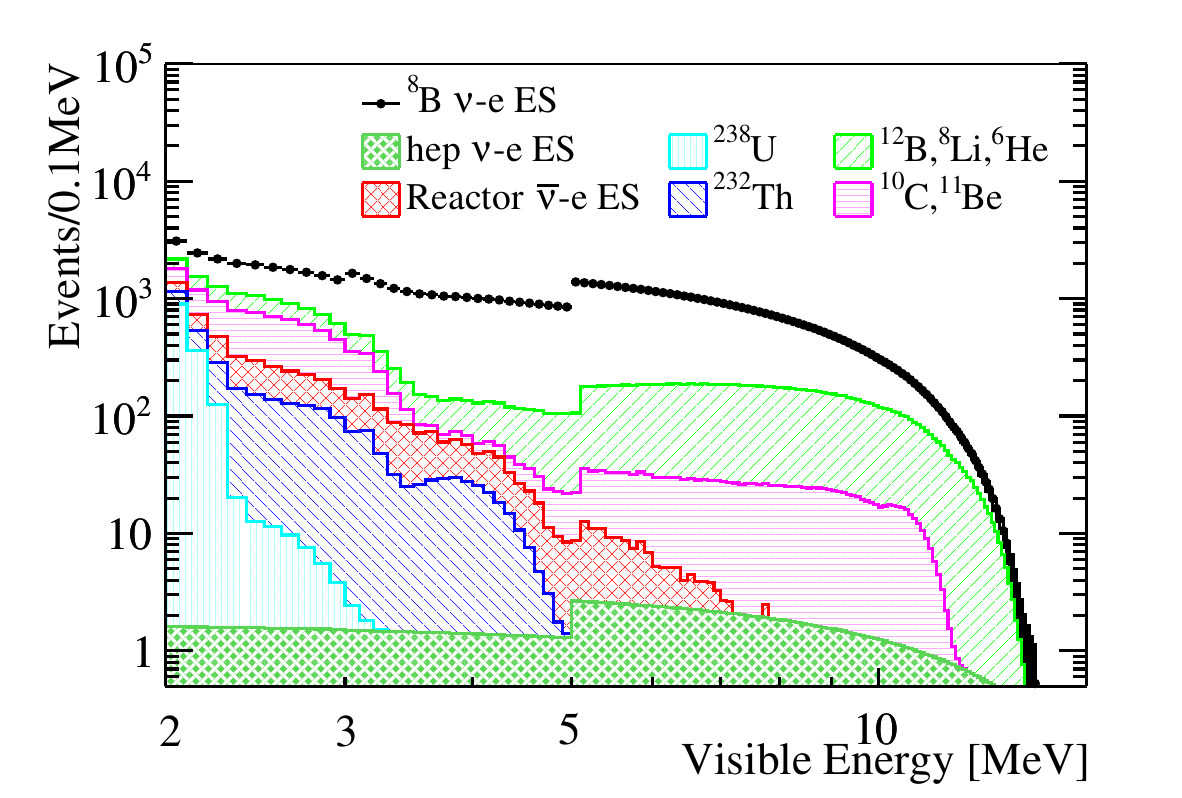}
\includegraphics[width=.42\textwidth]{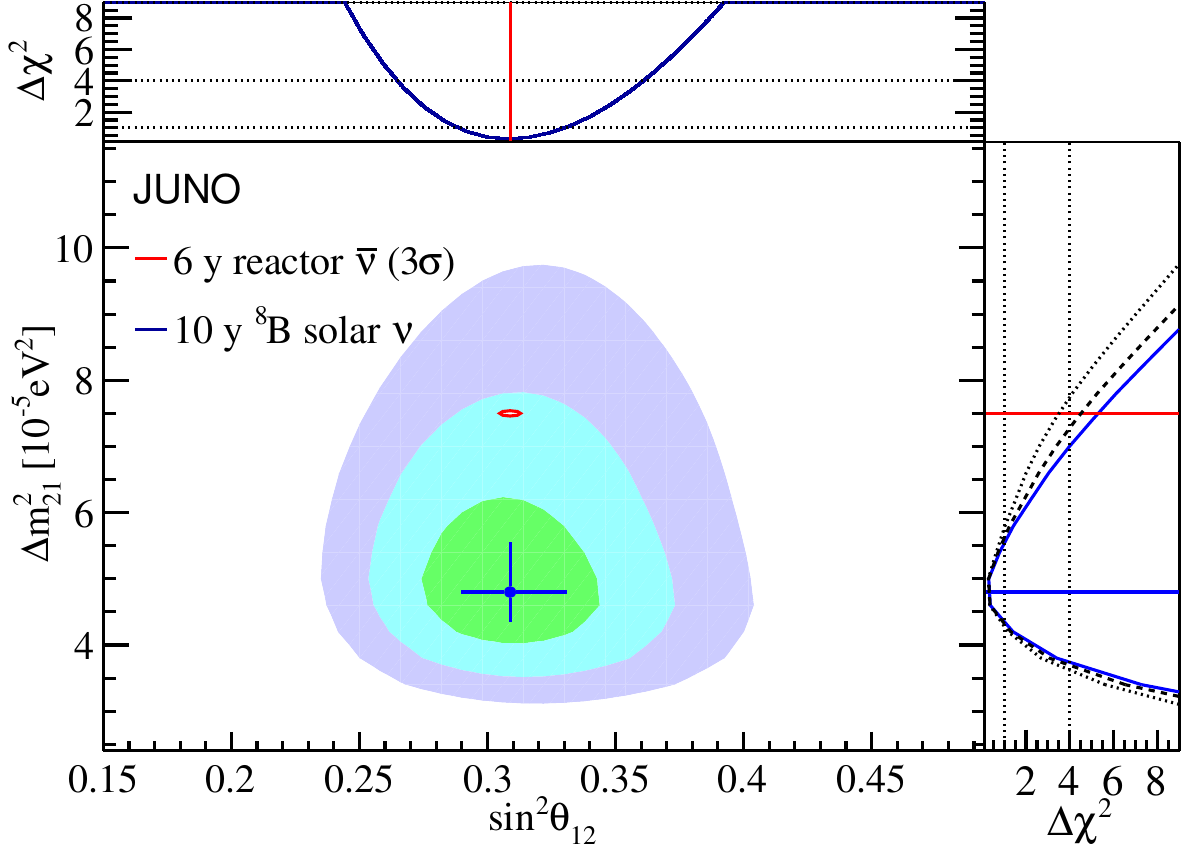}
\caption{\label{fig:B8osci} Left: signal and background spectra above 2\,MeV expected in JUNO in 10 years after the selection cuts. The energy-dependent fiducial volumes account for the discontinuities at~3 MeV and 5~MeV. Right: the expected precision of $\sin^2\theta_{12}$ and $\Delta m^2_{21}$ using $^8$B solar neutrinos and reactor antineutrinos at JUNO. Both plots are taken from Ref.~\cite{Abusleme:2020zyc}}
\end{centering}
\end{figure}

For JUNO as an LS detector, the primary detection channel for solar neutrinos is the elastic scattering off electrons.
The background consists of the intrinsic natural radioactivity in LS, gamma rays from external detector materials, and unstable isotopes produced by cosmic ray muons passing through the detector.
With sufficient shielding, the external background with energies larger than 2~MeV is suppressed to a negligible level.
This permits the reduction of the analysis threshold to 2~MeV in a spherical fiducial volume with a radius of 13~m.
However, the threshold can not be reduced further due to numerous cosmogenic $^{11}$C decays, more than 10,000 per day in the LS.
Above the $^{208}$Tl decay spectrum, the fiducial volume is further enlarged.
Muon-induced backgrounds are well suppressed by cylindrical volume veto cuts along all muon tracks, and spherical volume veto cuts around spallation neutron candidates.
Assuming a 10$^{-17}$ g/g level for the intrinsic $^{238}$U and $^{232}$Th contamination, the signal and background spectra are shown in Fig.~\ref{fig:B8osci}.
In ten years of data taking, after all cuts are applied, about 60,000 signal events and 30,000 background events are expected in the energy range above 2~MeV.
Details are reported in Ref.~\cite{Abusleme:2020zyc}.

Neutrino oscillation physics would benefit from a large low-threshold ES sample of $^8$B neutrinos.
For example, JUNO can set a statistical limit for rejecting the absence of an upturn in the $P_{ee}$ transition region of 5$\sigma$ if $\Delta m^2_{21}=7.5\times10^{-5}$~eV$^2$.
Taking systematic uncertainties into consideration, a $\sim$3$\sigma$ sensitivity is achievable.
Moreover, the expected precision of the day-night asymmetry measurement is 0.9\%, slightly better than the most recent result~(1.1\%) of Super-K~\cite{SK2020}.
In the standard neutrino oscillation framework, a joint fit of the ES reaction rate, the spectral distortion, the day-night asymmetry enables a simultaneous determination of $\sin^2\theta_{12}$ and $\Delta m^2_{21}$, as shown in Fig.~\ref{fig:B8osci}.
Here, the arrival flux of $^8$B neutrinos is constrained to (5.25$\pm$0.20)$\times10^{6}$~cm$^{-2}$s$^{-1}$ according to the NC measurement of SNO~\cite{Aharmim:2011vm}, and the flux of $hep$ neutrinos is fixed to 8.25$\times10^{3}$~cm$^{-2}$s$^{-1}$~\cite{Bahcall:2004pz}.
The $^8$B neutrino spectrum and its uncertainty are taken from Ref.~\cite{Bahcall:1996qv}.
In the fitting results, the precision of $\sin^2\theta_{12}$ is limited by the neutrino flux uncertainty in the SNO measurement.
Although limited by the statistical uncertainty of the day-night asymmetry measurement, the $\Delta m^2_{21}$ precision is similar to the solar global fitting results in Ref.~\cite{Esteban:2018azc}.

In addition to the ES channel, the possible CC and NC channels between neutrinos and carbon nuclei have been explored in Refs.~\cite{FUKUGITA1988139,Suzuki:2012aa,Suzuki:2019cra}.
The primary channels and the expected event numbers for a 200~kt$\cdot$year exposure are listed in Tab.~\ref{table:solarnuCCNC}.
The reactions between neutrinos and $^{12}$C nuclei can not happen due to high energy thresholds.
The 1.1\% natural abundance of $^{13}$C makes it possible to measure the $^8$B neutrino flux via both the CC and NC channels.
There are 9$\times 10^{30}$ $^{13}$C nuclei in a 20~kt LS. Using the cross-sections in Ref.~\cite{Suzuki:2019cra}, about 6,000 CC signals and 3,000 NC signals are expected in 10 years of data-taking.

In the CC reaction, the electron carries most of the neutrino energy, but it should be separated efficiently from the ES electrons.
For reaction 3 in Tab.~\ref{table:solarnuCCNC}, the CC electrons could be distinguished by tagging the subsequent $^{13}$N decays.
In the NC reaction, all neutrino flavors have the same cross section.
For reaction 5, the 3.685~MeV $\gamma$'s from the de-excitation of $^{13}{\rm C}^*$ could form a small peak on the ES electron spectrum.
Benefiting from the excellent energy resolution, the area of the $\gamma$ peak can be obtained from a joint fitting with the continuous spectrum of ES electrons.
The sensitivity would further profit from the efforts of discriminating $\gamma$ and $e^-$ in Ref.~\cite{PIDJUNO}.
Detailed studies will be reported later.

\begin{table}[!ht]
\begin{center}
	\begin{tabular}{c|c|c|c|c|c|c}
	\hline
	&Type & Reaction & Threshold & Final products & $^8$B & $hep$ \\ \hline
	1& ES   & $\nu+e^-\rightarrow\nu+e^-$               & 0           & $e^-$ & 3$\times 10^5$ & 640 \\
	2& CC   & $\nu_e+^{12}{\rm C}\rightarrow e^-+^{12}{\rm N}$ & 16.8~MeV   & $e^-$  & 0 & 0.41 \\
	3& CC   & $\nu_e+^{13}{\rm C}\rightarrow e^-+^{13}{\rm N}$  & 2.2~MeV     & $e^-$ and $^{13}$N decay & 3768 & 14 \\
	4& NC   & $\nu+^{12}{\rm C}\rightarrow\nu+^{12}{\rm C}^*~(1^+)$ & 15.1~MeV & $\gamma$ & 0.2 & 5 \\
	5& NC   & $\nu+^{13}{\rm C}\rightarrow\nu+^{13}{\rm C}^*~(\frac{3}{2}^-)$ & 3.685~MeV & $\gamma$ & 3165 & 13.5 \\
	\hline
	\end{tabular}
    \caption{ \label{table:solarnuCCNC} Reactions via ES, CC, and NC channels for $^8$B and $hep$ solar neutrinos in JUNO. The statistics correspond to a 200~kt$\cdot$year exposure and a 100\% detection efficiency.}
\end{center}
\end{table}

\subsubsection{Potential of measuring \texorpdfstring{$pp$}{pp} and \texorpdfstring{$^7$Be}{7Be} neutrinos}

The Compton-like edge (665\,keV) of the continuous spectrum of scattered electrons due to $^7$Be solar neutrinos is a very distinct feature of the energy spectrum.
Combined with the unprecedented energy resolution of JUNO, it should be possible to detect $^7$Be solar neutrinos considering a wide range of background levels.
With the optimal assumption of $^{238}$U and $^{232}$Th concentrations in Ref.~\cite{yellowbook}, i.e., 10$^{-17}$~g/g in secular equilibrium, these contaminations can be determined via easily detectable fast $^{214(212)}$Bi/$^{214(212)}$Po coincidences, respectively.
From the experience of Borexino~\cite{Agostini:2020mfq}, $^{210}$Po and $^{210}$Pb/$^{210}$Bi isotopes that are part of the $^{238}$U chain could have additional sources and break the secular equilibrium.
For $^{210}$Po, it is possible to identify such $\alpha$'s on an event-by-event basis with high efficiency and purity~\cite{PIDJUNO}.
In addition, $^{210}$Po has a lifetime of 200\,days and thus its contamination can be significantly reduced over the whole JUNO lifetime.
However, the contamination of $^{210}$Bi (Q = 1160\,MeV, $\tau$ = 7.23\,days), supported by the long-lived $^{210}$Pb (Q = 63\,keV, $\tau$ = 32.2\,years) can be disentangled only via a spectral fit of the data itself.
Similarly dangerous background is $^{85}$Kr (Q = 687\,keV, $\tau$ = 15.4\,years) that is present in the air.
Even though its electron spectrum is correlated with $^7$Be neutrino spectrum, $^{85}$Kr contamination can be constrained via its rare decay branch (0.43\%) providing fast coincidences~\cite{BxPhaseI}.
Due to these possible backgrounds, exceeding the 2.7\% precision of the flux measurement of $^7$Be neutrinos achieved in Borexino Phase-II~\cite{Agostini:2018uly} will be challenging and needs further studies.
The measurement of $^7$Be neutrino flux can be useful for solving the ``metallicity problem'' in solar physics, searching for neutrino magnetic moment~\cite{Borex-NMM} and flavor-diagonal non-standard neutrino interactions~\cite{Borex-NSI}.
Ideas on searching for the $^7$Be neutrino flux time variations at different frequencies have also been presented~\cite{LENA}.

Measurement of the $pp$ neutrinos with the highest flux and lowest energy ($<$ 420\,keV) is challenging due to the irreducible background from $^{14}$C (Q = 156\,keV, $\tau$ = 8270\,years).
Even in the ideal case of JUNO achieving the radio-purity of Borexino Phase-I, the $pp$ neutrino signal would be the dominating part of the spectrum only in a narrow interval between 160 to 230\,keV.
Additional complications arise from the random pile-up of $^{14}$C with itself or with other backgrounds~\cite{BX-pp}.
The energy resolution at low energies would also be affected by the relatively high dark noise of the 20-inch PMTs.
All these effects will need careful studies.
Borexino has measured the $pp$ neutrino flux with about 10\% precision~\cite{Agostini:2018uly}.
A more precise $pp$ measurement would improve the comparison between the solar photon and neutrino luminosities and would help to examine the energy dependence of neutrino oscillation in the vacuum-dominated region.

\subsection{Atmospheric neutrinos}
\label{subsec:atmos}

Atmospheric neutrinos are a very important source for studying neutrino oscillation physics. In 1998, the Super-Kamiokande experiment reported the first evidence of neutrino oscillation based on a zenith-angle dependent deficit of atmospheric muon neutrinos~\cite{Fukuda:1998mi}. Atmospheric neutrinos originate from the decays of $\pi$ and $K$ produced in extensive air showers initiated by the interaction of cosmic rays with the Earth's atmosphere~\cite{Barr:2004br,Guan:2015vja}. Since the Earth is mostly transparent to neutrinos below the PeV energy scale, an atmospheric neutrino detector is able to see neutrinos coming from all directions. They have a broad range in baseline (15 km $\sim$ 13000 km) and energy (0.1 GeV $\sim$ 10 TeV) and contain neutrinos and antineutrinos of all flavors. The Mikheyev-Smirnov-Wolfenstein (MSW) matter effect~\cite{Wolfenstein:1977ue,Mikheev:1986gs} acting on
neutrinos passing through the Earth will play a key role in determining the Neutrino Mass Ordering (NMO). It is worthwhile to stress that the JUNO NMO sensitivity from atmospheric neutrinos is complementary to that from the JUNO reactor neutrino results. Since the best fit occurs at different values of $|\Delta m_{31}^2|$ for the wrong-ordering hypothesis, the combined sensitivity from atmospheric and reactor neutrinos will exceed the purely statistical combination of their stand-alone sensitivities~\cite{Blennow:2013vta,Bezerra:2019dao}.

In Ref.~\cite{yellowbook}, we have investigated atmospheric neutrinos in JUNO and discussed their contributions to the NMO. In terms of the reconstruction potential of the JUNO detector, we conservatively use the atmospheric $\nu_\mu$ and $\bar{\nu}_\mu$ events with the track length $L_\mu$ larger than 5 m and $1^\circ$ angular resolution. These events have been classified into the fully contained (FC) $\nu_\mu$-like, FC $\bar{\nu}_\mu$-like, partially contained (PC) $\nu_\mu$-like and PC $\bar{\nu}_\mu$-like samples in terms of the
$\mu^\pm$ track and the statistical charge separation. Our numerical results have shown that the JUNO's NMO sensitivity can reach 0.9$\sigma$ for a 200 kton-years exposure and $\sin^2 \theta_{23} =0.5$. In addition, an optimistic sensitivity may reach 1.8$\sigma$ for 10-year data based on the following reasonable assumptions. Firstly, the $\nu_e/\bar{\nu}_e$ CC events can be identified and reconstructed very well if the $e^\pm$ visible energy $E^e_{\rm vis} > 1$ GeV and the visible energy of accompanying hadrons is smaller than $E^e_{\rm vis}$. Secondly, we extend the selection condition $L_{\mu} > 5 $m to $L_{\mu} > 3 $m for the $\nu_\mu/\bar{\nu}_\mu$ CC events. Finally, the charged lepton direction will be replaced by the neutrino direction with a $10^\circ$ angular resolution. A combined NMO sensitivity with both atmospheric and reactor neutrinos in JUNO will be evaluated.

\begin{figure}[!htb]
\centering
\includegraphics[width=0.8\textwidth]{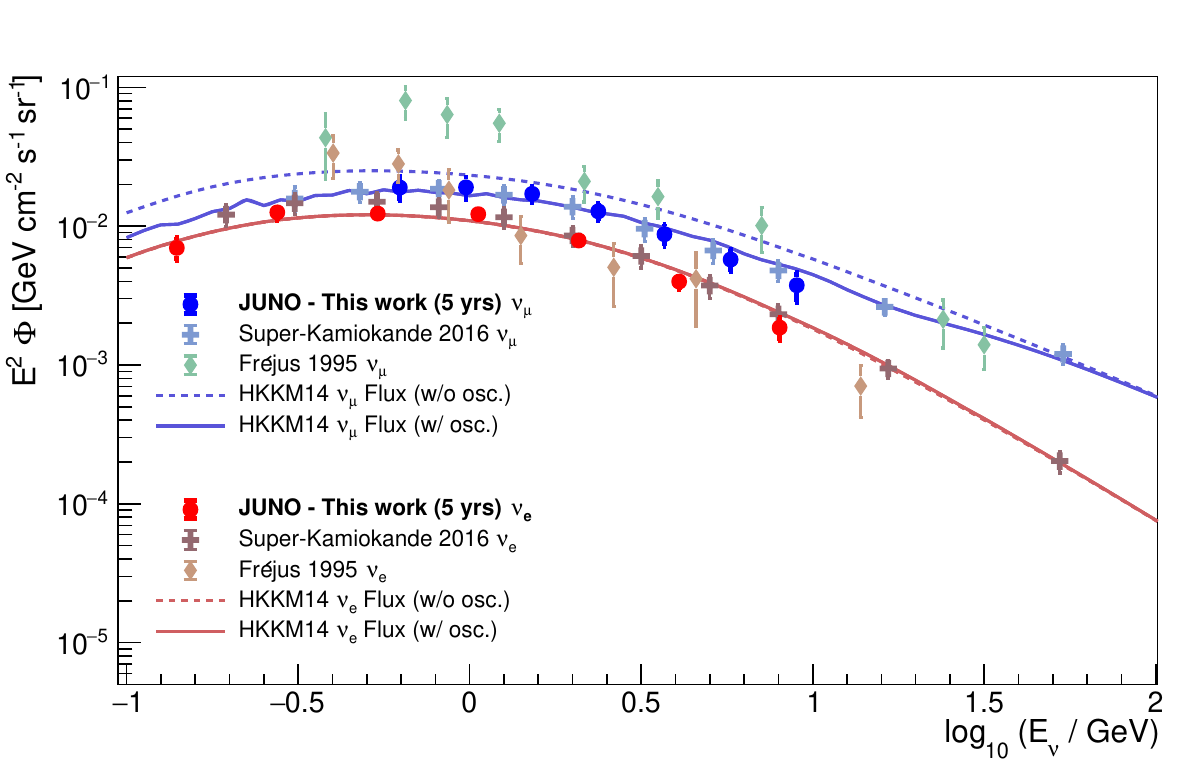}
\vspace{-0.0 cm} \caption{Atmospheric neutrino energy spectra reconstructed by the JUNO detector for $\nu_\mu$ (blue color) and $\nu_e$ (red color), compared with present Super-Kamiokande~\cite{Richard:2015aua} and ${\rm Fr\acute{e}jus}$~\cite{Daum:1994bf} measurements in the same energy region. The HKKM14~\cite{Honda:2015fha} model predictions are also reported, both at the source and including oscillation effects. The fluxes are multiplied by $E^2$ to give a better picture.
\label{fig:atmo_spec}}
\end{figure}

In Ref.~\cite{Settanta:2019ecp}, JUNO's performance in reconstructing the atmospheric neutrino spectrum has been evaluated. A Monte Carlo sample of the atmospheric neutrinos has been generated, the simulated spectrum has been reconstructed between 100 MeV and 10 GeV, showing a great potential of the detector in the atmospheric low energy region. The different hit time patterns caused by final-state electrons and muons allow discriminating the flavor of the primary $\nu_e$ and $\nu_\mu$ neutrinos. To reconstruct the time pattern of
events, the signals of 3-inch PMTs have been used, which will be operated in digital mode due to their small area. Furthermore, the Transit Time Spread (TTS) of 3-inch PMTs is of the order of the nanosecond, while the 20-inch PMTs one is larger, for most of them. By use of a probabilistic unfolding method, we have reconstructed separately the primary $\nu_e$ and $\nu_\mu$ spectra as shown in Fig.~\ref{fig:atmo_spec}. To remove low-quality events, preliminary cuts have been applied to the neutrino sample. A cut on the interaction vertex position is applied, to remove events that release their energy near the edge of the acrylic vessel. Furthermore, an additional cut on the total number of hits seen by the water pool veto PMTs has been used in the analysis to discard PC events and suppress the atmospheric muon background. The resulting $\nu_e$ and $\nu_\mu$ populations have been passed through the analysis procedure. The total uncertainty on the atmospheric neutrino spectrum reconstruction is evaluated in each energy bin, including both contributions from statistics and systematic effects, and it is reported in Fig.~\ref{fig:atmo_spec}. The effects from oscillation parameters and cross-section uncertainties, sample selection and unfolding procedure have been included in estimating the systematics. The total uncertainty ranges between $10\%$ and $25\%$, with the best performance obtained at around 1~GeV.

\subsection{Geoneutrinos}
\label{subsec:geoneutrino}

Geoneutrinos, antineutrinos from the decays of long-lived radioactive elements inside the Earth, are a unique tool to study our planet, particularly its radiogenic power, and bring insights into its formation and chemical composition. The inverse beta decay on protons with 1.8\,MeV threshold makes it possible to measure geoneutrinos from the $^{238}$U and $^{232}$Th decay chains. Only two experiments have measured geoneutrinos so far: KamLAND ~\cite{Araki:2005qa,Abe:2008aa,Gando:1900zz,Gando:2013nba,watanabe19} and Borexino~\cite{Bellini:2010geo,Bellini:2013geo,Agostini:2015cba,Borex-geo-2019}. Both experiments reached a precision of about 16-18\%, detecting in total $168.8^{+26.3}_{-26.5}$ events in KamLAND~\cite{watanabe19} and $52.6 ^{+9.4}_{-8.6}\,({\rm stat}) ^{+2.7}_{-2.1}\,({\rm sys})$ events in Borexino~\cite{Borex-geo-2019}. Both experiments do not have sufficient sensitivity to determine the Th/U ratio, a parameter important for understanding the Earth's formation, with good accuracy.

To facilitate the comparison among experiments, the normalized $\bar{\nu_e}$ geoneutrino signal rate is expressed in Terrestrial Neutrino Units (TNU)\footnote{One TNU corresponds to 1\,event detected with a detector of 100\% detection efficiency, containing 10$^{32}$ target protons (roughly 1\,kton of liquid scintillator) in 1\,year.}.
A total signal of $47.0^{+8.4}_{-7.7}\,({\rm stat)}^{+2.4}_{-1.9}\,({\rm sys})$ TNU is measured by Borexino~\cite{Borex-geo-2019}. The null-hypothesis of a mantle signal is excluded at 99\% C.L. The estimated Earth radiogenic power from U and Th is $31.7 ^{+14.4}_{-9.2}$\,TW.
Borexino sets tight upper limits on the power of a georeactor hypothesised~\cite{herndon1993feasibility,herndon1996substructure,rusov2007geoantineutrino,de2008feasibility} to be present inside the Earth.
KamLAND measured~\cite{watanabe19} geoneutrino signal of $32.1 ^{+5.0}_{-5.0}$\,TNU and estimated the Earth radiogenic power from U and Th as $12.4^{+4.9}_{-4.9}$\,TW.
 Contrary to Borexino, KamLAND prefers the geological models with a lower concentration of heat-producing elements in the Earth, which brings a weak discrepancy between the two results.

 A next-generation experiment like JUNO is needed to provide more precise results concerning the Earth's radiogenic power and to fully exploit geoneutrinos in order to understand the natural radioactivity of our planet. With its detector being at least 20 times larger than the existing detectors, JUNO will join the family of geoneutrino experiments and represent a fantastic opportunity to measure geoneutrinos. Within the first year of running, JUNO will record more geoneutrino events than all other detectors will have accumulated to that time. In the following, we predict the signal and the precision level of JUNO and introduce a local refined geological model. These models provide a possibility of extracting the mantle component.

\subsubsection{Potential of geoneutrino measurement in JUNO}

The JUNO detector will provide an exciting opportunity to obtain a high statistics measurement of geoneutrinos. The expected signal and background rates, corresponding systematic uncertainties, and the JUNO sensitivity to geoneutrinos are published in~\cite{Han:2016geoPotential, yellowbook}.

The expected geoneutrino signal depends on U and Th abundances and distribution in the Earth.
The published JUNO sensitivity studies~\cite{Han:2016geoPotential, yellowbook} consider a global composition of the Earth as described in Ref.~\cite{Strati:2015geosignal}, leading to an expected signal in JUNO of $39.7^{+6.5}_{-5.2}$~TNU ($\sim$400 geoneutrinos per year).
The 500\,km range of the crust around JUNO contributes more than 50\% to the total signal~\cite{Strati:2015geosignal}.
Thus, local refined geological models are needed for a precise estimation of the crustal signal and, consequently, for disentanglement of the mantle signal. Such an effort is also a good opportunity for geologists and particle physicists to work together and test our predictions of the planet's interior.

\begin{figure}[htbp]
\centering
\subfigure[]{
\begin{minipage}[t]{0.5\linewidth}
\centering
\includegraphics[width=8.5cm]{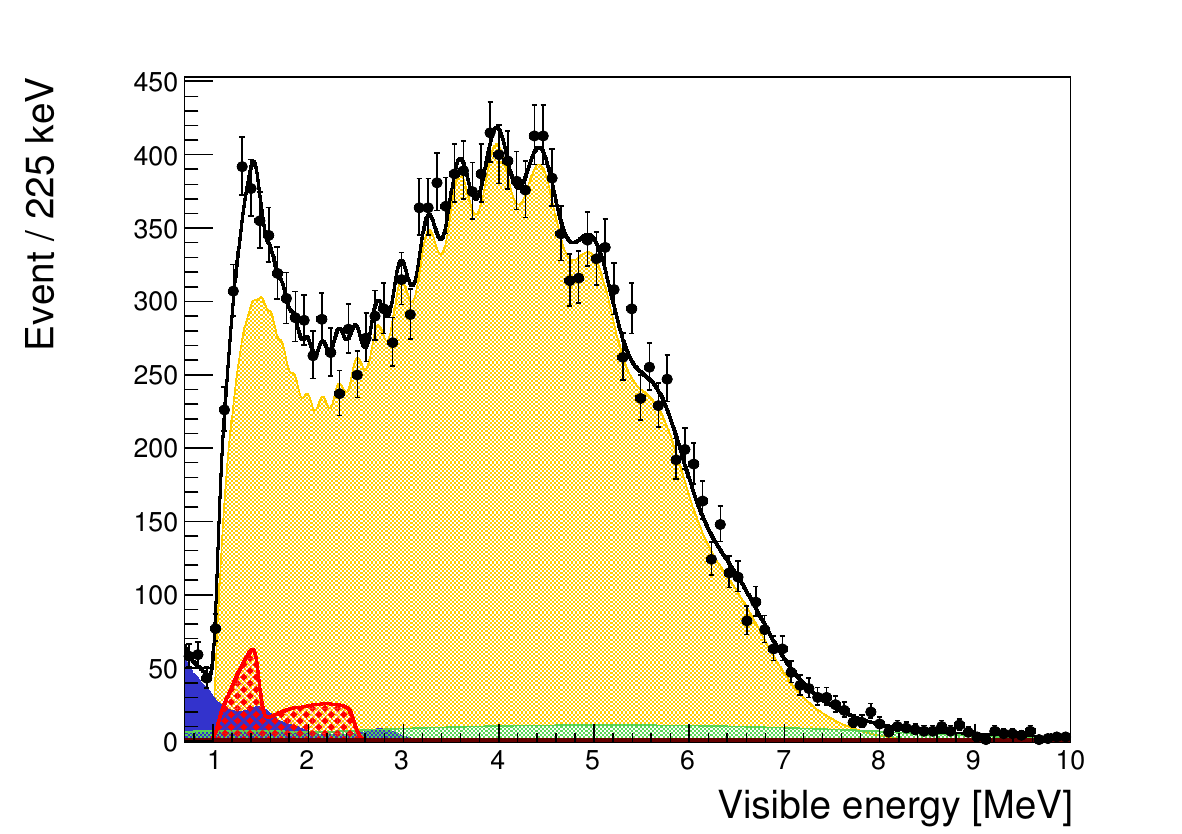}
\label{figspectral}
\end{minipage}%
}%
\subfigure[]{
\begin{minipage}[t]{0.5\linewidth}
\centering
\includegraphics[width=8.5cm]{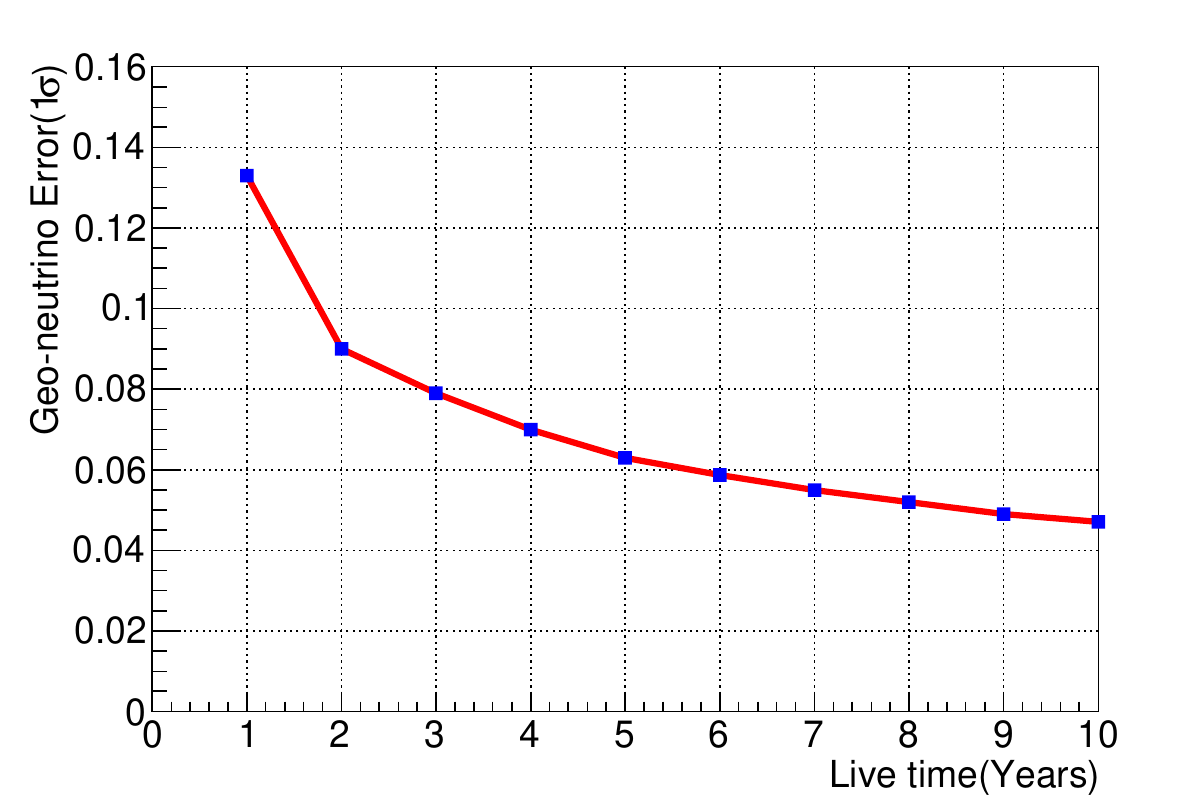}
\label{1sigmaerror}
\end{minipage}%
}%
\centering
\caption{(a) The energy spectra of prompt IBD candidates from geoneutrinos (red), reactor antineutrinos (orange), and other non-antineutrino backgrounds (accidental in blue, $^9$Li-$^8$He in green,$^{13}$C$(\alpha,n)^{16}$O in small magenta) for one year of data-taking after the selection cuts~\cite{yellowbook}. (b) The expected 1$\sigma$ uncertainty for geoneutrino measurement in JUNO with a fixed chondritic Th/U ratio as a function of live time~\cite{Han:2016geoPotential}.}
\end{figure}

The main challenge for the JUNO geoneutrino measurement is the large reactor antineutrino background. In addition, there are other non-antineutrino backgrounds relevant for geoneutrino detection, like cosmic muons, cosmogenic $^9$Li-$^8$He, fast neutrons,$^{13}$C$(\alpha,n)^{16}$O reactions, and accidental coincidences. The expected energy spectrum is shown in Fig.~\ref{figspectral}.

The potential of the JUNO geoneutrino measurement was estimated by two methods considering the same muon veto (17\% exposure loss), IBD selection (80\% efficiency), and detector response (3\% at 1\,MeV energy resolution). One is based on the generation of thousands of pseudo-experiments, in which the rate of each spectral component is randomly extracted from the Gaussian distributions reflecting the uncertainty of rate predictions~\cite{yellowbook}; each pseudo-experiment is then fit using a log-likelihood method and constraining the non-antineutrino backgrounds. The other method is using the Asimov data set, which employs the $\chi^2$ fit method to quantitatively assess the potential of geoneutrino measurements in JUNO~\cite{Han:2016geoPotential}. The theoretical spectral shapes are used for geoneutrinos~\cite{web:enomoto}. The first method did not consider the shape uncertainty of spectral components, while the second one did. The two methods gave compatible results. After 1, 3, 5, and 10 years, the first (second) method estimated the precision of geoneutrino measurement with a fixed chondritic mass Th/U = 3.9 in the fit as 17(13)\%, 10(8)\%, 8(6)\%, and 6(5)\%. The results of the second method are shown in Fig.~\ref{1sigmaerror}. After several years of measurements, JUNO also has the potential to constrain the Th/U ratio in the observed geoneutrino signal~\cite{Han:2016geoPotential,yellowbook}.

\subsubsection{Refined crustal models and expected signal for JUNO}

The fact that geoneutrinos signal produced by U and Th in the continental crust within $\sim$100\,km of the detector equals the whole mantle signal~\cite{Strati:2015geosignal} illustrates the importance of refined geological models around JUNO.
Involving the advanced know-how of the geophysical and geochemical communities, the South China Block surrounding JUNO can be characterized in terms of density, crustal layers thickness, and chemical composition.

The JULOC (JUNO Local Crust)\,\cite{Gao:2020JULOC} and GIGJ (GOCE Inversion for Geoneutrinos at JUNO)~\cite{Reguzzoni:2019GIGJ} models provide information about the thickness and density of the upper, middle, and lower crust around JUNO.
JULOC, a 3-D comprehensive high-resolution crustal model around JUNO, uses seismic ambient noise tomography~\cite{Zheng:2010NoiseSesmic} in the input and covers an area of $10^\circ \times 10^\circ$ around the detector. For the South China block, the crustal structure is pretty uniform. No significant large-scale velocity and density anomalies are found in the JULOC model. The density uncertainty is less than that of the global crustal model. According to GIGJ, a $6^\circ \times 4^\circ$ refined geophysical model obtained by a Bayesian inversion of gravimetric data of GOCE satellite (Gravity field and steady‐state Ocean Circulation Explorer), the overall thickness uncertainties of upper, middle, and lower crustal layers are 2\%, 3\%, and 1\%, respectively~\cite{Reguzzoni:2019GIGJ}.

On the basis of compiled U and Th abundances data from over 3000 rock samples, the JULOC model suggests that the local upper (middle and lower) crust has higher (lower) average U and Th abundances than the global average. The estimated crustal geoneutrino signal by JULOC is ($38.3 \pm 4.8$)\,TNU, to be compared with ($28.2^{+5.2}_{-4.5}$)\,TNU~\cite{Strati:2015geosignal} based on a global crustal model~\cite{Huang:2013RMmodel}. The significant increase of the signal is mainly the consequence of the upper crust of South China Block that is richer in U and Th than the global average.

\subsection{Nucleon decays}
\label{subsec:ndecays}

The baryon number $B$ violation is one of three basic ingredients to explain the observed cosmological matter-antimatter asymmetry~\cite{Sakharov:1967dj}. The baryon number is necessarily violated in the Grand Unified Theories (GUTs)~\cite{Georgi:1974sy,Nath:2006ut}, which can unify the strong, weak, and electromagnetic interactions into a single underlying force at a scale of $M_{GUT}\simeq 2\times 10^{16}$ GeV. A general prediction of the GUTs is nucleon decay. However, no experimental evidence to date for the proton decay or $B$-violating neutron decay has been found. With a 20 kton LS detector, JUNO is expected to have a good performance on the nucleon decay search, especially on the channel of $p \to \bar{\nu} K^+$~\cite{Svoboda,yellowbook}. This decay mode displays a clear three-fold coincidence feature in time, a prompt signal from the $K^+$ kinetic energy, a short delayed signal ($\tau=12.38$ ns) from its decay daughters, and a long-delayed signal ($\tau=2.2$ $\mu$s) from the final Michel electron.

\begin{figure}[htb]
\begin{center}
\includegraphics[scale=0.43]{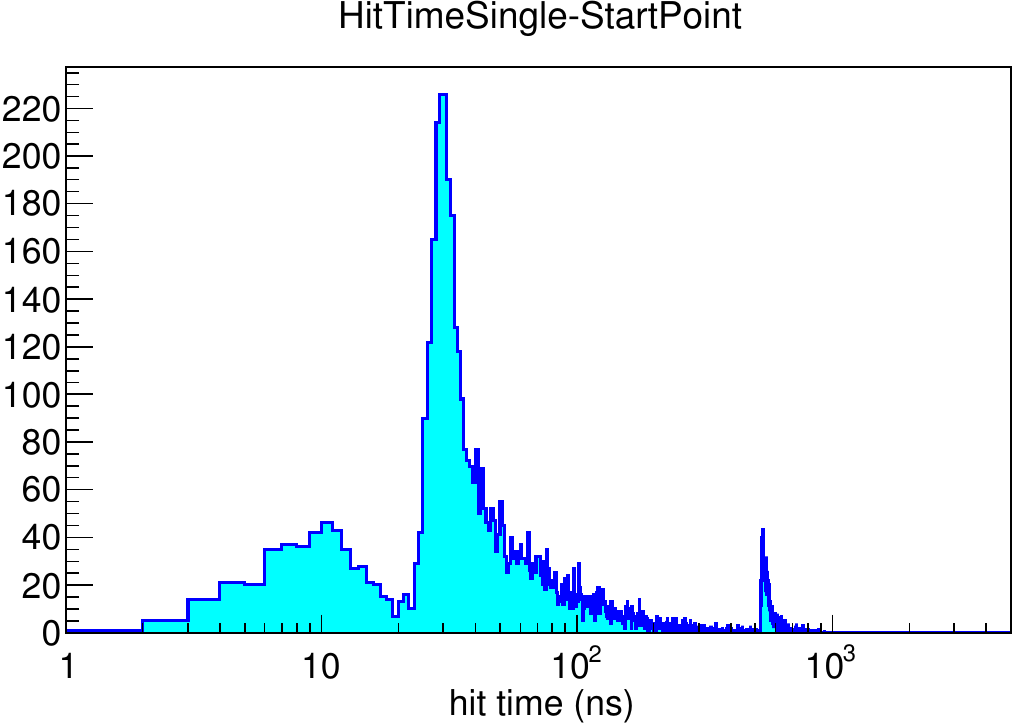}
\includegraphics[scale=0.37]{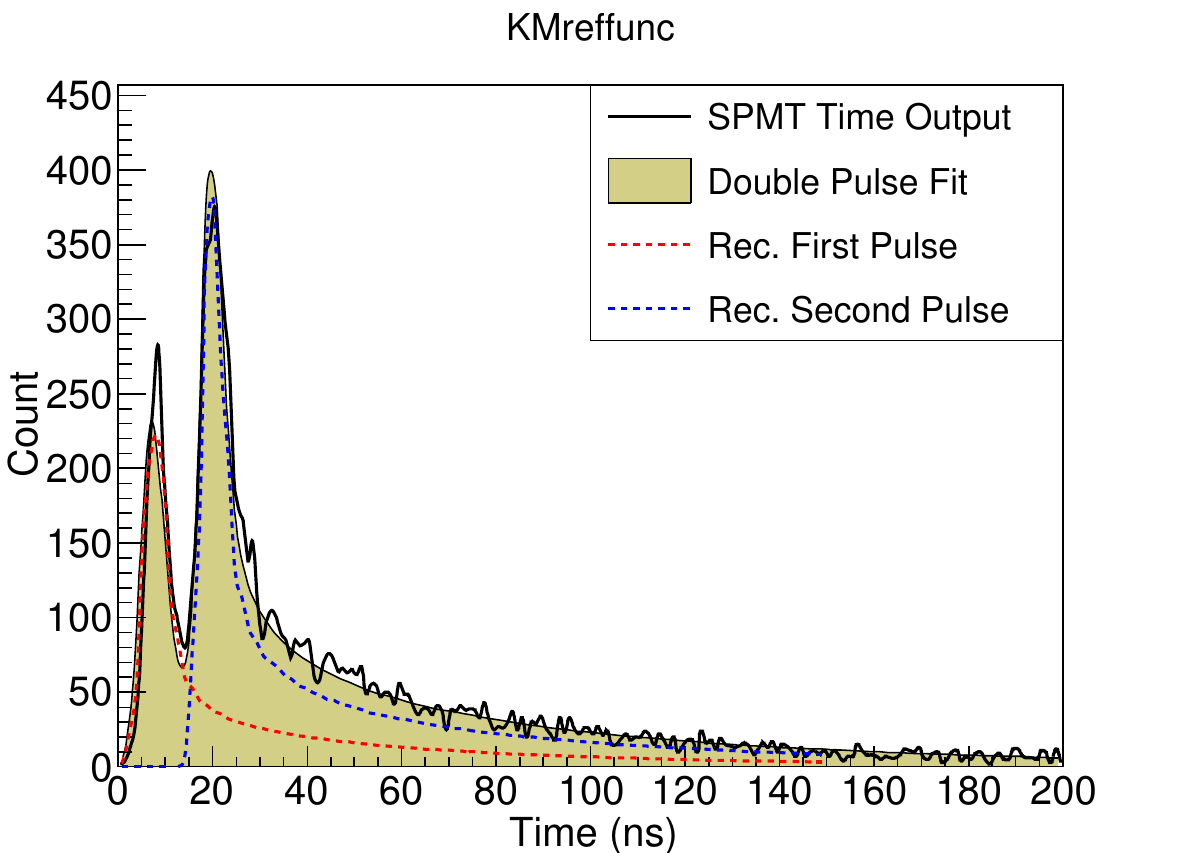}
\end{center}
\vspace{-0.5 cm}\caption{Left: A sample of hit time distribution after TOF correction. A double signal pile-up shape should be observed. The third signal in around 500 ns comes from the Michel election. Right: A sample of Multi-pulse fitting. The time difference of both signal is 11 ns.}
\label{fig:HitTime}
\end{figure}

The JUNO potential on  $p \to \bar{\nu} K^+$ using the JUNO Monte Carlo (MC) simulation framework~\cite{Zou:2015ioy} and considering the detector performances in detail. Generations of $p \to \bar{\nu} K^+$ and atmospheric neutrino backgrounds rely on GENIE 3.0~\cite{Andreopoulos:2015wxa}. We modified the GENIE code to take into account the final state interaction and the residual nucleus de-excitations. The time-correlated triple coincidence permits the JUNO detector to effectively identify the $p\rightarrow \bar{\nu} K^+$ signals and to reject atmospheric neutrino backgrounds. The primary production $K^+$ and its decay daughters will leave two deposition signals in the hit-time spectrum after TOF correction. There will be a pile-up of signals due to the short lifetime of $K^+$. With the help of the 3-inch PMT system (TTS 1.5 ns) of JUNO, we can observe this pile-up shape in the hit time spectrum, as shown in Fig.~\ref{fig:HitTime}. Due to this feature, the uprising time of $p \to \bar{\nu} K^+$ should be larger than that of atmospheric neutrinos~\cite{Undagoitia:2005uu}. Furthermore, the selection of $p \to \bar{\nu} K^+$ can be improved with the help of multi-pulse fitting~\cite{TheKamLAND-Zen:2015eva}, in which we can reconstruct both signals, and get the correlated time difference and the energy deposition of each signal, as shown in the right panel of Fig.~\ref{fig:HitTime}.

Beyond the timing analysis, additional selection criteria have been considered using visible energy, fiducial volume, muon veto, the Michel electron, and neutron capture. Using all these criteria, 31\% $p \to \bar{\nu} K^+$ remained while only 0.3 backgrounds survived in ten years. According to the Feldman-Cousins approach~\cite{Feldman:1997qc}, the preliminary sensitivity of $p \to \bar{\nu} K^+$ of JUNO is estimated as $8.34 \times 10^{33}$ years (90\% C.L.) for JUNO 10 years running. In addition to $p \to \bar{\nu} K^+$, other nucleon decay modes, such as $n\rightarrow 3 \nu$ and  $p\rightarrow \mu^+ \mu^+ \mu^-$ are being investigated; it is expected that JUNO will give more stringent limits on these decay channels in the future.

\subsection{Other physics}
\label{subsec:other}

JUNO has the potential to address a wide range of open questions in the fields of particle physics, nuclear physics, astrophysics and cosmology besides the topics covered in the previous parts. In this subsection, we shall explore several new directions, which are of fundamental importance in the field of particle physics and cosmology.

\begin{itemize}
\item Testing the Majorana nature of the neutrino is one of the most important tasks in the field of particle physics and nuclear physics. The only realistic way to probe this is the search for the neutrinoless double beta ($0\nu\beta\beta$) decay ${^{A}_Z}N \to {^A_{Z+2}}N + 2e^-$~\cite{Dolinski:2019nrj}, in which the half-life is possibly reachable based on the effective neutrino mass $m_{\beta\beta}$ constrained by the mass splitting measurements in neutrino oscillation, if it is mediated by light active Majorana neutrinos. Experimental studies of the $0\nu\beta\beta$ decays using $^{76}{\rm Ge}$, $^{136}{\rm Xe}$ or $^{130}{\rm Te}$ have been carried out in the past years, and the lower limits on the $0\nu\beta\beta$ half-life have reached the levels of $ 10^{25}~{\rm}$ to $10^{26}~{\rm yr}$~\cite{Dolinski:2019nrj}, corresponding to an effective neutrino mass $m_{\beta\beta}$ of $(60\cdots 300)~{\rm meV}$ depending on the values of nuclear matrix elements. Next-generation ton-scale $0\nu\beta\beta$ decay experiments will improve the half-life sensitivity to the levels of $ 10^{27}~{\rm}$ to $10^{28}~{\rm yr}$~\cite{Abgrall:2017syy,Kharusi:2018eqi},
    covering the $m_{\beta\beta}$ range of $(10\cdots 50)~{\rm meV}$ for the inverted mass ordering.

    However, the parameter space for the normal mass ordering is beyond the sensitivity of next-generation experiments. Suitable experimental strategies are still under development. The JUNO experiment has great potential to be upgraded to a $0\nu\beta\beta$ phase~\cite{JUNObetabetaLOI} after the primary goal of the neutrino mass ordering is accomplished. JUNO's sensitivity reaches the meV scale. Preliminary studies have shown that, within 5 years the $m_{\beta\beta}$ sensitivity could reach $(5\cdots 12)~{\rm meV}$ with 50 tons of fiducial $^{136}{\rm Xe}$~\cite{Zhao:2016brs}, and $(2.3\cdots 6.0)~{\rm meV}$ with 400 tons of fiducial $^{130}{\rm Te}$~\cite{Cao:2019hli}, respectively. The effective energy resolution near the $Q$ values is at the level of 2\%. The significant backgrounds arise from $^{8}{\rm B}$ solar neutrinos, two-neutrino double beta decays, natural radioactivity, and cosmogenic radionuclides. It is shown in Ref.~\cite{Zhao:2016brs} that the background index could be comparable to the next generation liquid Xeon experiment. A more careful evaluation of the expected background levels is required and the investigation of isotope doping and background reduction is still on-going.

    Finally, one should emphasize the importance for future $0\nu\beta\beta$ decay experiments to reach the sensitivity of $m_{\beta\beta}$ around $1~{\rm meV}$. Such a measurement can not only probe the Majorana nature of neutrinos and lepton number violation, but also determine the lightest neutrino mass, fix the neutrino mass spectrum and place restrictive constraints on one of two Majorana CP phases. All of these go beyond the expected sensitivities of future $\beta$ decay experiments~\cite{Wolf:2008hf,Esfahani:2017dmu} and cosmological observations~\cite{Dvorkin:2019jgs}.

\item The existence of Dark Matter (DM) has been well confirmed in the scales of the galaxies, galaxy clusters and cosmological distance. However, the particle nature of DM is still an open problem in the field of particle physics and cosmology. Neutrinos produced from the annihilation or decays of DM inside the Sun or Galactic Centre are a promising indirect signature of DM, and have been searched for in large neutrino detectors, such as Super-Kamiokande~\cite{Choi:2015ara} and IceCube~\cite{Aartsen:2016pfc}. Sensitivity studies on the neutrino signals from the DM annihilation inside the Sun have been obtained for JUNO using electron neutrinos~\cite{Guo:2015hsy} and muon neutrinos~\cite{yellowbook}. Within ten years of data taking, sensitivities for the spin-independent cross-section can reach the level of $(10^{-41}\cdots 10^{-42})~{\rm cm}^2$ for both electron and muon neutrino channels. These sensitivity levels will be relevant for the DM mass smaller than 10 GeV. On the other hand, JUNO can provide a sensitivity at the level of $(10^{-39}\cdots 10^{-40})~{\rm cm}^2$ for DM-nucleon spin-dependent cross section~\cite{yellowbook,Guo:2015hsy}, which will surpass the existing limits for light DM particles.

    Putative primordial black holes (PBH) are a viable candidate for the cold DM~\cite{Green:2020jor}, and detectable in the mass range of $(10^{15}\cdots 10^{16})~{\rm g}$ via neutrinos and antineutrinos that are emitted as part of the Hawking radiation~\cite{Hawking:1974sw}. A search for PBH via low energy antineutrinos~\cite{Dasgupta:2019cae} has been carried out using Super-Kamiokande data~\cite{Bays:2011si}. JUNO can improve the current limit on PBH abundance by one order of magnitude~\cite{WangPBH} because of its excellent neutron tagging and background reduction techniques.

\item Light sterile neutrinos~\cite{Gariazzo:2015rra}, non-unitarity of the neutrino mixing matrix~\cite{Antusch:2006vwa} and non-standard neutrino interactions~\cite{Ohlsson:2012kf} are representative new-physics scenarios beyond the standard three neutrino oscillation framework.
    JUNO can search for light sterile neutrinos in several sensitive oscillation channels. Precise reactor antineutrino measurements of the JUNO detector
    can set severe limits on light sterile neutrinos for a $\Delta m^2$ range of $(10^{-5}\cdots 10^{-3})~{\rm eV}^2$~\cite{yellowbook}.
    Moreover, the JUNO-TAO detector will be able to test the reactor antineutrino anomaly and search for light sterile neutrinos at the eV scale~\cite{junotaocdr}.
    Precision measurement of oscillation parameters at JUNO, in particular for the mixing angle $\theta_{12}$, is crucial in the global effort of leptonic unitarity tests~\cite{yellowbook,Li:2018jgd}. Finally different completive limits for new physics beyond three-flavor oscillations can be obtained with the high-precision oscillation measurements of reactor antineutrinos~\cite{Li:2014mlo,Li:2014rya}, solar neutrinos~\cite{Li:2019snw}, atmospheric neutrinos~\cite{Antonelli:2018fbv,Antonelli:2020uui} and supernova neutrinos~\cite{Tang:2020pkp}.

\item Large neutrino detectors also are excellent tools to observe exotic particles produced in cosmic-ray interactions or the early Universe. The magnetic monopole was first predicted by Dirac in 1931, and is a general prediction of the Grand Unified Theories~\cite{tHooft:1974kcl}. Searches for signals of magnetic monopoles have been performed by many experiments, such as MACRO~\cite{Ambrosio:2002qq}, Soudan-2~\cite{Thron:1992ri} and NO$\nu$A~\cite{Acero:2020isd}. Q-balls~\cite{Coleman:1985ki} are predicted as non-topological solitons in theories with a global $U(1)$ symmetry and a complex scalar field, which can act as a viable candidate of DM~\cite{Kusenko:1997si}. They have been searched for by several experiments, including Super-Kamiokande~\cite{Takenaga:2006nr}, Baksan~\cite{Arafune:2000yv} and IceCube~\cite{Kasuya:2015uka}. Strange Quark Matter (SQM)~\cite{Witten:1984rs} with approximately equal numbers of up, down and strange quarks is predicted as the ground state of quantum chromodynamics. Stable SQM objects with masses larger than $10^{10}$ GeV are known as nuclearites~\cite{DeRujula:1984axn}.
    Searches for nuclearites have been carried out in MACRO~\cite{Ahlen:1992mq}, SLIM~\cite{Cecchini:2008su}, and ANTARES~\cite{Pavalas:2015nab}, with no experimental evidence. All these exotic particles can be searched for using the JUNO detector with rather similar detection strategies. Taking the nuclearites for instance, it has been shown that the sensitivity can be improved by one order of magnitude for the masses of nuclearites between
    $10^{13}$ GeV and $10^{15}$ GeV~\cite{Guo:2016kyj}.

\end{itemize}

Many more exciting topics in JUNO with leading sensitivities or unique features might be added to the list, given the large target mass, low background, low energy threshold and excellent energy resolution of the detector.

\section{Detector design and R\&D}
\label{sec:detector}
\subsection{Detector design overview}
\label{subsec:detector}

The JUNO detector system includes a 20 kton liquid scintillator (LS) detector, a water Cherenkov detector in which the liquid scintillator detector is submerged, and a plastic scintillator array on top of them. A schematic view of the detector is shown in Fig.~\ref{fig:junodetector}.

The 20 kton LS is contained in a spherical acrylic vessel with an inner diameter of 35.4~m. The Main Structure, including a spherical stainless steel structure and its bearing, supports the acrylic vessel via 590 connecting bars. The light emitted by the LS is watched by 17,612 20-inch PMTs (referred to as ``large PMT") and 25,600 3-inch PMTs (referred to as ``small PMT"), which are installed on the inner surface of the Main Structure. The photocathode coverage is 75.2\% for the 20-inch PMTs and 2.7\% for the 3-inch PMTs. All PMTs have been tested. The average photon detection efficiency is 29.1\% for large PMTs in the Central Detector and $>24\%$ for small PMTs. The large PMTs feature special protection in case of implosion. The entire LS detector is submerged in a cylindrical water pool. A water buffer of 1.42~m thickness between the acrylic vessel and the PMT surface protects the LS from the PMT glass's radioactivity. The water buffer is connected with the outer water Cherenkov detector but is optically separated.  A chimney for calibration operations is connected to the top of the acrylic vessel. Special radioactivity shielding and a muon detector will cover the chimney and the calibration system on the top.

The LS has a similar recipe as the Daya Bay LS but without gadolinium loading. Linear alkylbenzene (LAB), a straight alkyl chain of 10--13 carbons attached to a benzene ring, is used as the detection medium due to its excellent transparency, high flash point, low chemical reactivity, and good light yield. The JUNO LS will contain 2.5 g/L 2,5-diphenyloxazole (PPO) as the fluor and 3~mg/L p-bis-(o-methylstyryl)-benzene (bis-MSB) as the wavelength shifter. The composition has been optimized in dedicated studies with a Daya Bay detector~\cite{DYBJUNOLS,Abusleme:2020bbm}. The LS will be purified before filling to improve its radiopurity and transparency, which is crucial for such a gigantic detector. A pilot plant test shows that the LS attenuation length can reach $>20$~m. Combined with the PMT detection efficiency and other detector parameters, a yield of 1345 photoelectrons per MeV is obtained in simulations. The targeted LS radiopurity of JUNO is $10^{-17}$~g/g for U/Th/K.

The water pool has a diameter of 43.5 m and a height of 44 m, providing sufficient buffer in all directions to protect the LS from the surrounding rock radioactivity. The water pool is equipped with 2,400 20-inch PMTs, mounted on the Main Structure, acting as a Cherenkov detector for the cosmic muons. The muon detection efficiency is expected to be 99.8\%, similar to that of the Daya Bay water Cherenkov detector. Compensation coils are installed on the Main Structure to protect the PMTs from the Earth magnetic field.

On top of the water pool, a muon tracker will be installed to measure the muon directions. Plastic scintillator strips decommissioned from the Target Tracker of the OPERA experiment~\cite{Adam:2007ex} will be reused as the JUNO Top Tracker (TT). The TT is composed of 62 walls with a sensitive area of 6.7$\times$6.7 m$^2$ each. Radioactivity from the surrounding rock will induce very high noise rates in the plastic scintillator strips. These accidental backgrounds will be suppressed with a multi-layer design with at least three x-y layers. The TT covers more than 25\% of the area of the top surface of the water pool.

Given the gigantic size and unprecedented energy resolution requirement, construction of the JUNO detector is very challenging in mechanics, LS radiopurity and transparency, PMT photon detection efficiency, reliability of electronics, etc.  In the following sections, we will describe the design and R\&D achievements. 
\subsection{Central Detector}
\label{subsec:cd}

\subsubsection{Central Detector scheme and requirements}
\begin{figure}[hbt]
    \centering
    \includegraphics[width=0.4\columnwidth]{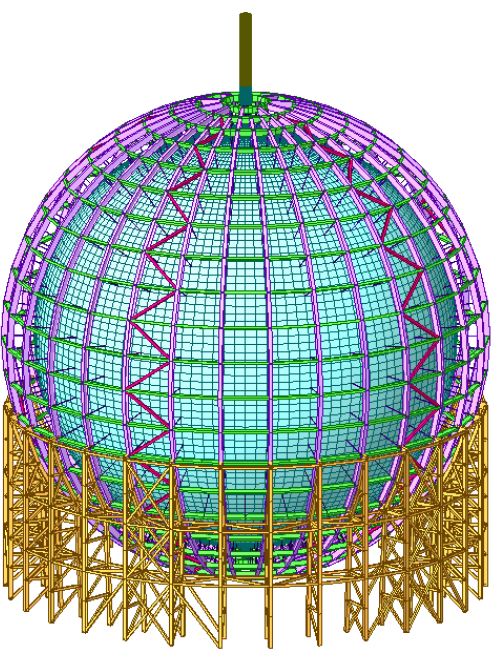}
    \caption{Scheme of the Central Detector (CD). The spherical Acrylic Vessel is supported by a spherical stainless steel shell structure via 590 Connecting Bars. The shell structure sits on a bearing consisting of 30 pairs of Supporting Legs made of stainless steel truss structures.
    \label{fig:CD-structure}}
\end{figure}

The 20 kton liquid scintillator detector, referred to as the Central Detector (CD), is one of the most challenging parts of JUNO, especially in mechanics, given its huge size. The LS will be contained in a spherical Acrylic Vessel with an inner diameter of 35.4~m and a thickness of 120~mm. The Acrylic Vessel will be supported by a spherical stainless steel shell structure with an inner diameter of 40.1~m via 590 Connecting Bars. The shell structure sits on a bearing consisting of 30 pairs of Supporting Legs, made of stainless steel truss structures and rooted on the concrete floor of the Water Pool. The shell structure and its bearing is called the Main Structure and is shown in Fig.~\ref{fig:CD-structure}. A total of 17,612 20-inch and 25,600 3-inch PMTs are installed on the Main Structure and point inward to detect the scintillation light. Another 2,400 20-inch PMTs of the water Cherenkov detector are also installed on the Main Structure and point outward to detect the water Cherenkov light. The CD and the water Cherenkov detector are optically separated. The Main Structure also provides support for the front-end electronics, cables, and the anti-geomagnetic field coil. The whole structure must be stable and reliable for 30 years of the designed JUNO lifetime.

To satisfy the requirements for the reactor and solar neutrino program, the CD must be made of selected radiopure materials. The surfaces of the CD components, including the Acrylic Vessel, pipes, pumps, and valves, are required to be extra clean and LS/water-compatible. Besides, the air-tightness of the CD sealing system must be very high to prevent Radon leakage from the air.

\subsubsection{The LS container: Acrylic Vessel}
The Acrylic Vessel will be made by bulk polymerization of 265 pieces of spherical acrylic panels. The acrylic recipe is carefully tuned with extensive R\&D, which does not include plasticizer and anti-UV material. It increases the light transparency significantly and improves the anti-aging and creep resistance performance. A dedicated acrylic panel production line is established specially for JUNO to reduce the radioactive background and increase transparency. It features fully closed pipelines and additional filters for the liquid Methyl Methacrylate (MMA), the acrylic monomer. A cleanroom of class 10000 and a high-purity water system are used to ensure the mold's cleanliness for the acrylic panel and the process of filling MMA into the mold. The radiopurity of acrylic reaches a level of $<0.5$~ppt for uranium and thorium. The process of unmolding and thermoforming at high temperatures is optimized to increase the spherical acrylic panels' transparency to $>96$\% in the water.

To ensure a lifetime of 30 years, the maximal pressure on the acrylic is required to be less than 3.5~MPa in the long term and 7~MPa in the short term, based on extensive dedicated experiments at IHEP. The Acrylic Vessel is supported by connection structures linked to the Main Structure. The supporting nodes on the Acrylic Vessel are critical to controlling the stress. The design is optimized with many prototype tests. The ultimate bearing capacity of the supporting nodes can reach 100~t, which is more than 6 times the designed load.

A new bonding method has been developed for JUNO to reduce the onsite construction time of the Acrylic Vessel. All the acrylic panels in the same layer will be bonded simultaneously. This technique has been verified with many tests and has been successfully applied in other projects.

\subsubsection{Main Structure}
The stainless steel Main Structure is designed as a shell structure of 30 longitudinal H-beams and 23 latitudinal H-beams. Five in-plane supports are uniformly arranged along the longitudinal direction to improve the structure's torsional stiffness and stability. The shell structure is supported by 30 pair Supporting Legs and 60 base plates, rooted on the concrete floor of the Water Pool. By optimizing the planar truss column's geometry, the reaction force of the inner and outer rings of bearings is nearly uniform. Under the working condition of about 3000~ton buoyancy applied on the Acrylic Vessel, the reaction force of each support base is kept below 60~tons. The Main Structure is designed to be safe for earthquakes of a horizontal acceleration of 46.5 Gal, corresponding to a 25\% exceedance probability in 50 years.

A total of 590 Connecting Bars are designed to connect the Main Structure and the Acrylic Vessel. At the Acrylic Vessel side, the Connecting Bars end in hinged connections. At the Main Structure side, 220 Connecting Bars between the 1st and 9th latitudinal layers end in disc springs to ensure that the Acrylic Vessel node's stress is less than 3.5 MPa with the buoyancy applied. Another 370 Connecting Bars from the 10th to 23rd layers use the rigid connection to control the structure's overall stability.

Pre-assembly of the supporting structure and part of the Main Structure have been tested in the factory. Key technologies for manufacturing and installation of the Main Structure have been developed, including the friction-type high-strength connecting technology, the anti-corrosion coating scheme for the disc spring, the monitoring system of the axial force and temperature of the Connecting Bars, etc.

\subsubsection{The Filling, Overflow, and Circulation (FOC) system}
\begin{figure}[hbt]
    \centering
    \includegraphics[width=0.6\columnwidth]{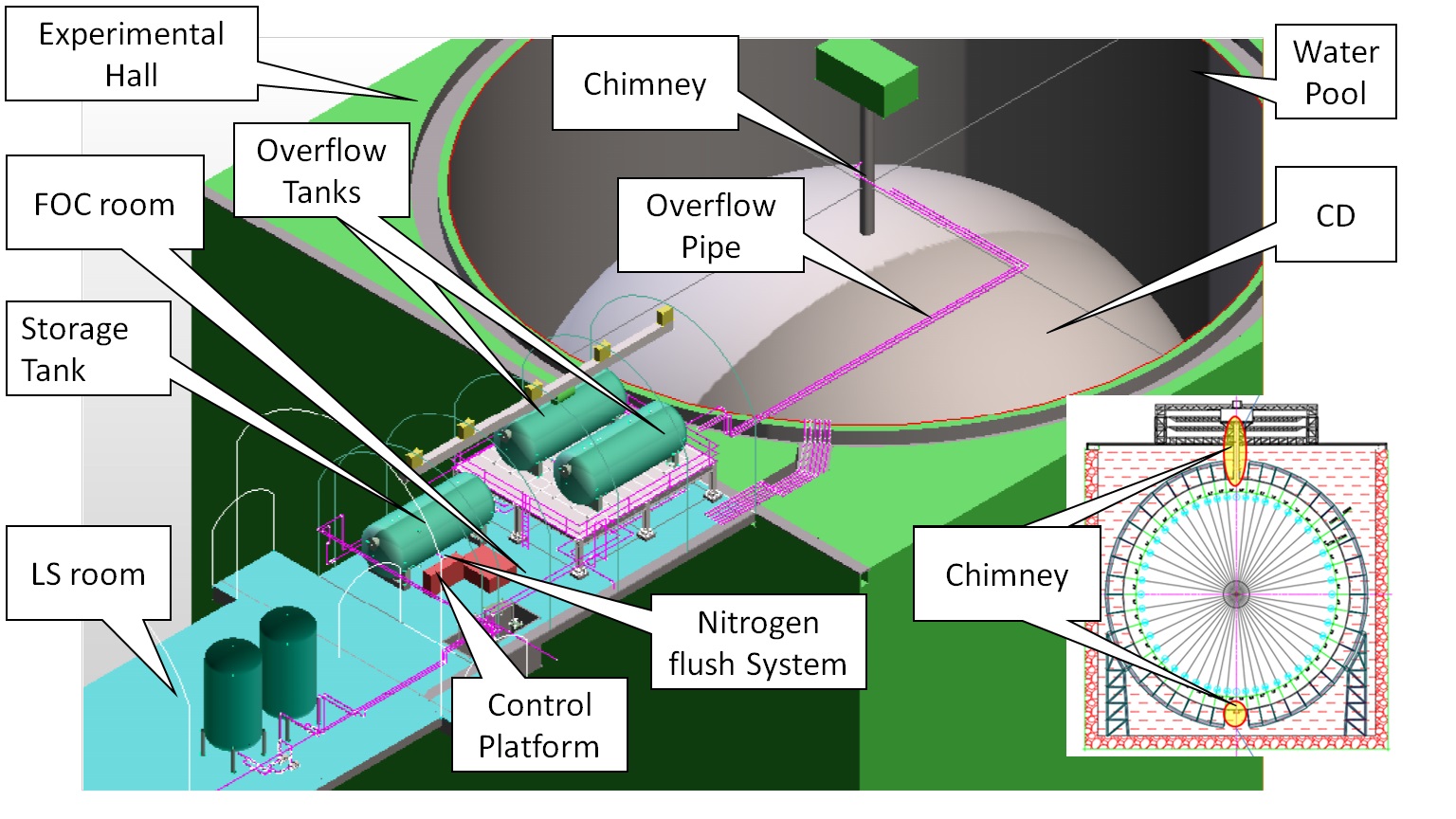}
    \caption{Scheme of the Filling, Overflow, and Circulation (FOC) system.
    \label{fig:foc}}
\end{figure}

The FOC system has one storage and two overflow tanks with a volume of 50 m$^3$ each. It also includes a nitrogen flushing system, the piping connected to the chimney of the CD, and a control platform, as shown in Fig.~\ref{fig:foc}. It has three primary functions: 1) filling pure water into the Acrylic Vessel synchronously with the filling of the Water Pool, then replacing water with LS during the 6-month LS production period; 2) stabilizing the LS level in the Acrylic Vessel within 20 cm when the liquid temperature varies within $21\pm1.4^\circ$C during operation; 3) circulating the LS in the detector through the underground LS purification systems (See Fig.~\ref{fig:LSpurification}) for the online purification.

\subsection{Veto detector}
\label{subsec:veto}

The experiment will be equipped with two veto systems providing both an efficient background reduction towards the environmental radioactivity and the residual cosmic muon flux crossing the detector.

\subsubsection{Water Cherenkov Detector (WCD)}

As shown in Fig.~\ref{fig:junodetector}, the Water Cherenkov Detector (WCD) is a cylinder of 43.5~m in diameter and 44~m in height, which is filled with 35 ktons of ultrapure water. The Cherenkov light produced by muons passing through the volume is detected by 2400 20-inch Microchannel Plate Photomultipliers (MCP PMTs) installed on the outer surface of the stainless steel Shell Structure of the CD. Tyvek reflective foils provide a coating for the pool walls and stainless steel support structure to increase light collection efficiency~\cite{Dayabay:2014vka}. The passive water shielding significantly reduces the gamma background in the CD due to radioactivity in the rocks.

The performance of the WCD requires stable working conditions over the future duration of the experiment.
A water system will provide and monitor the ultra-high purity of the water and will also ensure temperature uniformity over the whole volume.

The water system includes two plants for water production and circulation, one on the surface and one underground. The surface plant is the production stage with a rate of 100~tons/hour. High-purity water will be transported by stainless steel pipes through the 1300~m long slope tunnel to the underground plant for further purification in circulation in the WCD.

The detector consists of materials with different thermal properties, for example, the Acrylic Vessel and the stainless steel Main Structure. The water temperature around the Acrylic Vessel has to be stabilized at (21$\pm$1~\textcelsius{}) to maintain the detector's mechanical stability.
The heat from the electronics box for PMTs ($\sim$260 kW) on the stainless steel support structure needs to be taken away by water circulation. The requirement to control the temperature at such a precision level in a vast volume is very challenging. The underground water plant was optimized for high precision temperature control in the water circulation, which occurs separately in two volumes, an inner part in the Central Detector and an outer part in the WCD volume.
The relevant parameters are optimized, including the temperature and flow rate of inlets, the size and the number of holes on the water distributor, the flow rate ratio between the top and bottom inlets, etc. The water inlets are located on the top and bottom of the water pool, whereas the outlet is on the equator. Water distribution systems are designed for both the top and bottom inlets to make the water flow uniform. The temperature stability requirement has been achieved for the entire surface of the Acrylic Vessel with extensive simulations.

The radon concentration in tap water is about 10~Bq/m$^{3}$~\cite{radon}. It must be reduced to $<0.2$~Bq/m$^3$ for reactor antineutrino oscillation studies. A factor of 1000 ($\sim$10~mBq/m$^{3}$) reduction is possible with our R\&D, ideal for other low energy physics. A radon removal system with a 3-stage degassing membrane system is integrated into the water system. In a prototype with a 3-stage degassing membrane, the radon concentration in water has been controlled to $<50$~mBq/m$^3$~\cite{Zhang:2017tsd:addjournal,CGuo:2018zzzzz:bibtex_by_hand}.
The latest results show that the radon concentration could be controlled at $\sim 10$~mBq/m$^{3}$ in the prototype water.
The radon removal efficiency of the first stage reaches 90\%. The second and third stages were found to be less efficient due to a lower gas content in water. Therefore, a micro-bubble system fed with high-purity nitrogen is added to the second and third stages and improves the removal efficiency to above 90\%. For JUNO water system, we adopt the multi-stage degassing membrane to reduce radon in the water. Simultaneously, it should have strict requirements on the cleanliness of the detector installation to prevent excessive dust from accumulating inside the detector to introduce radioactive contamination(such as radium).

An HDPE film will be used as a liner to cover the pool walls and prevent radon's incursion from external rocks. According to the emanation measurements on several rock samples, including those from JUNO and Daya Bay sites, an HPDE film thickness of 5 mm
is required to keep the radon concentration in the water below 10~mBq/m$^{3}$. A 5~mm thick HDPE film will be used. The gaps between adjacent pieces of HDPE film will be closed by HDPE strips attached by hot-melt welding. During the HDPE installation, a network of electrodes will be buried in the cement to perform a leakage test after the liner installation.

The pool will be covered with a black rubber sheet that is sealed by a gas-tight zipper. This type of cover has been used in the Daya Bay experiment and shows excellent sealing performances.
Due to the chimney interface in the center of the pool, the cover will be divided into two semicircles and sealed by gas-tight zippers withstanding a maximum pressure of 100~kPa. Gas tightness is very difficult to maintain due to the large surface of more than 1000~m$^2$. Therefore,
the cover is placed 1~meter higher than the water surface. The volume in between will be filled with high-purity nitrogen with a positive pressure of a few hundred Pa to prevent radon contamination from outside.

Since the Top Tracker on top of the pool has a good efficiency in detecting muons, the WCD is optimized to have 1045 and 1355 PMTs on the upper and lower hemisphere of the Main Structure, respectively. As such, the upper hemisphere's effective detection efficiency plus the Top Tracker equals that of the lower hemisphere. The WCD PMTs are grouped into ten zones: 5 in the top and 5 in the bottom hemisphere. A local trigger with a lower threshold is used instead of the global trigger. If adopting global trigger, the muon detection efficiency is anticipated to be 98\% with a trigger threshold of 54 PMTs. The local triggers include two cases. One case is the fired PMT number in one zone above the threshold ($\sim$19 PMTs). Another is the fired PMTs number of two adjacent zones simultaneously above a lower threshold ($\sim$13 PMTs).
The detection efficiency can reach 99.5\% while keeping the noise at an acceptable level.

The 20-inch PMTs in the Central Detector and the Water Cherenkov Detector may be affected by the geomagnetic field, losing up to 60\% detection efficiency. A set of 32 circular coils surrounding the detector is designed to compensate for the geomatic field. The residual magnetic field will be less than 0.05~G in the Central Detector PMT region and lower than 0.1~G in the WCD PMT region, satisfying the experimental requirements.

The muon flux in the experimental hall is 0.004~Hz/m${^2}$. The average muon energy is 207~GeV. Based on the JUNO detector simulation, the average track length of muons going through the pool is about 13~m. The muon tagging efficiency can achieve 99.5\%. Most of the un-tagged muons have short tracks ($<$0.5~meter) and are far from the Central Detector, making little contribution to the fast neutron background. The fast neutron background from muons spallation can be controlled at a rate of about 0.1~event/day.

\subsubsection{Top Tracker (TT)}
\label{subsec:veto:tt}

\begin{figure}[ht]
  \begin{center}
  \begin{minipage}{0.6\textwidth}
    \begin{tikzpicture}[thick,x=0.15\textwidth,y=0.15\textwidth]
      \node (0,0) {\includegraphics[width=\textwidth]{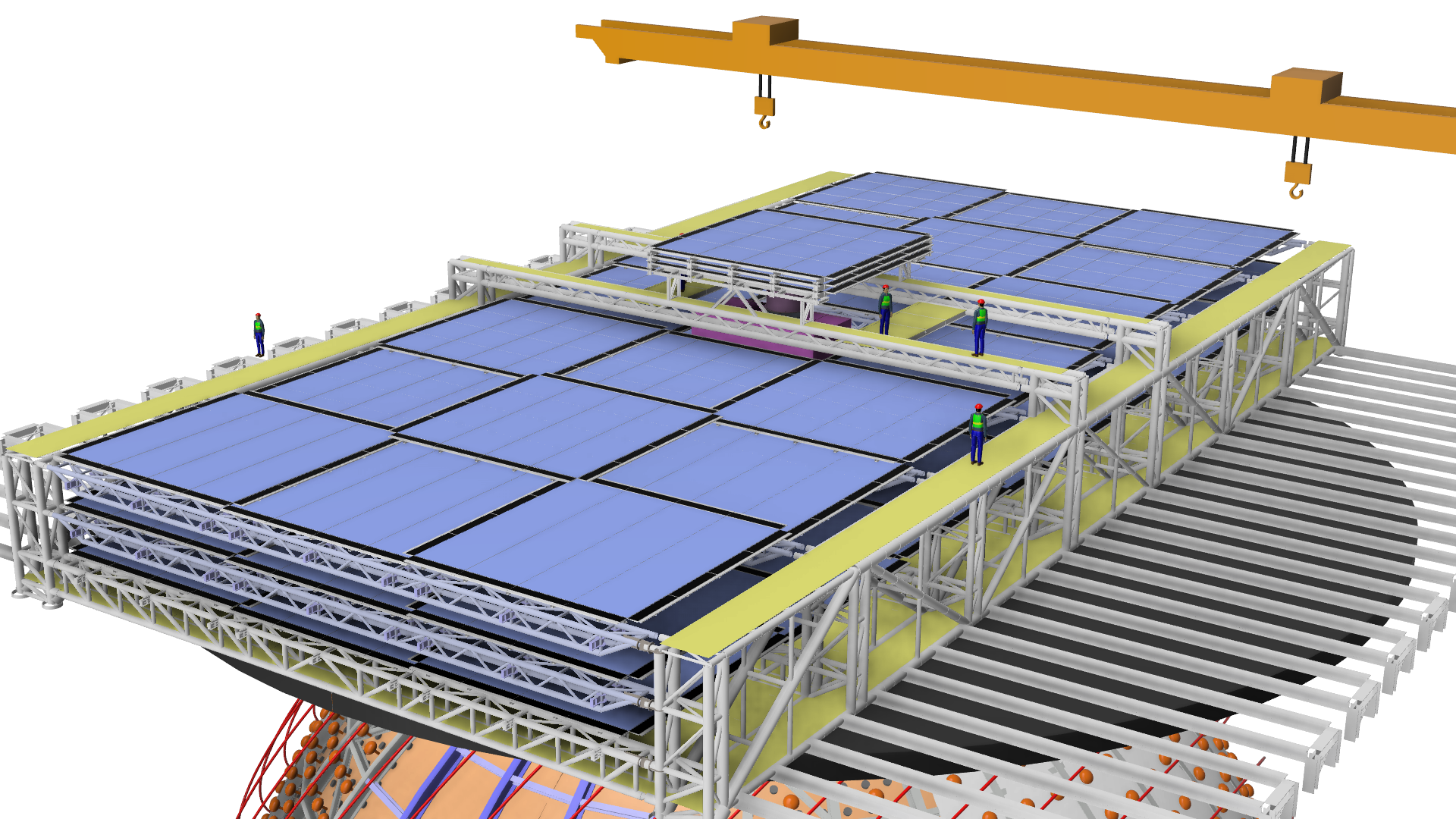}};
      \draw[<->] (-0.46,-1.49) -- (-2.98,-0.65) node [midway, below left=0.5em and 1em, sloped] {\bfseries 20~m};
      \draw[<->] (-3.02,-0.20) -- (0.66,1.08) node [midway, above, sloped] {\bfseries 47~m};
      \draw[<->] (-2.98,-0.65) -- (-3.02,-0.20) node [midway, above=-1em, sloped, rotate=180] {\bfseries 3~m};
      \draw[thin,color=red,fill,opacity=0.4] (-0.71,-0.32) --
        (-1.54,-0.11) node [midway, below left=-0.25em and -1.1em, sloped, opacity=1] {\bfseries \scriptsize 6.7~m} --
        (-0.89,0.16) --
        (-0.09,-0.02) --
        (-0.71,-0.32) node [midway, below right=-0.25em and -1.1em, sloped, opacity=1] {\bfseries \scriptsize 6.7~m};
      \draw[color=green] (-0.8075,-0.0725) node {\bfseries \footnotesize TT wall};
    \end{tikzpicture}
  \end{minipage}
  \end{center}
    \caption{The JUNO Top Tracker.}
    \label{fig:veto:tt:top_tracker}
\end{figure}
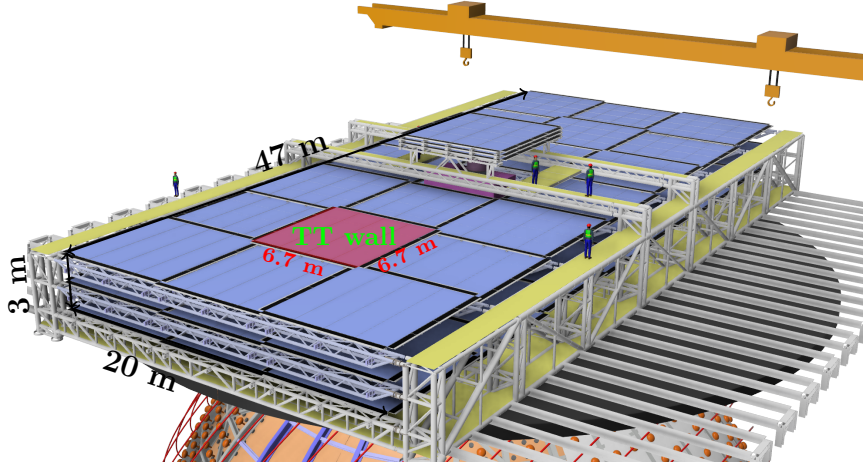

The Top Tracker will be located above the WCD as shown in Fig.~\ref{fig:veto:tt:top_tracker}.
As discussed in Ref.~\cite{Djurcic:2015vqa}, the JUNO Top Tracker (TT) will use
existing scintillating strips from OPERA's Target Tracker~\cite{Adam:2007ex}.
Since the dismantling of the OPERA detector (LNGS, Gran Sasso, Italy), the quality of the TT modules has been monitored.
We do not observe any extra deterioration of the plastic scintillator properties from the current aging monitoring. At least $4.9\pm1.5$~p.e.\@{} are observed for muons crossing the middle (most disfavored position) of the plastic scintillator strips. The expected overall efficiency of a TT module, as extrapolated from the previous experience,
is at the level of  $(98.0\pm0.5)\%$ for a 1~p.e.\@{} threshold. This number is due to the scintillator properties, the threshold, and the geometry of the TT modules. It has to be noted that in the experiment, a 1/3~p.e.\@{} threshold will be used.

As shown in Fig.~\ref{fig:veto:tt:top_tracker}, ``walls'' of plastic scintillator, called TT walls, will be placed in horizontal layers on top of the JUNO Central Detector rather than vertically as it was done in OPERA. Given the TT walls are flexible, a supporting structure was developed to limit the sag to less than 6~mm.

The TT walls will be distributed on 3 horizontal layers, separated by 1.5~m, on a ${3 \times 7}$ horizontal grid, as shown in Fig.~\ref{fig:veto:tt:top_tracker}.
There is a 15~cm overlap between adjacent walls of the same layer to avoid dead zones in the detector. The 3 walls in the center of the TT are moved up to leave enough space for the calibration house and the Central Detector chimney. To fit in the available space below the cranes, these walls are also closer, with a separation of only 23~cm.

With this geometry, about 1/3 of all atmospheric muons passing through the CD will also cross 3 layers of the TT so that they can be tracked.
In these cases, the TT can be used as a veto, but for the remaining muons crossing the CD, the veto strategy depends only on the WCD and CD. One notable region where the TT veto is particularly effective is for
atmospheric muons entering the detector through the chimney region, which is well covered by the TT, and might present difficulties for the other subsystems in some classes of events.

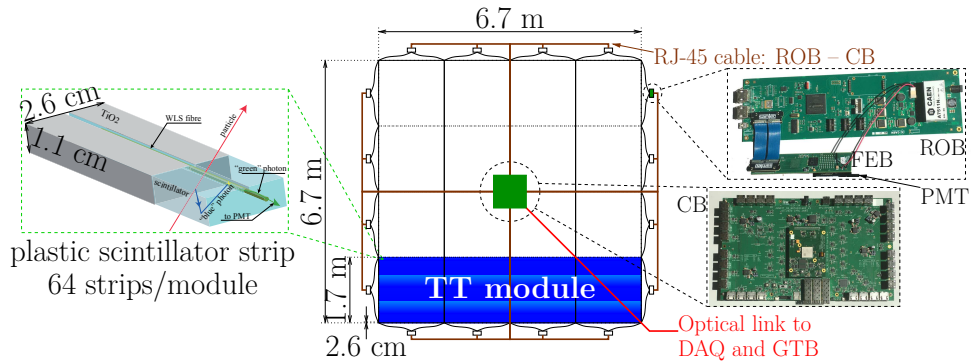
\begin{figure}[hb]
  \centering
    \resizebox{0.7\textwidth}{!}{%
\begin{picture}(0,0)%
\includegraphics{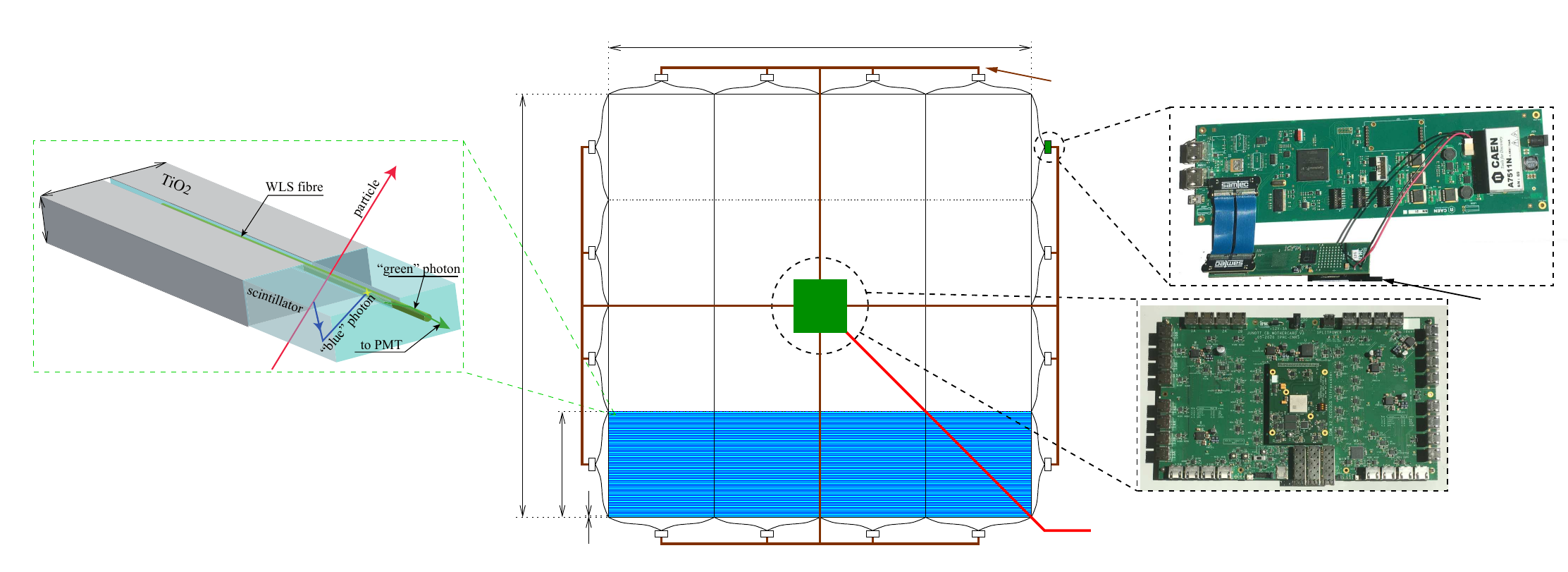}%
\end{picture}%
\setlength{\unitlength}{3947sp}%
\begingroup\makeatletter\ifx\SetFigFont\undefined%
\gdef\SetFigFont#1#2#3#4#5{%
  \fontsize{#1}{#2pt}%
  \fontfamily{#3}\fontseries{#4}\fontshape{#5}%
  \selectfont}%
\fi\endgroup%
\begin{picture}(17800,6480)(-5708,-5776)
\put(151,-2761){\rotatebox{90.0}{\makebox(0,0)[b]{\smash{{\SetFigFont{34}{40.8}{\rmdefault}{\mddefault}{\updefault}{\color[rgb]{0,0,0}6.7 m}%
}}}}}
\put(3601,239){\makebox(0,0)[b]{\smash{{\SetFigFont{34}{40.8}{\rmdefault}{\mddefault}{\updefault}{\color[rgb]{0,0,0}6.7 m}%
}}}}
\put(-4574,-1261){\rotatebox{17.0}{\makebox(0,0)[b]{\smash{{\SetFigFont{34}{40.8}{\rmdefault}{\mddefault}{\updefault}{\color[rgb]{0,0,0}2.6 cm}%
}}}}}
\put(-5192,-1895){\rotatebox{340.0}{\makebox(0,0)[lb]{\smash{{\SetFigFont{34}{40.8}{\rmdefault}{\mddefault}{\updefault}{\color[rgb]{0,0,0}1.1 cm}%
}}}}}
\put(7201,-3136){\makebox(0,0)[rb]{\smash{{\SetFigFont{25}{30.0}{\rmdefault}{\mddefault}{\updefault}{\color[rgb]{0,0,0}CB}%
}}}}
\put(11926,-2086){\makebox(0,0)[rb]{\smash{{\SetFigFont{25}{30.0}{\rmdefault}{\mddefault}{\updefault}{\color[rgb]{0,0,0}ROB}%
}}}}
\put(9826,-2311){\makebox(0,0)[lb]{\smash{{\SetFigFont{25}{30.0}{\rmdefault}{\mddefault}{\updefault}{\color[rgb]{0,0,0}FEB}%
}}}}
\put(11101,-2986){\makebox(0,0)[lb]{\smash{{\SetFigFont{25}{30.0}{\rmdefault}{\mddefault}{\updefault}{\color[rgb]{0,0,0}PMT}%
}}}}
\put(6226,-436){\makebox(0,0)[lb]{\smash{{\SetFigFont{25}{30.0}{\rmdefault}{\mddefault}{\updefault}{\color[rgb]{.5,.17,0}RJ-45 cable: ROB~--~CB}%
}}}}
\put(6676,-5311){\makebox(0,0)[lb]{\smash{{\SetFigFont{25}{30.0}{\rmdefault}{\mddefault}{\updefault}{\color[rgb]{1,0,0}Optical link to}%
}}}}
\put(601,-4561){\rotatebox{90.0}{\makebox(0,0)[b]{\smash{{\SetFigFont{34}{40.8}{\rmdefault}{\mddefault}{\updefault}{\color[rgb]{0,0,0}1.7 m}%
}}}}}
\put(976,-5761){\makebox(0,0)[b]{\smash{{\SetFigFont{34}{40.8}{\rmdefault}{\mddefault}{\updefault}{\color[rgb]{0,0,0}2.6 cm}%
}}}}
\put(-2924,-4636){\makebox(0,0)[b]{\smash{{\SetFigFont{34}{40.8}{\rmdefault}{\mddefault}{\updefault}{\color[rgb]{0,0,0}64 strips/module}%
}}}}
\put(6676,-5761){\makebox(0,0)[lb]{\smash{{\SetFigFont{25}{30.0}{\rmdefault}{\mddefault}{\updefault}{\color[rgb]{1,0,0}DAQ and GTB}%
}}}}
\put(-2924,-4036){\makebox(0,0)[b]{\smash{{\SetFigFont{34}{40.8}{\rmdefault}{\mddefault}{\updefault}{\color[rgb]{0,0,0}plastic scintillator strip}%
}}}}
\put(3601,-4711){\makebox(0,0)[b]{\smash{{\SetFigFont{34}{40.8}{\rmdefault}{\mddefault}{\updefault}{\color[rgb]{1,1,1}\bfseries TT module}%
}}}}
\end{picture}%
        }
        \caption{Schematics of a TT wall, with detail of a plastic scintillator strip (left), and photos of the TT wall electronics (right). Dimensions of a TT wall, a TT module (one of the modules is shown in blue) and a plastic scintillator strip are also drawn.}
        \label{fig:veto:tt:wall}
\end{figure}

Each TT wall is made using a total of 512 strips of plastic scintillator.
Every 64 plastic scintillator strips, with dimensions
${6.7~\textrm{m} \times 2.6~\textrm{cm} \times 1.1~\textrm{cm}}$,
are grouped in a TT module.
Wavelength-shifting fibers are placed in every strip, as shown on the left part of
Fig.~\ref{fig:veto:tt:wall}.
These fibers are read from both sides by 64-channel multi-anode photomultipliers
placed at each end of the TT modules, with every fiber directly in front of a channel
of the multi-anode photomultiplier.
A group of 4 TT modules are placed side by side to form a ${6.7 \times 6.7}$~m$^2$ square,
forming a TT plane.
Two TT planes are placed on top of each other, rotated by 90$^\circ$ to form a TT wall, as shown
in the central part of Fig.~\ref{fig:veto:tt:wall}.

The natural radioactivity of the rock in the JUNO cavern is expected to
create a ${\sim 50}$~kHz per PMT background rate in the TT, based on
the $^{238}$U, $^{232}$Th, and $^{40}$K concentrations found in the
rock prospected from the cavern site, which are two orders of magnitude higher than
in the LNGS underground laboratory.
This rate per PMT corresponds to a
${\sim 2}$~MHz full detector rate, which is to be compared with the expected
${\sim 3}$~Hz of atmospheric muons passing through the TT.
The new TT readout electronics is designed to work under these conditions, and
remove quickly large fractions of events produced by this background.

The TT readout electronics chain is composed of 4 different boards.
The Front-End Board (FEB) and the Read-Out Board (ROB) are located next to the
multi-anode 64 channels PMT, at each end of the modules.
These cards perform the PMT digitization, the slow control and provide the high-voltage (HV) supply.
Digitization of any charge detected with the ROB takes about 7~$\mu$s.
In the center of every wall, grouping 16 FEB~--~ROB pairs is placed a Concentrator Board (CB),
as shown in Fig.~\ref{fig:veto:tt:wall}.
The CB does the timestamping of the hits observed and is responsible for
the first level trigger (L1) by requiring 2D coincidences in the same wall.
If the CB does not find a coincidence, it can reset the acquisition being performed
on the FEB~--~ROB pairs within 300~ns in order to reduce the acquisition dead time.
Finally, the Global Trigger Board (GTB) will be placed near the center of the detector to
provide a global (L2) trigger to the TT by requiring at least 3 aligned walls on different layers
to have an L1 trigger before the event is accepted.
As is the case of the CB, the GTB can also reset the acquisition in case no coincidence is found. However, this feedback is provided within about 1~$\mu$s  in this case.

The L1 trigger produced by the CB reduces the detected rate by about one order of magnitude,
while the L2 trigger produced by the GTB further reduces the rate by two orders of magnitude, reaching a ${\sim 2}$~kHz detection rate
for radioactive events. As discussed above, both triggers reset the acquisition in the FEB~--~ROB
after about 300~ns and 1~$\mu$s, respectively, to reduce the detector's dead time in case that no coincidence is found. While this rate is still about 3 orders of magnitude larger than the expected atmospheric $\mu$ rate crossing the TT, the TT off-line reconstruction is able to further reject this background by
requiring the alignment of the crossing points between individual strips, rather than of the alignment of walls, which is done by the L2 trigger.

Thanks to the small granularity of the TT of ${2.6 \times 2.6}$~cm$^2$,
using this trigger logic and the TT reconstruction,
the efficiency for the TT to reconstruct atmospheric muons
is of about 93\% and their median angular resolution is of 0.20$^\circ$.
By also requiring a corresponding signal to be seen in the CD or WCD, the TT provides a well
reconstructed muon sample with purity $>99$\% that can be used to calibrate and to tune
reconstruction algorithms in the CD and WCD.

Well-reconstructed muons from the TT will also be used to measure the distribution of the distance
between the muon and the cosmogenic isotopes produced in the CD, which is a key component in
deciding the size of the muon veto cylinder used in the analysis.
For these well-tagged isotopes ($^9$Li and $^8$He), another important parameter,
the energy distribution, can also be measured and used in the JUNO physics analysis.
In addition to that, the TT should contribute to the estimation of the
fast-n rate experimentally, thus avoiding uncertainties in the simulation of neutron production, propagation and interaction in the detector.

\subsection{Liquid scintillator}
\label{subsec:ls}

The preparation of the liquid scintillator (LS) requires high quality of all the chemicals to satisfy the strict requirements of JUNO. There are three components in the JUNO LS recipe: Linear Alkyl Benzene (LAB), 2,5-diphenyloxazole (PPO), and 1,4-bis(2-methylstyryl)benzene (bis-MSB). The optimal LS composition was determined to be the purified solvent LAB with 2.5 g/L PPO and 3 mg/L bis-MSB~\cite{Abusleme:2020bbm}. More than five years of R\&D efforts were devoted to raw material procurement and the design of the purification plants in order to produce the best possible LS in terms of optical and radio-purity properties. For the raw materials, one LAB factory succeeded in providing high-quality LAB with a long attenuation length at 430~nm. This company already produced 20~tons special LAB for JUNO LS pilot production, achieving an attenuation
length of around 25~m at 430~nm. The LAB will be transported by dedicatedly cleaned ISO-tanks with nitrogen covering, before being stored in a 5000~ton stainless steel tank and purged by nitrogen onsite. Similar long-term cooperation was established with one PPO supplier, which was able to provide purified PPO with residual contamination in $^{238}$U and $^{232}$Th around 1~ppt after a great effort. 

The requirement on the U/Th radiopurity of the LS is $1\times10^{-15}$~g/g  for the reactor neutrino studies and $1\times10^{-17}$~g/g for the solar neutrino studies. It is difficult to quantitatively demonstrate that the LS could be purified to such a low level with our purification systems, until we can use the data of the JUNO detector. Therefore, the minimum requirement is determined to be $1\times10^{-15}$~g/g upon filling, with a target radiopurity of $<1\times10^{-16}$~g/g. The LS will be further purified online, with the water extraction and gas stripping systems, to reach the goal of $1\times10^{-17}$~g/g.

\subsubsection{Purification systems}

\begin{figure}[hbt]
    \centering
    \includegraphics[width=0.9\columnwidth]{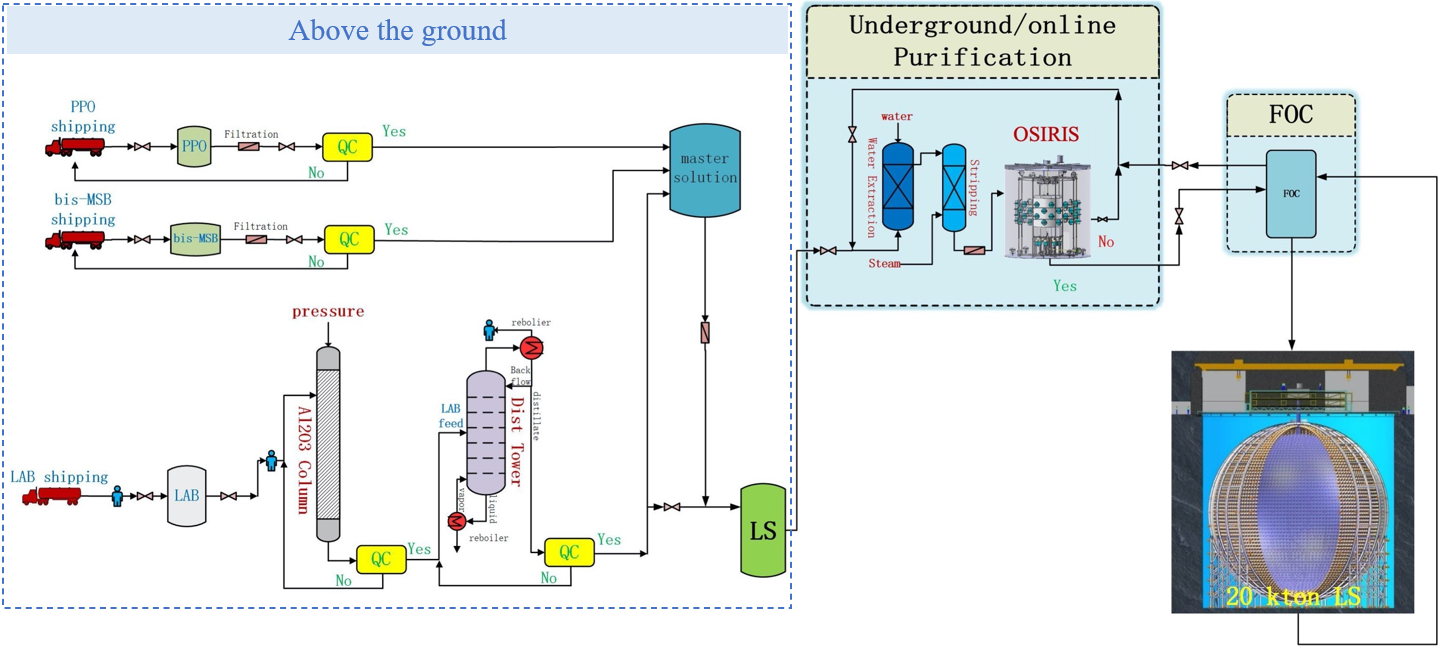}
    \caption{Flowchart of the liquid scintillator processing system.
    \label{fig:LSpurification}}
\end{figure}

To further improve optical and radio-purity properties of LS, a combined system of purification plants
has been designed as shown in Fig.~\ref{fig:LSpurification}. The plants will be operated in a sequence
following a strategy optimized with a successful test done at Daya Bay using pilot plants \cite{Lombardi:2019epz}.
In the JUNO surface area, raw LAB will pass through the alumina columns and the distillation plant, before being mixed with PPO and bis-MSB. The resulting LS will be sent underground through an electro-polished pipe for the last two purification stages, water extraction and stripping plants.
After the final quality check in the Online Scintillator Internal Radioactivity Investigation System (OSIRIS) detector, the LS will be sent to the FOC system.

In order to reduce radon emanation and diffusion, the internal surface roughness of all the equipments
is kept below 0.4~$\mu$m with mechanical polishing followed by electrochemical polishing.
Furthermore, all joints are realized with double O-ring flanges with internal nitrogen purging and
a tested overall leakage rate less than 10$^{-6}$~mbar$\cdot$l/s. In addition, besides normal
nitrogen, a high purity nitrogen system is designed for blanketing and running of all the plants.
The system consists of two Low Temperature Absorber (LTA) columns filled with ultralow radioactive background activated carbon. One LTA column can continuously work for more than 7~days at 50~Nm$^3$/h
flux rate before its activated carbon needs to be regenerated. The radon content of the high purity
nitrogen can be less than 10~$\mu$Bq/m$^3$. The two-stage filters (220~nm and 50~nm) are used in each sub-system to remove small particulates in LS and to reduce the $^{238}$U and $^{232}$Th radioactive background.

\paragraph{Alumina column system}
The LAB will be purified by the alumina (Al$_2$O$_3$) columns to improve transparency. Based on the pilot plant experiments, we determined the final design of the alumina filtration system. The results show that the system can significantly improve the attenuation length of LAB without apparent increase of the radioactive background (below the measurement sensitivity of $\sim 10^{-15}$~g/g). The system consists of eight filtration columns, two alumina filling tanks, and three buffer tanks. The height of the column is 2.6~m, and the diameter is 60~cm. The column can work at a height-to-diameter ratio of 3:1 or 4:1 to keep a high purification efficiency. The alumina will be packed in vacuum aluminum film bags for protection from radon. The alumina will be filled in the filling tank via a glove box to minimize possible radon contamination.

\paragraph{Distillation system}
Distillation in a partial vacuum (5~mbar) is used in the second stage of the purification process to remove the heaviest radio-impurities ($^{238}$U, $^{232}$Th and $^{40}$K) from the LAB and to
further improve its optical property in terms of absorbance and attenuation length.
After a preheating up to 190$^\circ$C inside a counter-current heat
exchanger (heat recovery), the LAB is delivered to the distillation column (7~m height and 2~m wide
with 6 sieve trays) where the purification process is carried out with a counter-current flow
of gaseous LAB produced inside a tube-bundle hot oil reboiler.

The gas stream is collected in the top of the column and liquefied in a condenser cooled with chilled water at 35$^\circ$C from a water-cooling tower. The purified LAB flow is partially sent back on the top of the column as reflux to increase the purification efficiency (up to 40\%). The distillation plant is designed to operate with a nominal discharge flow from the column bottom of 1--2\% of the input stream, in order to get a better compromise between the product purity and reasonable throughput.

The LAB distillation technique was proved effective in terms of radio-purity and optical transparency in the pilot plant experiments performed at Daya Bay~\cite{Lombardi:2019epz}.

\paragraph{Mixing system}
The mixing system is located in the surface LS hall. It is responsible for dissolving PPO and bis-MSB into LAB to form LS master solution batch by batch. Acid/water washing is also performed to the master solution to purify PPO and bis-MSB at this stage. Possible master solution distillation techniques to remove $^{238}$U/$^{232}$Th from PPO are under study. Finally, the master solution will be diluted with LAB to form LS and transported to underground facilities. Besides double O-ring flanges, glove box feeding and magnetic fluid seal are used to ensure the good sealing performance of the system.

\paragraph{Water extraction system}
The water extraction system is used to remove $^{238}$U and $^{232}$Th backgrounds from the LS, especially impurities
introduced by PPO and bis-MSB. The water extraction system has been designed based on laboratory studies, pilot plant experiments, and theoretical calculations. It is a turbine extraction tower of 13 meters high and 1 meter in diameter, with five theoretical stages. Water and LS are fully mixed by turbine stirring. The LS flow rate is 7~m$^3$/h, while the water flow rate is adjustable within 1.4--2.3 m$^3$/h. The purified LS is first fed into a static separation tank before the final product tank. Continuous nitrogen purge in the LS in the product tank ensures the water content below 200 ppm.
In order to reduce the interfacial fouling produced by extraction, there is a sewage discharge port at the LS-water interface. The water with ultra-low $^{238}$U/$^{232}$Th and Rn concentration will be provided by an Ultrapure Water (UPW) system, which is based on the highest UPW requirements and technologies for the semiconductor industry. The $^{238}$U and $^{232}$Th of UPW will be $<1\times10^{-15}$~g/g, which corresponds to $<1\times10^{-16}$~g/g LS radiopurity given the water/LAB partition coefficient being 10, and could be upgraded in the future.

\paragraph{Stripping system}
The stripping plant is the final stage of the purification procedure. The main purpose of the plant is the removal of radioactive gases and gaseous impurities from the liquid scintillator phase, using a gaseous stream of nitrogen and/or superheated steam in counter-current flow mode. The purification process is performed inside a 9-meters-high stripping column, filled with AISI~316 L metal pall rings. LAB entering the plant is pumped at 7~m$^3$/h by a magnetic driven pump through a set of input fine filters (50~nm pore size) and heated up by tube-bundle heat exchangers up to 90$^\circ$C, to reduce the LAB viscosity and increase the process efficiency. The stripping column is fed at the top with pre-heated liquid LAB while at the bottom with an adjustable mixture of ultra-pure nitrogen and UPW steam (30--60 kg/h). The stripping process is performed in a partial vacuum, at about 250~mbar. During pilot plant tests, carried out at the Daya Bay laboratory in 2018, this combination of parameters allowed to achieve 95\% purification efficiency in Rn removal \cite{Lombardi:2019epz}. The full-size plant has been improved with a 50\% higher stripping column; therefore we expect better efficiency.

\subsubsection{OSIRIS}
The Online Scintillator Internal Radioactivity Investigation System (OSIRIS) is a stand-alone detector to monitor the radiopurity of the LS while the JUNO CD is filled and to confirm the proper operation of the purification plants. The aim is to guarantee that the concentrations of $^{238}$U and $^{232}$Th in the LS do not exceed the given limits of 10$^{-15}$~g/g or 10$^{-16}$~g/g for the IBD or solar neutrino measurement, respectively. The measurement is based on observing the easy-to-identify fast coincidence decays of $^{214}$Bi-$^{214}$Po ($\tau\sim164$~$\mu$s) and $^{212}$Bi-$^{212}$Po ($\tau\sim0.43$~$\mu$s) in $\sim20$~m$^3$ of LS, or $\sim17$~tons of LS. The decays indicate the presence of $^{238}$U and $^{232}$Th, respectively. Based on identifying a handful of coincidence events per day, OSIRIS can reach the IBD sensitivity level within a few days and the solar level within 2--3 weeks.

\begin{figure}[hbt]
    \centering
    \includegraphics[width=0.5\columnwidth]{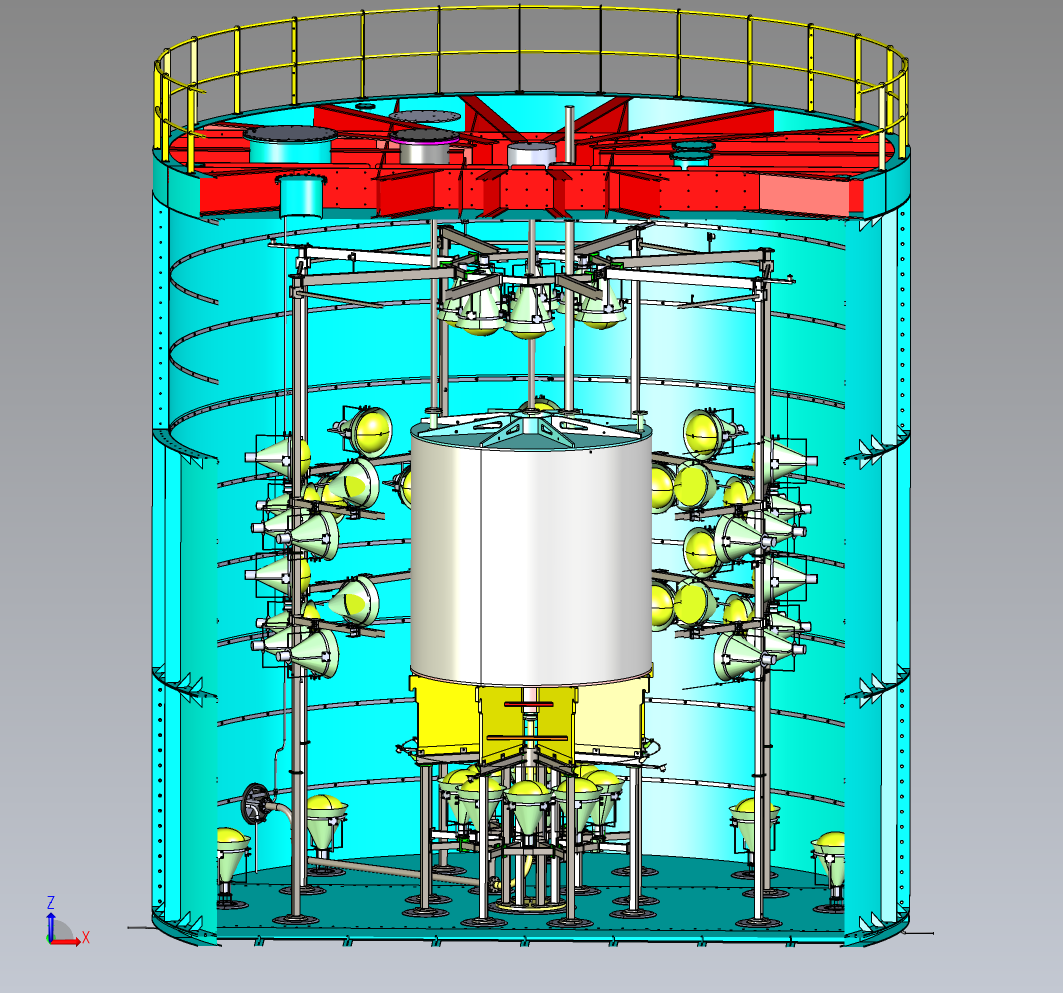}
    \caption{Scheme of the OSIRIS detector.
    \label{fig:OSIRIS}}
\end{figure}

OSIRIS is located in a bypass to the main LS line, permitting to sample a relevant fraction ($\sim1/6$) of the LS processed in the purification plants. The detector design is illustrated in Fig.~\ref{fig:OSIRIS}. It consists of a cylindrical Acrylic Vessel (AV) with a height and diameter of 3~m holding the LS. The AV is mounted inside a water-filled steel tank (height and diameter 9~m), which provides 3~m.w.e.\ of shielding against external radiation. This is sufficient to reduce accidental coincidences of external $\gamma$ events to a negligible level. The main background to the measurement will be formed by the radon contamination of the LS that stems from the emanation of components during the LS purification process.

The scintillation light from the decay events is collected by 64 20-inch PMTs (Hamamatsu R12860) that are mounted to a stainless-steel frame surrounding the AV at a distance of 1.3~m facing towards the AV. In addition, 12 20-inch PMTs are directed outwards and act as a muon veto detector. These two subsets of PMTs will be optically separated by a PET foil installed to the stainless-steel frame. The expected photoelectron yield is $\sim$275 p.e.\ at 1~MeV, corresponding to an energy resolution of $\sim6$\% at 1\,MeV for the inner detector. The outer PMTs provide a muon tagging efficiency of $>85$\%.

One design feature of OSIRIS is that the data acquisition electronics is mounted directly to the back of each PMT, a novel concept called intelligent PMT (iPMT). Each iPMT acts as an individual detector with a large dynamic range (1--10$^3$ p.e.). The analog signal of the PMT is continuously sampled inside the iPMT using a highly integrated receiver chip (VULCAN). In case the trigger threshold of an individual PMT is exceeded, the waveform is sent via the network to the DAQ system. There, the data of all PMTs is time-sorted, and a software trigger extracts the physical events, enabling a very flexible implementation.

The OSIRIS detector holds two calibration systems. The source-insertion system re-uses one of the Automatic Calibration Units (ACU) of the Daya Bay experiment. It will be mounted at an off-axis location on the top of the OSIRIS tank and is connected to the AV via an LS-filled pipe. The system is equipped with two capsules containing weak radioactive $\gamma$-sources and one LED to enable energy and position calibration. A second system feeds the pulses of an external picosecond laser system via optical fibres to an array of diffusers mounted to the steel frame holding the PMTs. It will be used for precise timing and single p.e.\ charge calibration.

OSIRIS will be operated in two measuring modes: During commissioning of the LS systems, single batches of LS will be measured for periods of up to 3 weeks for a precise determination of scintillator quality and purification efficiency down to the solar neutrino sensitivity level. During JUNO filling, a continuous flow mode is foreseen where new LS is constantly added at the top of the AV while the already-monitored LS is drained from below. Heating the inflowing LS will help a stratification of the liquid inside the AV and thus prevent intermixing of the materials in the stable temperature layers. Sensitivity will be close to the IBD level and thus sufficient to discover sudden spikes in radon levels due to the occurrence of leaks in the LS lines or malfunctions of the purification plants.

\subsection{20-inch PMT Test and Instrumentation}
\label{subsec:PMTInstrum}

The ambitious goal of the JUNO detector, to determine the neutrino mass ordering, requires an excellent energy resolution of 3\% at 1 MeV. From the perspective of PMTs, the key ingredients to reaching the design energy resolution are a large photodetector area coverage, high photon detection efficiency, low dark noise, and stable operation of the whole PMT system. Requirements for the PMTs were formulated in the production contracts with NNVT and Hamamatsu companies to produce 15,000 MCP PMTs and 5,000 dynodes PMTs, respectively. Based on that, procedures and devices were developed to test the PMTs after they were received from the producers, and extensive tests were carried out to ensure the PMTs' performance and quality. After then, the PMTs were instrumented with high voltage dividers, waterproof sealings, and protection covers to eventually work in water for the JUNO experiment. Below we present in brief the PMT test and its instrumentation.

\subsubsection{20-inch PMT Test}

\paragraph{Acceptance Test}
Many tests are helpful for PMT evaluation. Of those, the acceptance test is the first test after delivery and also the most important one. The purpose of this test is to check whether the delivered PMTs from the producers can meet the JUNO requirements or not. The acceptance test consists of a visual inspection and a performance test. The visual inspection aims to identify the bubbles, cracks, or any other defects on the bare PMT, which are weak points for pressure in the water, with the dimension and weight being measured. The PMT performance test features a measurement covering all the typical parameters: Photon Detection Efficiency (PDE), Dark Count Rate (DCR), Gain, Peak-to-Valley ratio (P/V), rise time, fall time, Transit Time Spread (TTS), prepulse, and afterpulse rate, etc. Only those PMTs that pass the acceptance test are accepted as the JUNO PMTs. The other PMTs will be returned to the producers.

\begin{figure}[ht]
\begin{center}
  \begin{minipage}[ht]{0.272\linewidth}
  \centering
          {\includegraphics[width=\linewidth]{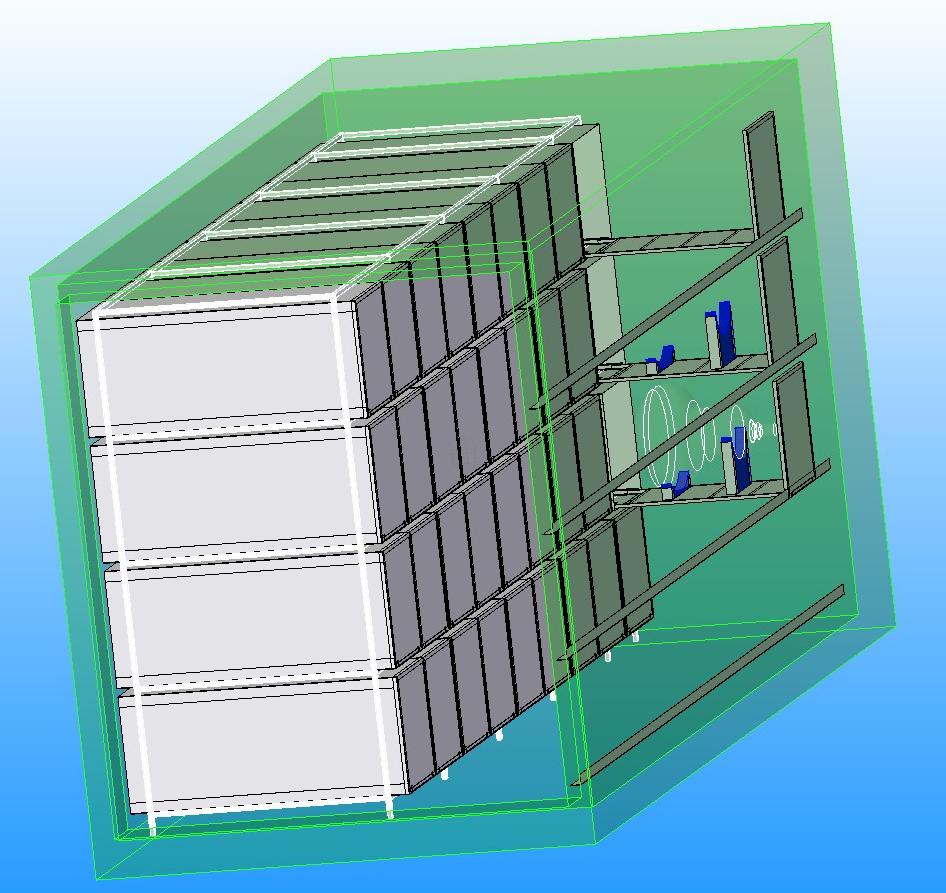}}
	  \label{fig:Container_Schematics}
  \end{minipage}
  \begin{minipage}[ht]{0.512\linewidth}
  \centering
    {\includegraphics[width=\linewidth]{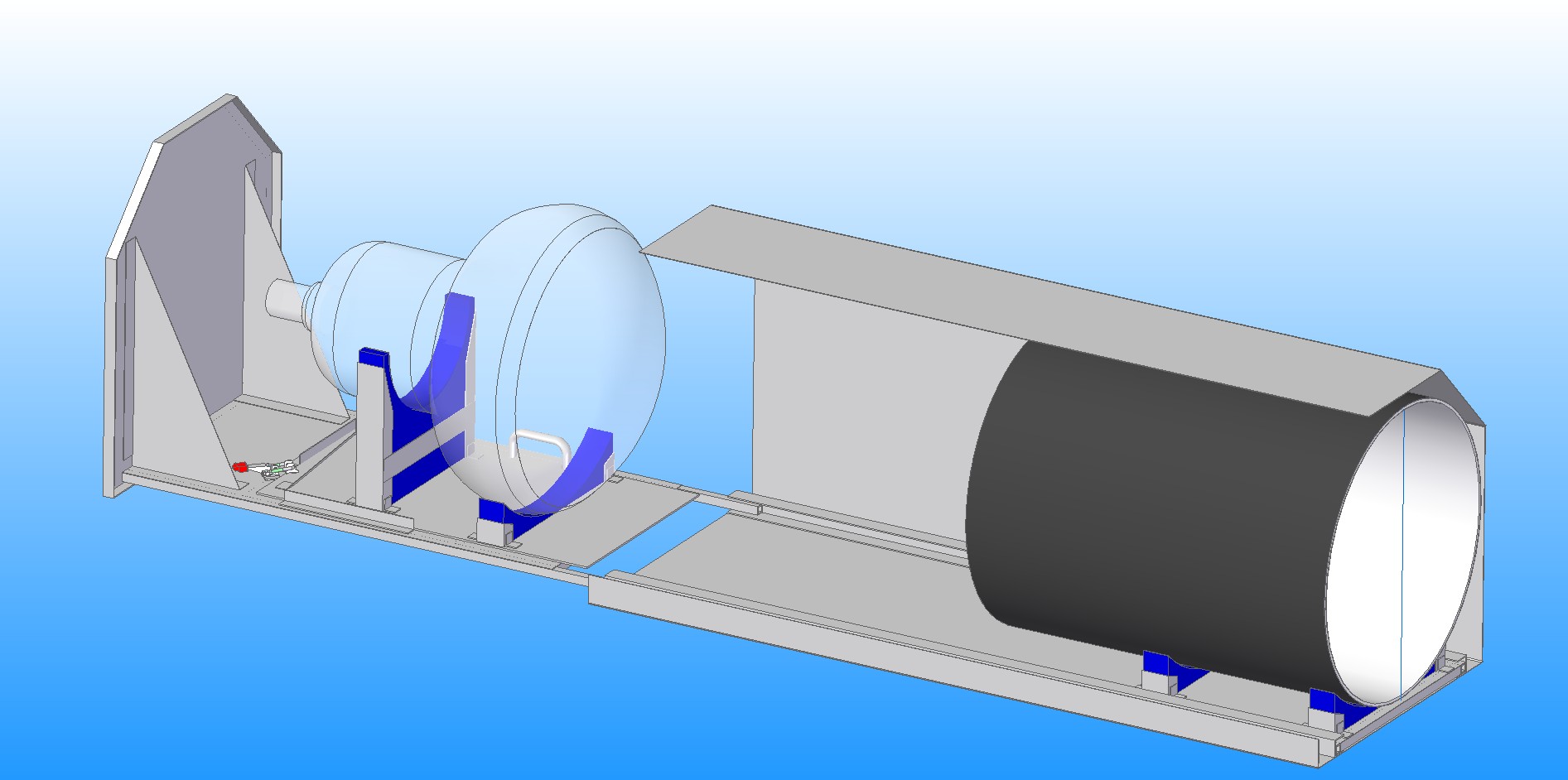}}
    \label{fig:Layout_drawer_box}
  \end{minipage}
	\caption{\textit{Left:} Conceptual picture of the container with the shelf system including 36 drawer boxes. \textit{Right:} Schematic view of the final design of the drawer system (for illustrative reasons one side of the drawer box is not depicted).}
	\label{fig:ContainerConcept}
\end{center}
\end{figure}

\paragraph{Container System}
This system consists of four 20-foot containers equipped with a high-powered Heating, Ventilation and Air Conditioning (HVAC) unit. The interior of the containers is protected from the Earth's magnetic field by a silicon-iron shielding providing a residual magnetic field of less than 5$\;\mu$T. Inside the containers, a shelf structure houses 36 drawer boxes (see Fig.~\ref{fig:ContainerConcept}). The drawer boxes act as the individual measurement devices for the mass testing of the PMTs. Each of them employs two light sources; a self-stabilized LED illuminating the whole photocathode and fibre connected to a picosecond laser used only for the TTS measurement. The containers are operated in a 24-hour cycle. After loading the PMTs in the container, an automatic measurement sequence is started. To monitor the system, 5 reference PMTs per container are selected in different drawer boxes during each of these measurements.

The whole system (mechanics, light sources, electronics) has been designed to provide stable and comparable measurement conditions for all drawer boxes. This has been confirmed by several consistency checks showing that the PDE can be measured with an absolute accuracy of about 1\%. A detailed description of the system can be found in Ref.~\cite{Wonsak:2021uum}.

Of the four container systems, the first and second containers are used for mass testing of the PMTs, the third container is used to test the long-term stability of a sampled PMTs, and the fourth container is equipped with the JUNO electronics and to test the combined performance of PMTs and electronics.

\paragraph{Scanning Station}
The scanning station is a system designed for the large photocathode PMTs' zonal testing and characterization. This system provides data on the PDE non-uniformity along the photocathode. It is also used to obtain complementary information to reveal potential problems for PMTs coming from the container system. It consists of the base, support, and rotating frame with 7 self-stabilized light sources (LEDs). The system is set up in a dark room, and Helmholtz coils are mounted on the walls to compensate for the Earth's magnetic field. Therefore the scanning station can also be utilized to study the PMT sensitivity to the magnetic field.

\begin{figure}[ht]
\begin{center}
  \begin{minipage}[ht]{0.265\linewidth}
  \centering
          {\includegraphics[width=\linewidth]{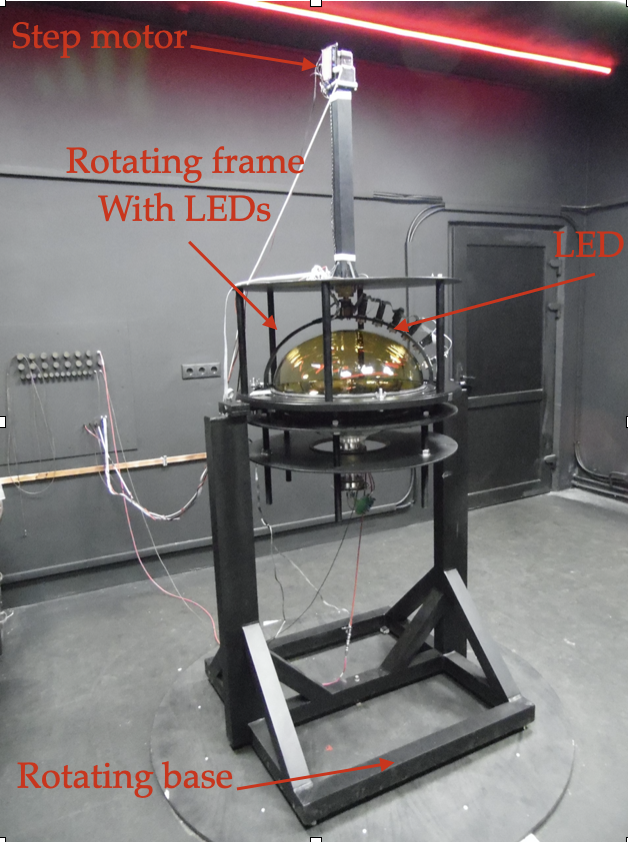}}
      \label{fig:scan_station}
  \end{minipage}
  \begin{minipage}[ht]{0.4\linewidth}
  \centering
    {\includegraphics[width=\linewidth]{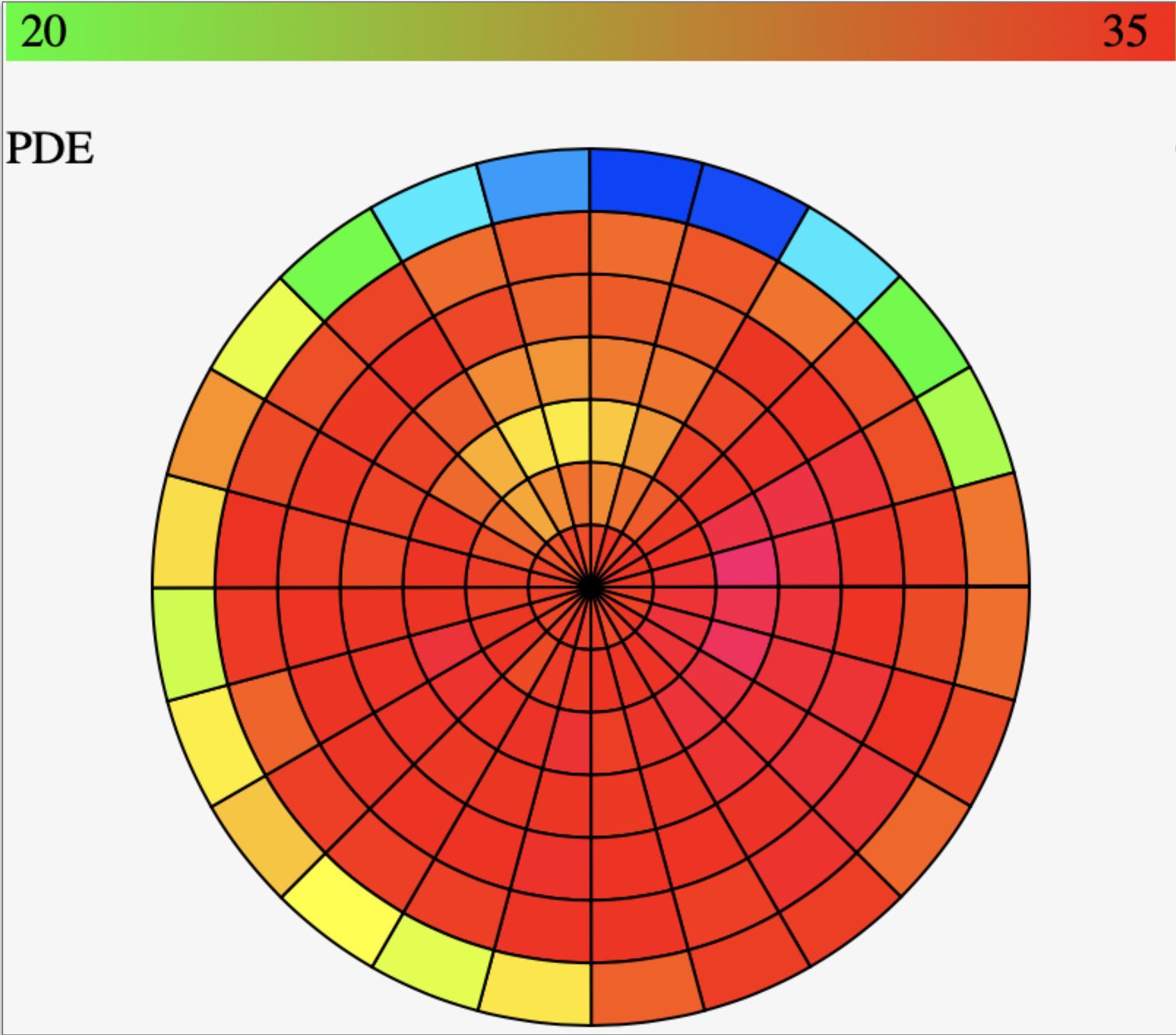}}
    \label{fig:scan_pmt}
  \end{minipage}
  \caption{20-inch PMT scanning: left - Overall view of the scan station; right - Scanning map for a dynode PMT}
  \label{fig:scanning}
\end{center}
\end{figure}

PMTs are put into the scanning station and illuminated by short, low-intensity light flashes from LEDs installed on a movable arc, rotated by a step motor. Routine scanning is being done for 24 azimuthal positions, while 7 LEDs allow testing the PMT at 7 defined zenith angles. Therefore in total, there are 168 test points along the photocathode surface. At each point, charge spectra are obtained from which PDE and gain are extracted. LEDs are individually calibrated by a reference small PMT regularly. Fig.~\ref{fig:scanning} shows the scanning station and a scanned PDE map of a dynode PMT.

\paragraph{Afterpulse System}
PMT afterpulsing will deteriorate the PMT time response and affect the event identification and reconstruction in JUNO. An afterpulse test facility is located inside the darkroom of the scanning station and shares the mechanical and optical system but uses an independent readout by a high-speed waveform digitizer with a capture window of more than 21 $\mu$s.

\paragraph{Database System}
Besides the test systems mentioned above, a database system is important to manage the PMT test data and other useful data. Users can access the corresponding data through the Graphical User Interface (GUI) of each software subsystem. The data will be fed to data tables and statistical charts on various web pages to achieve data visualization.

\begin{figure}[htbp]
    \centering
  \includegraphics[width=.45\textwidth]{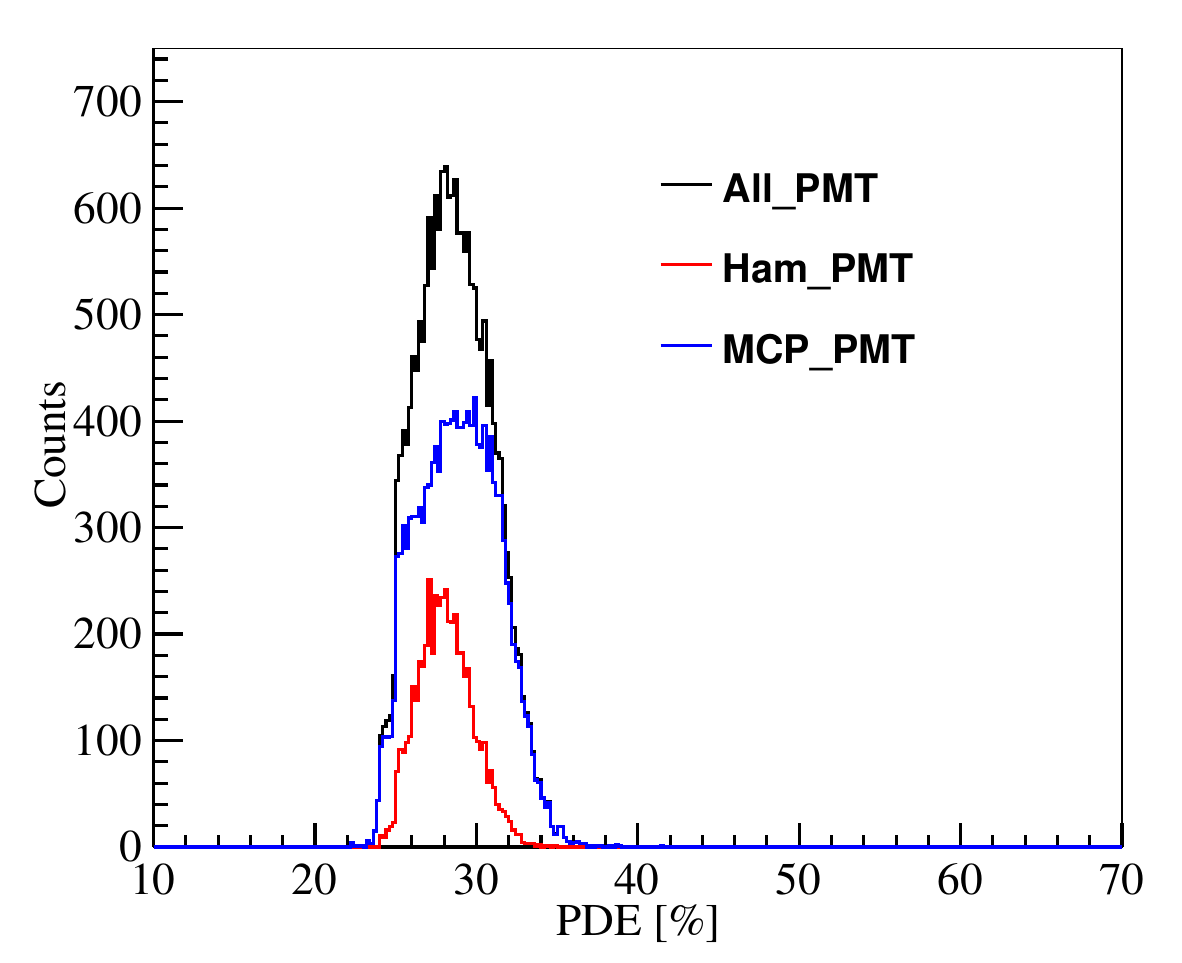}
    \caption{PDE distributions for all PMTs, MCP PMTs and dynode PMTs}
    \label{fig:bare_PDE}
\end{figure}

\paragraph{Test Results}
The acceptance tests started in mid-2017 at the Pan-Asia Testing and Potting Laboratory in Zhongshan city, Guangdong Province in South China. All the 20-inch PMTs have been tested. The average PDE reaches 28.7\%. The average PDEs for the MCP PMTs and dynode PMTs are
28.9\% and 28.1\%, respectively, as shown in Fig.~\ref{fig:bare_PDE}. The energy resolution and vertex reconstruction of the Central Detector are critical. Therefore, all 5,000 dynode PMTs, which have better timing resolution than the MCP PMTs, and 12,612 (out of 15,000) MCP PMTs with better performance are selected for the Central Detector. The average PDE of all 20-inch PMTs in the Central Detector will be 29.1\%, higher than the JUNO requirement of 27\%. A detailed description of the PMT selection strategy for the JUNO Central Detector can be found in Ref.~\cite{Wen:PMTSelections}, where the PDE, DCR, TTS, radioactivity, pre-pulse and afterpulse are evaluated with a single figure of merit based on their impacts to the energy resolution and equivalent physical sensitivity. Since most of these parameters are either not sensitive or in a narrow range for the procured PMTs, only the PDE and DCR are considered when selecting the PMTs for the Central Detector.

\begin{figure}[htbp]
    \centering
  \includegraphics[width=.45\textwidth]{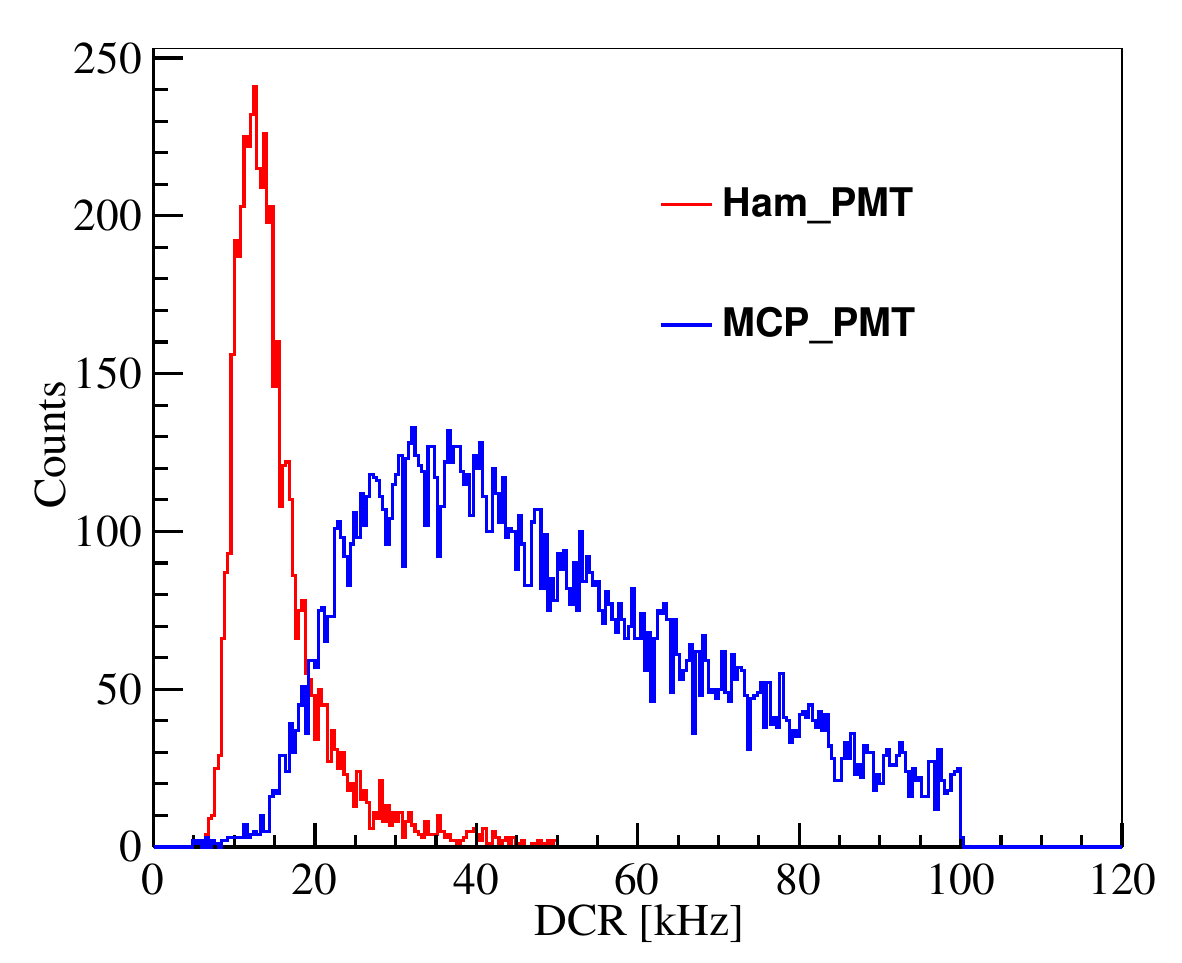}
    \caption{DCR distributions for MCP PMTs and dynode PMTs}
    \label{fig:bare_DCR}
\end{figure}

Fig.~\ref{fig:bare_DCR} shows the DCR for all PMTs. The mean value of DCR is 48.6~kHz for MCP PMTs and 15.3~kHz for dynode PMTs, which meet the JUNO requirement of less than 50~kHz average DCR for bare PMTs. When the PMTs are potted, a DCR reduction of up to 40\% is found. The DCR reduction might come from the better base connection (soldering before potting compared to the socket connection during tests) and smaller noise pickup.

\begin{figure}[htbp]
    \centering
  \includegraphics[width=.7\textwidth]{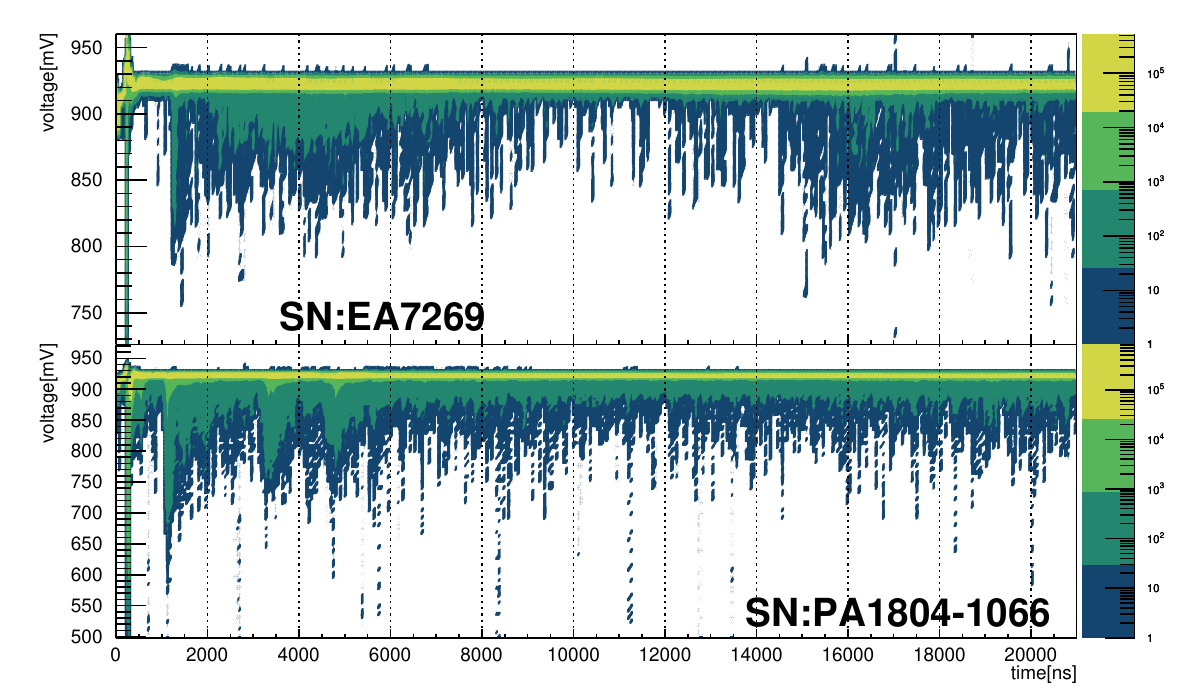}
    \caption{Two dimensional plot of the afterpulse waveforms, for a dynode PMT with serial number of EA7269 (upper) and a MCP PMT with serial number of PA1804-1066 (lower)}
    \label{fig:apwave2d}
\end{figure}

The afterpulse test mainly focuses on the ion-induced afterpulse signals, which appear at  500-20000 ns after the primary pulse.  We have observed 4 groups of afterpulse for MCP PMTs and 3 groups of afterpulse for dynode PMTs. Fig.~\ref{fig:apwave2d} shows the typical time structure of afterpulses, with about 20,000 events stacked into a 2-D plot. Based on the current measurements, the total afterpulse charge ratio for MCP (dynode) PMT is 6.7\% (11.6\%), which both meet the JUNO requirement of 15\%.

Other tests like the characterization after instrumentation, the long-term test of the sampled PMTs, and tests with the JUNO electronics are ongoing, without any apparent problems are observed.

\subsubsection{20-inch PMT Instrumentation}

Once a PMT passes the acceptance test, it will be installed with a high voltage divider, then a potting process is started. After that, the PMT will be protected by a cover to avoid chain implosion underwater.

\paragraph{PMT High voltage divider}
The High Voltage (HV) divider will provide the working voltage to PMT and collect the PMT signal to electronics. The JUNO HV divider is designed to operate at a positive voltage with less than 3000~V and 300~$\mu$A current considering the performance and stability for both types of PMTs. Following the JUNO requirements, metal film resistors and ceramic capacitors are selected for high reliability. The designed divider has been checked for Single Photoelectron (SPE) performance and linearity response up to about 1000 photoelectrons. All the produced HV dividers have undergone a burn-in selection procedure with 75$^{\circ}$C and maximum design voltage for around 300 hours. The HV divider will be directly soldered to the PMT pins.

\paragraph{PMT Potting}
To work reliably in water, the PMT tube end and its HV divider have to be potted to ensure a full waterproof tightness. According to JUNO requirements, the total PMTs loss in the first 6 years should be $<1$\%. A 1\% PMT loss at random positions and asymmetric positions will degrade the detector energy resolution from $3.02\%/\sqrt{E}$ to  $3.04\%/\sqrt{E}$ and  $3.06\%/\sqrt{E}$, respectively, with help of the detector calibration~\cite{Abusleme:2020lur}. The failure rate due to PMT potting is required to be  $<0.5$\% after 6-years of operation in the water of 40~m deep. A design with multiple waterproof sealings is applied to reach this goal, including a butyl tape layer, a polyurethane layer, and an epoxy layer, from the outside to the inside. A stainless-steel shell is used as the supporting frame, and a heat-shrinkable tube is used to fix the outmost layer of sealing. Extensive tests have been done to validate the design, including a long-term test with 0.8~MPa water pressure (corresponding to twice of the JUNO water depth), an accelerating aging test with heated water up to 70$^\circ$C ($\sim 100$ accelerating factor), and tests with pressurized air and vacuum. Test results have shown that the design meets JUNO requirements. A potting workshop has been built, and potting of PMTs has started.

\begin{figure}[htbp]
    \centering
  \includegraphics[width=.5\textwidth]{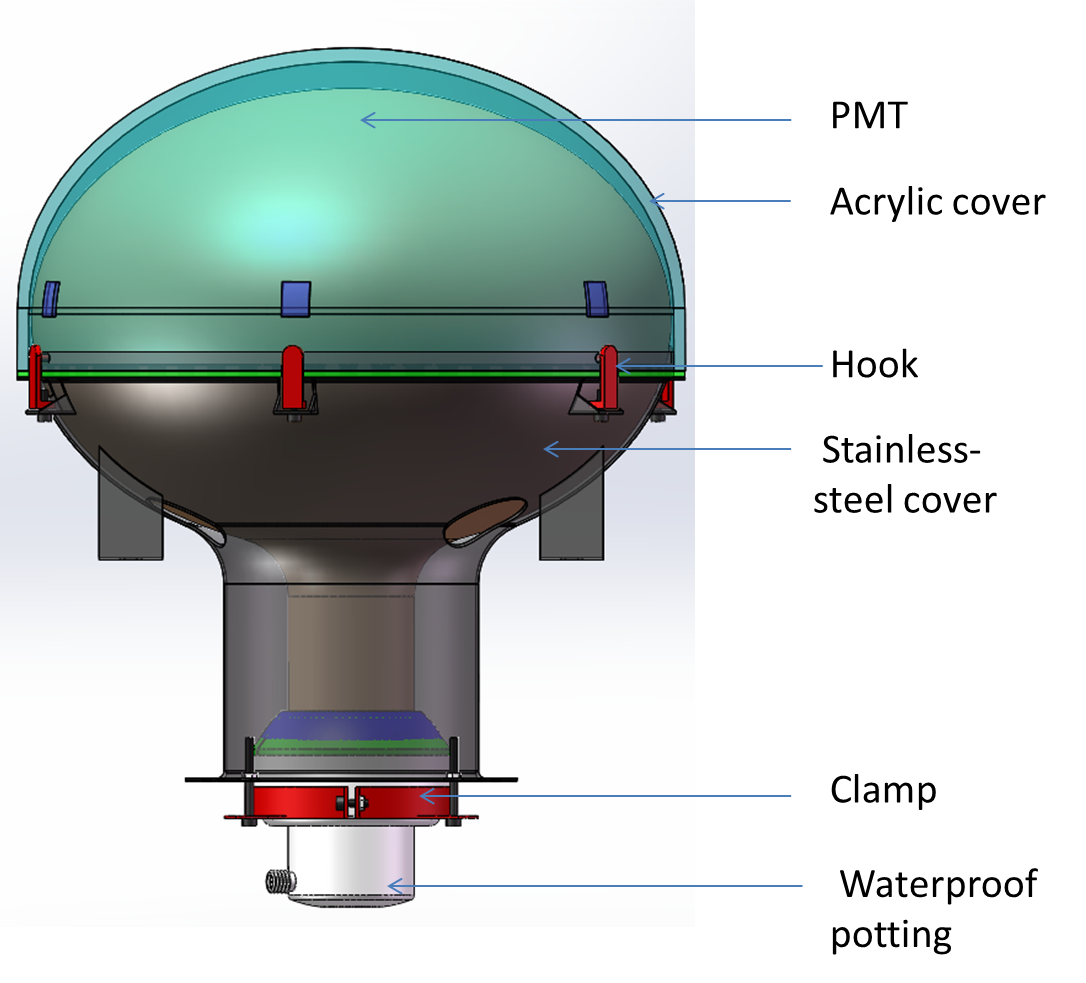}
    \caption{Schematics of an instrumented MCP PMT}
    \label{fig:PMT_assembly}
\end{figure}

\paragraph{PMT Protection}
There is a small chance that a PMT glass bulb will break underwater during long-term operation at a pressure of up to 0.5~MPa. In this case, a shockwave will be generated and will propagate in the water creating a chain reaction that leads to the implosion of a large number of PMTs, which happened years ago in SuperK. Therefore, a protection system was designed. Each PMT will be equipped with a cover, consisting of a $\sim$10~mm thick acrylic semisphere at the top and a $\sim$2~mm thick stainless steel semisphere at the bottom, connected by six hooks. The acrylic cover was prototyped by injection molding, and the transparency was measured to be better than 91\% at 420~nm in the air. The stainless steel cover was made by stamping with additional welding for the support and connection structures. The minimum distance between two neighboring protection covers is 3~mm. More than 20 underwater tests were done at a pressure of 0.5~MPa, and the protection system successfully prevented the generation of chain reactions.  Fig.~\ref{fig:PMT_assembly} shows a schematic view of an instrumented MCP PMT.

\subsection{The large PMT readout and trigger electronics}
\label{subsec:elec}

The initial design of the large PMT electronics~\cite{Djurcic:2015vqa} and the following R\&D program~\cite{bib:juno:bxscheme} have been driven by the main requirements of reconstructing the deposited energy in the LS with high resolution and a good linearity response over a wide dynamic range: from 1~p.e.\ for low energy events to 1000~p.e.\ for showering muons and muon bundles. In addition, a precise measurement of the photon's arrival time is mandatory~\cite{Djurcic:2015vqa}.

\begin{figure}[htbp]
\centering
  \includegraphics[width=0.7\columnwidth]{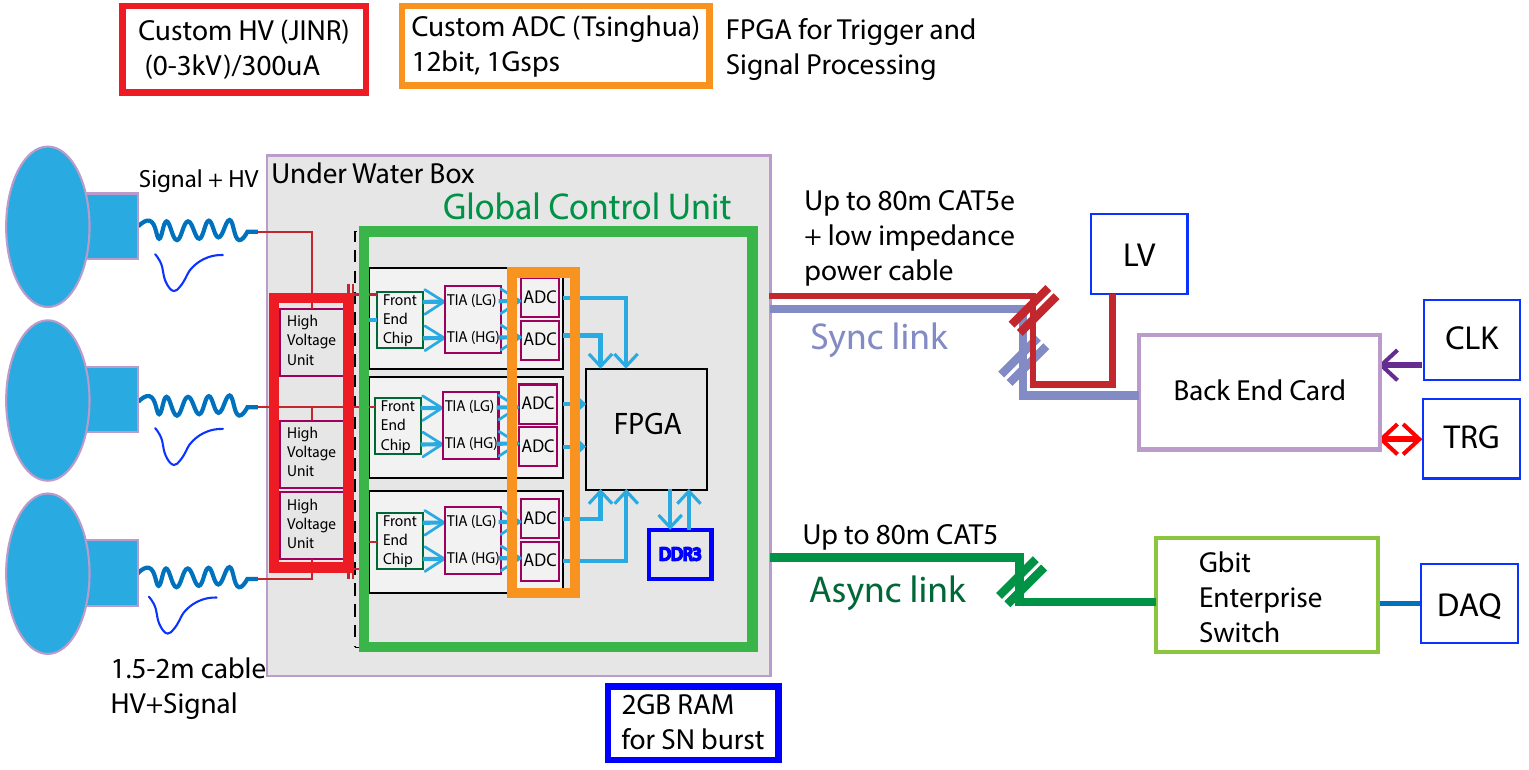}
  \caption{\label{fig:elec:scheme}Large PMT electronics scheme: the 'wet' electronics (left) is connected to the 'dry' electronics by means of Ethernet cables and a dedicated low impedance cable for power distribution.}
\end{figure}

A scheme of the large PMT electronics is reported in Fig.~\ref{fig:elec:scheme}. The current design is an optimization of previous developments~\cite{bib:juno:bxscheme} where the large PMT electronics is split into two parts: the 'wet' electronics located at few meters from the PMTs inside a custom stainless steel box (UWbox), and the 'dry' electronics in the electronics rooms. Since it is almost impossible to repair the 'wet' electronics underwater, its loss rate is required to be $< 0.5\%$ in 6 years. This requirement has generated important constraints on the reliability of the 'wet' electronics (see the discussion in~\cite{bib:juno:bxscheme}).

Three PMT output signals are fed to the front-end and readout electronics located inside the UWbox (see Fig.~\ref{fig:elec:scheme}). The PMT high voltage is provided for each PMT by a custom High Voltage module~\cite{bib:juno:bxscheme} located inside the UWbox. The analog signal is amplified and converted to digital with a 12~bit, 1~GS/s, custom ADC. The signal is further processed (local trigger generation, charge reconstruction and timestamp tagging) and stored temporarily in a local memory before being sent to the data acquisition (DAQ), once validated by the global trigger electronics. Besides the local memory available in the readout-board FPGA, a 2~GBytes DDR3 memory is available and used to provide a larger memory buffer in the exceptional case of a sudden increase of the input rate, which overruns the current data transfer bandwidth between the 'wet' electronics and the DAQ. This situation will certainly happen in case of a Supernova explosion not very far from JUNO, when the expected rate of neutrino interactions in the
LS is expected to increase suddenly and for a short time.

The readout electronics is connected to the 'dry' electronics through a so-called 'synchronous link', which provides the clock and synchronization to the boards and handles trigger primitives, and an 'asynchronous link' which is fully dedicated to the DAQ.
\begin{figure}[htbp]
\centering
  \includegraphics[width=0.7\columnwidth]{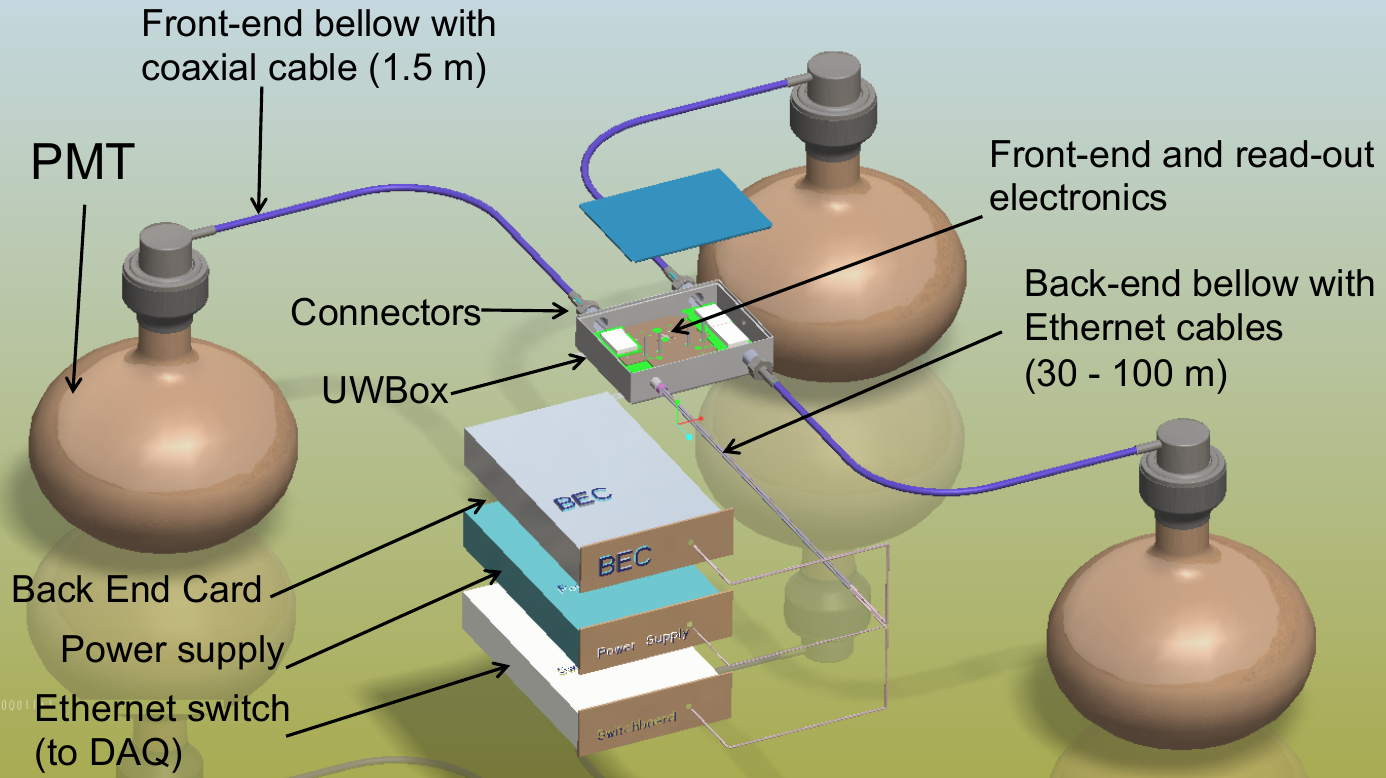}
  \caption{\label{fig:elec:techdraw}Large PMT electronics technical drawing.}
\end{figure}

As can be seen in Fig.~\ref{fig:elec:techdraw}, both cables connecting the UWbox to the PMTs and the UWbox to the 'dry' electronics run inside a stainless steel bellow, which completely isolates the cables from the water in which the electronics is immersed. The length of the bellows (and cables) is fixed to about 1.5~m from the PMTs to the UWbox and will have a variable length of 30~m to 100~m from the UWbox to the back-end. The range of lengths has been introduced to optimize and fulfill the installation constraints.

The large PMT electronics can work with a centralized 'global trigger' mode, where the information from the single 'fired' PMTs is collected and processed in a Central Trigger Unit (CTU). The latter validates the trigger based on a simple PMT multiplicity condition or a more refined topological distribution of the fired PMTs. Upon a trigger request, validated waveforms are sent to the DAQ event builder through the Gigabit Ethernet asynchronous link. The IPBus Core~\cite{bib:elec:ipbus} protocol is used for data transfer, slow control monitoring, and electronics configurations.

An alternative scheme is possible where all readout boards send their locally triggered waveforms to the DAQ, independently of each other. With this approach, all the digitized waveforms, including those generated by dark noise photoelectrons, will be sent to the DAQ. The two approaches are complementary, but the 'global trigger' mode is the default scheme during the operation of the JUNO detector.

\subsection{Small PMT system}
\label{subsec:SPMT}

\subsubsection{Overview and motivation}

\begin{figure}[htb]
\begin{center}
\includegraphics[width=50mm]{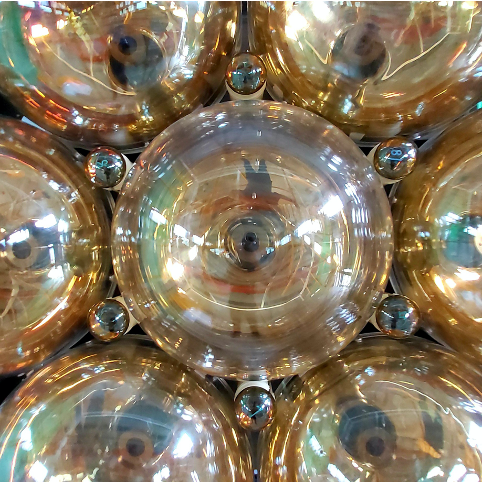}
\caption{\label{fig:spmtlpmt} Photograph showing a real size mock-up of the 3-inch photomultipliers interlaced between the 20-inch ones.}
\end{center}
\end{figure}

JUNO will be instrumented with a total of 25,600 3-inch PMTs installed in the space between the 20-inch PMTs, as demonstrated in Fig.~\ref{fig:spmtlpmt}. These small PMTs (SPMTs) will face the central detector and be uniformly distributed around it. They will be powered and read out in groups of 128 by front-end electronics and high-voltage splitter cards contained in a total of 200 Underwater Boxes (UWBs).

 The primary goal of the SPMT system is to provide a complementary set of sensors looking at the same events as the 20-inch PMTs. Due to their size, the SPMTs will consistently operate in a photon-counting mode in the energy region of interest for JUNO's neutrino oscillation goals (0-10 MeV), providing a measurement of the charge of every event in the detector with essentially no instrumental non-linearities. Even though they will see significantly less light than the large PMTs, the accumulation of statistics over time will allow them to serve as a calibration reference, reducing the impact of any systematic uncertainties in the energy response of the large PMT system and improving the energy resolution overall. Likewise, the SPMTs will be used to perform some physics measurements in a semi-independent way to the 20-inch PMT system, most notably the estimation of the solar oscillation parameters $\theta_{12}$ and ${\rm{\Delta }}{m}^2_{21}$. This will allow independently cross-checking the systematic uncertainties that are not common to the two systems.

 The SPMTs will also improve other aspects of the JUNO experiment. For example, they will extend the energy dynamic range of the detector and mitigate any potential non-linearities or saturation effects in the 20-inch PMTs at high energies. Due to their relatively good time resolution, they will also help improve event reconstruction, particularly for cosmic-ray muons. In the case of a supernova burst, they will provide complementary information about the rate of these events thanks to the much smaller pile-up.

\subsubsection{3-inch photomultipliers production and instrumentation}

26,000 3-inch PMTs (XP72B22) were ordered from Hainan Zhanchuang Photonics Technology Co., Ltd (HZC) in 2017 after an international bidding. The mass production of the PMTs started in January 2018, and finished in December 2019, averaging a production speed of $\sim$1,000 pieces per month. The characterization of the PMTs was performed concurrently with production in the factory as a joint effort between HZC and JUNO. Fifteen performance parameters were tracked at different sampling ratios, and the results showed good consistency with the specifications throughout the full production. In the end, only 15 PMTs were found to be unqualified and thus rejected. The radioactivity of the glass bulb was also monitored continuously to meet the experiment's stringent requirements~\cite{Cao:2021wrq}.

A divider was designed to distribute HV to the dynodes of the PMT. Considering that groups of 16 SPMTs are powered by the same HV source with a maximum current of 300~$\mu$A, fourteen 15-M$\Omega$ resistors were chosen to limit the current below 6.2~$\mu$A at a working positive HV up to 1,300~V. A coaxial cable with a high-density polyethylene jacket and a length of either 5~m or 10~m is soldered on each divider to transmit both HV and signal. The divider, the PMT pins, and the front-end of the cable are protected with polyurethane in an ABS plastic shell for better electric insulation. Butyl tape fixed by a shrinkable tube is used to seal the glass bulb in the shell, whereas the cable going through the shell is sealed by an inner layer of epoxy and an outer layer of low-pressure injection molding polymer plastic. The back-end of each cable is equipped with an MCX connector, and 16 cables are sealed together by mold-injected polyethylene, constituting a 16-channel connector, developed through a partnership with the Axon Cable company. This allows separating the SPMTs and the electronics, greatly relaxing the production, testing, transportation, and installation constraints.

\subsubsection{Underwater front-end electronics}

\begin{figure}[htbp]
\includegraphics[width=\columnwidth]{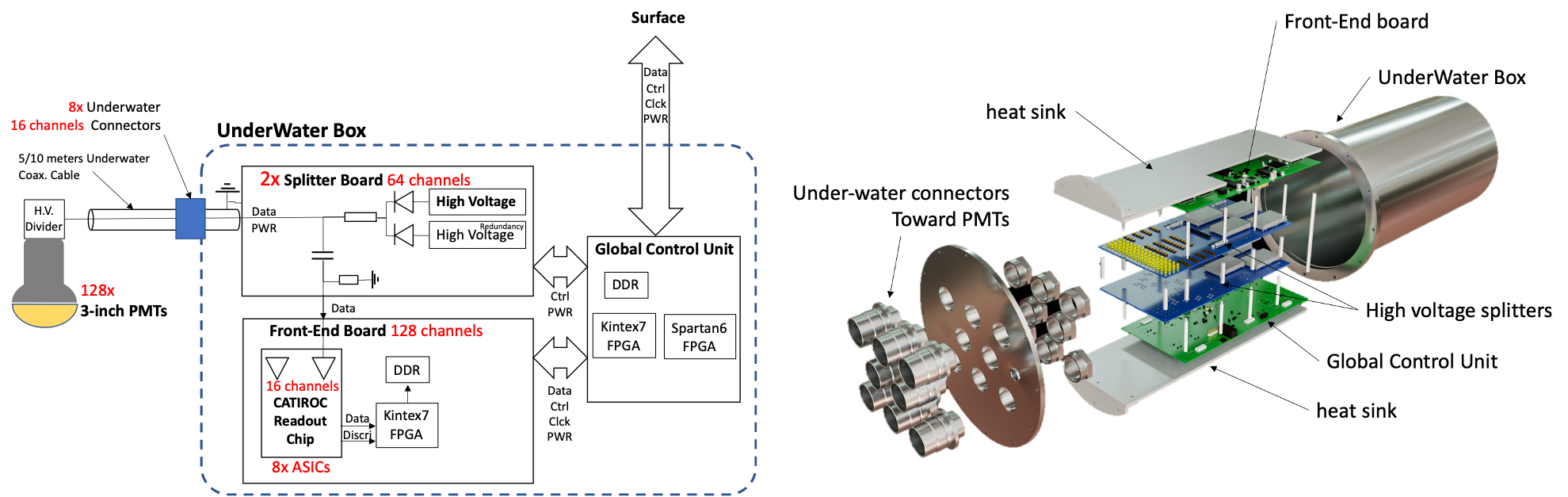}
\caption{\label{fig:spmtDiagram} Schematic overview of the JUNO SPMT system (left), and exploded view of the main components of the UWB (right). 128 3-inch PMTs are connected to a single UWB through eight 16-channel waterproof connectors. Each UWB contains two high-voltage splitter boards, one readout front-end board, and one control board (GCU). The latter two, which are shown in green, are thermally coupled to heat-sinks that dissipate the heat through the lid of the UWB. The cables that connect the multi-channel receptacles to the splitter boards are not shown.}
\end{figure}

Schematics illustrating the main components of the SPMT system and their arrangement are shown in Fig.~\ref{fig:spmtDiagram}. A total of 200 UWBs service the SPMTs in groups of 128. The UWBs are made of passivated stainless steel and consist of a cylindrical body, a flange, and two lids. The body is a 500~mm long Schedule 10 (SHC10) pipe with an internal diameter of 264.5~mm. The flange consists of a 15~mm long cylinder with roughly the same internal diameter as the body but an external diameter of $318$~mm. This piece is very carefully machined to achieve a circular cross-section with essentially no eccentricity, and is welded to one end of the body. A 15~mm thick disk is welded to the other end and serves as a fixed lid. The other lid is carefully machined from a 20~mm thick disk to hold 3 O-rings (2 axial, 1 radial) that make the seal with the flange. This removable lid is also machined to hold 8 receptacles that couple to the 16-channel connectors mentioned above. Each receptacle is sealed on the lid with two O-rings and interfaces with the connector through a stainless steel shell with another two O-rings and one interfacial seal.

Each UWB provides all the necessary electronics to operate the 128 photomultipliers connected to it: supply of high voltage and separation from the signal, readout and digitization of the signal, acquisition and communication with the surface electronics room.

Since the signal from the SPMTs travels on the same coaxial cable that carries the bias voltage, it is necessary to separate the two. This is accomplished by two high-voltage splitter boards servicing 64 channels each. These boards connect to the 20~cm long cables detaching from the 16-channel receptacles on the UWB via MCX connectors, greatly easing the integration process. The high-voltage is produced by the same high-voltage units (HVUs)~\cite{bib:juno:bxscheme} used in the 20-inch PMT system. A total of 16 HVUs are hosted in the two splitter boards of each UWB, although only 8 actively feed the 128 SPMTs connected to that UWB at any one time. The remaining eight serve as a one-to-one backup in case of failure, ensuring the longevity of the system.

The readout and digitization of the 128 channels are carried out by a single electronics front-end board holding 8 CATIROC~\cite{Conforti:2020ibx} ASICs of 16 channels each and controlled by a Kintex-7 FPGA. At the gain the SPMTs will operate ($3\times 10^6$), the CATIROC ASIC provides a charge measurement over a dynamic range from well below 1 photoelectron to several hundreds. It also provides a timing measurement with an accuracy of 200~ps (RMS) per channel. A gain adjustment per channel (over 8~bits) allows compensating the non-uniformity of the 16 PMTs operating at the same high voltage. Sixteen discriminator outputs are also available to cope with event pile-up, as expected when a muon goes through the scintillator close to the PMTs or in case of a nearby supernova burst. The board operates in a trigger-less mode for the 128 channels and the resulting complex management of the event flow is handled by the FPGA and a 1-GB DDR RAM.

A Global Control Unit (GCU) electronics board takes care of powering and controlling the HVUs and the front-end board, distributing the clock and slow control to the same front-end board, and transferring the data to the electronics at the surface. As shown in Fig.~\ref{fig:spmtDiagram}, the GCU and the front-end board are thermally coupled to a heat sink that dissipates the small amount of heat  produced in them ($<$ 40~W) through at least one of the two lids of the UWB. The connection to the surface for the controls, clock distribution, and data flow is done through two CAT5e Ethernet cables, and an additional twisted pair bring the low voltage power. All those cables are embedded in a stainless steel bellow tube that is tens of meters long.

\subsection{Calibration}
\label{subsec:calibration}

To determine the neutrino mass ordering, the JUNO central detector requires a better than 1\% energy linearity and a 3\% effective energy resolution. Multiple calibration sources and multiple dimensional scan systems are
developed to correct the energy non-linearity and spatial non-uniformity of the detector response.

\subsubsection{Calibration system design}
\begin{figure}[!htb]
    \centering
    \includegraphics[width=0.4\linewidth]{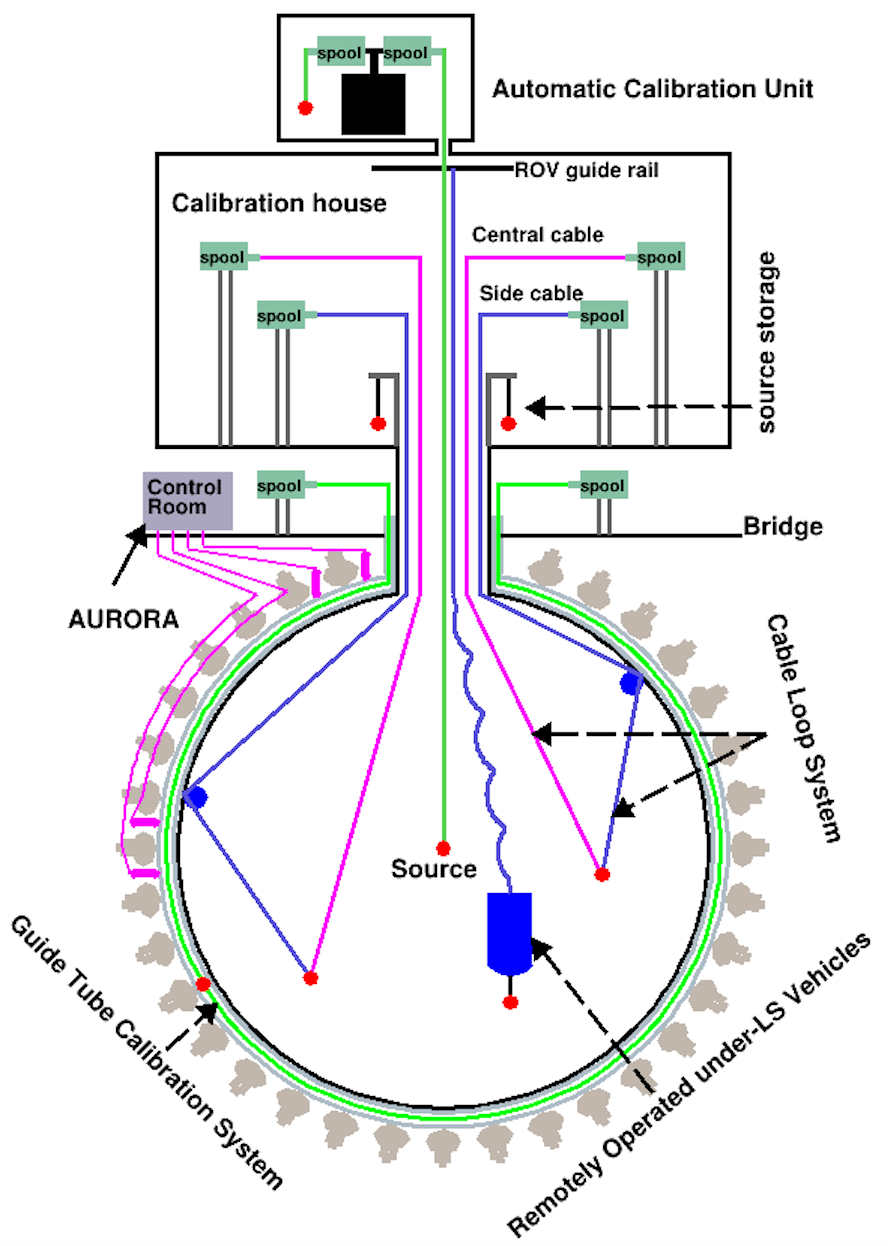}
    \caption{Overview of the calibration system (not drawn to scale).
      \label{fig:overview_calibration_system} }
\end{figure}

The system design is shown in Fig.~\ref{fig:overview_calibration_system}. Functions of sub-systems and auxiliary systems are introduced below.

\begin{table}[!htb]
\begin{center}
    \caption{The radioactive sources and radiation types \label{tab:radiation}}
    \begin{tabular}{c|c|c}\hline
          Source & Type & Radiation\\\hline
          $^{137}$Cs& $\gamma$ & 0.662 MeV\\
          $^{54}$Mn & $\gamma$ & 0.835 MeV\\
          $^{60}$Co & $\gamma$ & 1.173 + 1.333 MeV\\
          $^{40}$K  & $\gamma$ & 1.461 MeV\\
          $^{68}$Ge & e$^{+}$ & annihilation 0.511 + 0.511 MeV\\
          $^{241}$Am-Be & n, $\gamma$ & neutron + 4.43 MeV \textcolor{black}{($^{12}$C$^{*}$)} \\
          $^{241}$Am-$^{13}$C &n, $\gamma$ & neutron + 6.13 MeV \textcolor{black}{($^{16}$O$^{*}$)}\\
          (n,$\gamma$)p & $\gamma$ & 2.22 MeV \\
          (n,$\gamma$)$^{12}$C & $\gamma$ & 4.94 MeV or 3.68 + 1.26 MeV\\\hline
	\end{tabular}
\end{center}
\end{table}

For a one-dimensional scan, the Automatic Calibration Unit (ACU) can deploy multiple radioactive sources or a pulsed laser diffuser ball along the central axis of the CD~\cite{Liu:2013ava}. Four sets of spools
on a turntable can deliver sources independently and automatically through the chimney of the CD. The  calibration positions can be controlled with a precision of better than 1 cm.

Off-axis calibration positions are important to investigate the non-uniformity of the detector response. A calibration source attached to a Cable Loop System (CLS) can be moved on a vertical half-plane by adjusting the lengths of two connection cables~\cite{ZYY_CLS_paper}. Two sets of CLSs will be deployed to provide a 79\% effective coverage on an entire vertical plane. A Guide Tube (GT), which surrounds the outside of the CD and runs in a longitudinal loop, will be used to calibrate non-uniformity of the energy response at the CD boundary. A source will be pulled in the Teflon GT with servomotors to reach the desired locations.

For a full volume scan, a source attached to a Remotely Operated under-LS Vehicle (ROV) can be deployed to the desired locations inside the LS. The ROV motion is driven by two jet pumps controlled through an umbilical cable. The ROV also contains a camera, which can be used to inspect the interior of the CD.

The ACU, CLS and ROV systems will be moving inside the CD, so they will be installed and deployed from an air-tight stainless steel calibration house located on the top of the detector to avoid diffusion of air-borne radon into the LS. The motion and source changing can be performed completely automatically via Programmable Logic Controllers (PLCs) control software. All parts in contact with the LS are assayed and selected for radiopurity and chemical compatibility.

\subsubsection{Calibration sources and program}
The envisioned calibration program is shown in Tab.~\ref{tab:pragram}, with the frequency of calibration runs, purposes and duration. The radioactive sources provide gammas and neutrons to calibrate energy responses in the energy range of the IBDs. A pulsed UV laser with 266 nm wavelength will be used to calibrate PMT timing response~\cite{Zhang_2019}.
\begin{table}
\begin{center}
    \caption{The envisioned calibration program}
	\begin{tabular}{cccc}
          \hline
          Program & Purpose & System & Duration [min] \\
          \hline
          \multirow{2}{*}{Weekly calibration} &
          Neutron (Am-C) & ACU & 63 \\
	      & Laser 	 & ACU    & 78\\
          \hline
          \multirow{4}{*}{Monthly calibration} &
          Neutron (Am-C) & ACU &  120 \\
          &Laser & ACU    &   147 \\
          &Neutron (Am-C) &  CLS  &  333 \\
          &Neutron (Am-C) &  GT   &  73 \\
          \hline
          \multirow{8}{*}{Comprehensive calibration} &
          Neutron (Am-C)  & ACU, CLS and GT  	   & 1942\\
	      & Neutron (Am-Be) & ACU  	   & 75  \\
          & Laser 	  & ACU 	  	   & 391   \\
          & $^{68}$Ge 	  & ACU      & 75 \\
          & $^{137}$Cs	  & ACU 	 & 75   \\
          & $^{54}$Mn 	  & ACU 	 & 75 \\
          & $^{60}$Co 	  & ACU      & 75  \\
          & $^{40}$K  	  & ACU      & 158  \\
          \hline
	\end{tabular}
	\label{tab:pragram}
\end{center}
\end{table}

\subsubsection{Auxiliary systems}
\label{sec:auxiliary}
There are a few auxiliary systems to monitor the detector, for example, to determine the source positions, and to measure the temperature and the optical properties of the LS.
\begin{itemize}
    \item An array of eight customized low radioactivity Ultrasonic Sensor System (USS) receivers are configured to reconstruct source positions based on signals emitted from transmitters on the source attachment fixture and the ROV.
    \item Four CCDs will be mounted on the equator facing the CLS plane to take photos with infrared light to monitor the source positions.
    \item A temperature sensor can be attached to the standard source attachment fixture to measure temperature in various positions.
    \item Several independent laser devices known as AURORA, located inside the ultrapure water shield, make {\it in situ} measurements of the attenuation length and Raleigh scattering length of the LS using detected light pattern on the PMTs.
\end{itemize}

\subsection{Data acquisition and detector control}
\label{subsec:daq}

\subsubsection{Data Acquisition System (DAQ)}
The DAQ system should readout about 40 GByte/s triggered waveform data and trigger-less time and charge data via Ethernet, from the 20,012 20-inch PMTs (17,162 in the Central Detector and 2,400 in the water Cherenkov detector) and the 25,600 3-inch PMTs, and build and process the events by the Online Event Classification algorithm and the software trigger to reduce the data rate by $\sim500$ times. The data will then be transferred to IHEP via Internet with a bandwidth of 1~Gbps (bit per second).

\begin{figure}[htbp]
\centering
\includegraphics[width=0.4\textwidth]{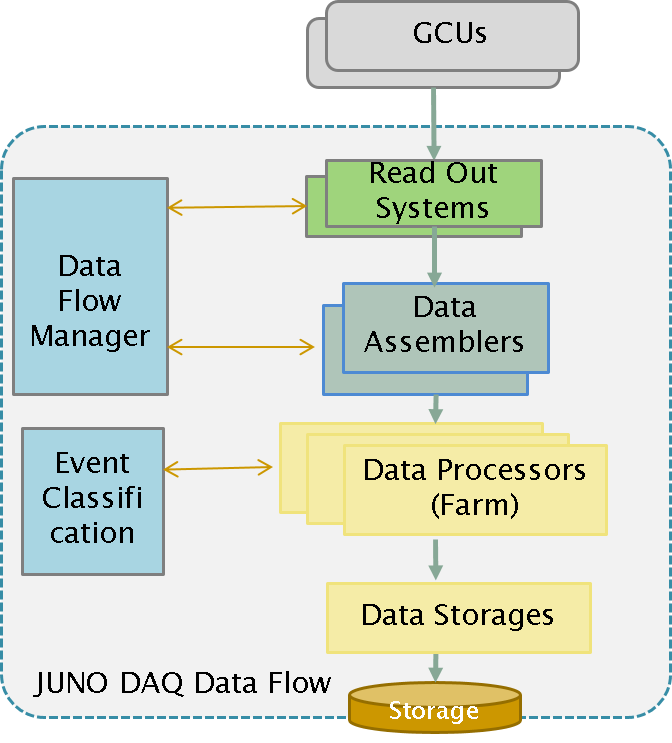}
\caption{The DAQ data flow consisting of four levels of scalable and distributed data processing including the readout systems, data assemblers, data processers, and data storages, and two centralized data processing including the data flow manager and the event classification. \label{fig:dataflow} }
\end{figure}

The JUNO electronics can completely handle the Supernova explosion beyond 0.5~kpc (more than 2 M events in $\sim10$ seconds). A dedicated multi-messenger trigger system is designed to achieve an ultra-low detection threshold of $\mathcal{O}$(10 keV). Lossless readout of these data put large challenges on the DAQ system, which is also required to monitor continuously the Supernova burst during the detector calibration. A unified data flow was designed to readout the two types of data streams from the Global Control Unit (GCU) electronics boards, process, and write to disks without any loss. Fig.~\ref{fig:dataflow} shows the data flow design. There are four levels of scalable and distributed data processing, including the readout systems, data assemblers, data processers, and data storages. Data flow manager and event classification are two centralized processing.

The comprehensive data flow performance tests have been done~\cite{TZeng:2019RT}. An online software design based on the Service-Oriented Architecture (SOA) was proposed~\cite{JLi:2019RT}. A graphics-integrated readout test program has been developed for the electronics and PMT tests. The test system supports the readout of 48 20-inch PMTs and shows waveform graphics in real-time. It has been operating at the JUNO PMT test site for months.

\subsubsection{Detector Control System (DCS)}
\begin{figure}[htbp]
\centering
\includegraphics[width=0.8\textwidth]{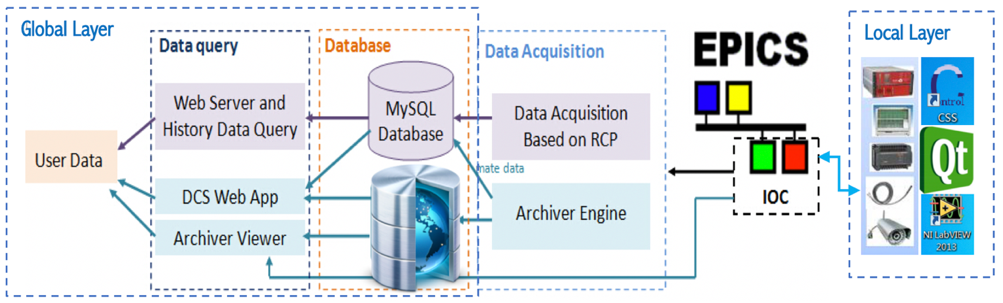}
\caption{Scheme of the Detector Control System. The DCS framework is based on EPICS and is divided into the Local Layer, the Data Acquisition Layer, and the Global Layer. The Local Layer includes the firmware of the embedded devices and the Industrial PCs.  The Data Acquisition Layer reads data from IOCs and save it into the databases. The Global Layer includes the databases and the user interfaces.
\label{fig:dcs1}}
\centering
\end{figure}

The main task of the DCS is to monitor and control the working condition of the detector and to raise alarms if a specific monitored value goes out of range. Given the distribution characteristics of the experimental equipments, the distributed data exchange platform will be used for the development.  The JUNO DCS framework is based on the Experimental Physics and Industrial Control System (EPICS)~\cite{EPICS}. It is divided into three layers: the Local Layer, the Data Acquisition Layer, and the Global Layer, as shown in Fig.~\ref{fig:dcs1}. The firmware of the embedded devices and local control of the Industrial PCs (IPC) form the Local Layer. They are managed by the Input-Output Controllers (IOC). The Data Acquisition Layer reads data from IOCs and saves it into databases, consisting of the Realtime Data Sharing Pool and the MySQL database. The operating status of the system devices is monitored in real-time and recorded in the database. The global control system shares the data and the interactive control commands by IOCs. The safety interlock between sub-systems is realized by the Programmable Logic Controller (PLC). Web-based data query applications provide the user interfaces to the databases.

A prototype of the DCS framework has been realized and tested, including the high voltage, power supply, temperature, and humidity monitoring, etc. More detector subsystems will be integrated when their local layer interfaces developed.

\subsection{Offline software and computing}
\label{subsec:offline}

\subsubsection{Framework}
For the processing and analysis of JUNO data, a general-purpose software framework called SNiPER~\cite{Zou:2015ioy} (Software for Non-collider PhysIcs expERiment) has been developed based on the experience gained with NuWa~\cite{DayaBay:2012aa}, the offline software of the Daya Bay experiment. With SNiPER, all functional modules are implemented in C++ and the process flow is controlled with Python configuration scripts. These scripts dynamically control the loading of various modules to build processing chains flexibly connected using a data buffer. Data objects within a pre-defined time window are retained in memory enabling time correlation analysis within reasonable memory budgets. A navigator object acts as a container for event data objects~\cite{Li:2017zku} representing the results of processing steps such as detector simulation, digitization, calibration and reconstruction. Event data objects can be shared by several navigators making it possible to construct physics events that combine consecutive readout events. As typically only a subset of the event information is needed during data analysis a lazy loading mechanism~\cite{Li:2016psi} using lightweight headers and data objects was been implemented to only load objects required by the analysis. SNiPER uses Threading Building Blocks (TBB)~\cite{10.5555/1461409} to execute algorithms on multiple CPU threads in parallel. By following the predefined conventions it is straightforward for developers to implement thread-safe algorithms for use within the framework.

A geometry management system has been developed within the JUNO offline software~\cite{Li:2018fny} to provide a consistent detector description for simulation, reconstruction, event display and data analysis. The geometry system is based on the Geant4 GDML geometry serialization and the automated conversion between Geant4 and ROOT representations of detector geometry provided by the ROOT GDML tools. It has been successfully used with various detector designs for optimization and performance evaluation with Monte Carlo data.

A ROOT-based event display system has been developed~\cite{you2018root} for JUNO which is integrated with the JUNO offline software, providing an intuitive way to examine detector structure, particle trajectories and hit time distributions. In addition, an independent Unity based event display system~\cite{zhu2019method} for JUNO is under development, providing vivid representations of events with the potential for straightforward porting to many hardware platforms.

\subsubsection{Simulation}
A complete simulation chain has been designed and developed~\cite{Lin:2017usg} based on the SNiPER framework and the Geant4 simulation toolkit~\cite{Agostinelli:2002hh} in order to support the JUNO detector design and performance optimization studies. The simulation chain consists of four components: various primary particle generators, detector simulation, electronics simulation and trigger simulation.

The reactor antineutrino generator operates from antineutrino spectra read from a database. Cosmic muons are simulated using a digitized topographic map of the JUNO site and the MUSIC Muon Simulation Code~\cite{Antonioli:1997qw}. Generators for natural radioactive decay are customized using data from the ENDF database~\cite{Li:2015cqa}. Meanwhile, more generators have also been developed to fulfill the requirements of the JUNO physics program, such as generators for geoneutrinos, atmospheric neutrinos, proton decays, supernova neutrinos, etc.

An accurate detector geometry is constructed with the geometry tools provided by Geant4 based on inputs of the detailed mechanical designs of Central Detector (CD), Water Pool (WP) and Top Tracker (TT). Measurements of the optical properties of all detector components are used to define materials and surfaces within the Geant4 based detector simulation allowing the generation of optical photons from scintillation and Cherenkov processes. A new LS optical model has been implemented to propagate the optical photons that properly accounts for scattering, absorption, re-emission within the scintillator. A new PMT optical model is under development, aiming to precisely handle photon detection on PMTs. The Geant4 implementations of all physics processes relevant to JUNO have been validated.

PMT arrival times and numbers of photoelectron (p.e.) on PMTs from the detector simulation are inputs to the next stage electronics simulation. An analog signal pulse for each PMT is generated and tracked through the digitization process, taking into account effects such as non-linearity, dark noise, pre-pulsing, after-pulsing, Transit Time Spread (TTS), and ringing of the waveform.

The electronics simulation is designed with a ``data-driven" strategy that is able to provide hit-level background mixing enabling accurate modeling of time correlations between different signals without requiring very large memory resources and with the ability to operate in pine-line mode.

The large size and high photon yield of the JUNO scintillator make optical photon simulation for cosmic muons computationally extremely challenging with regard to both processing time and memory resources. Research into eliminating these bottlenecks by offloading optical photon simulation to GPU co-processors with the Opticks package is well advanced~\cite{SimonC:2017hge,Blyth:2019yrd}.

\subsubsection{Reconstruction}
Particles that deposit energy in the liquid scintillator yield optical photons which result in PMT signals. Flash ADCs convert PMT signals in 1~$\mu$s DAQ windows with 1~GHz frequency yielding waveforms which are filtered to remove high-frequency white noise and then deconvoluted with the single p.e.\ PMT response to yield charge and time information of photon hits. These are then used as inputs to reconstruct the event vertex and energy using various algorithms.

The simple and fast charge center method provides a rough estimation of the vertex from the sum of the charge-weighted PMT positions. A more complex likelihood method using the time information of PMT hits achieves a better vertex resolution~\cite{Liu:2018fpq,Li:2021oos}. The residual between the PMT hit time and the time of flight of photon propagation mainly depends on the liquid scintillator decay time profile and the PMT TTS. Modeling these effects with a probability density function permits to perform a likelihood fit to reconstruct the event vertex. In addition, an optical model-independent method has been developed to reconstruct the event energy in JUNO~\cite{Wu:2018zwk}. It utilizes the calibration data to construct the expected charge response of PMTs. A maximum likelihood fit is built based on the observed and expected charge of all PMTs to reconstruct the event energy. The choice of calibration sources as well as the arrangement of calibration points have been investigated in order to reduce the impact of energy non-uniformity on energy resolution. In addition to these traditional reconstruction methods, novel techniques such as Machine Learning have been developed to reconstruct the vertex and energy~\cite{Qian:2021vnh}.

One of the main backgrounds for anti-neutrino detection in JUNO is the cosmogenic isotopes $^9$Li/$^8$He created when cosmic muons pass through the scintillator. In order to reduce this background, precise muon track information is required to define the veto region around the muon trajectory. For muons traversing through the Top Tracker, a tracking algorithm has been developed using all the hits on the TT wire planes to reconstruct the trajectory. For muons that go through the Water Pool and the Central Detector, the trajectory is reconstructed using the PMT charge and time information. Several central detector muon reconstruction algorithms have been developed including a geometrical approach~\cite{Genster:2018caz} and a fastest light technique~\cite{Zhang:2018kag}. Due to the importance of muon reconstruction, other novel methods using deep learning techniques are also being studied.

\subsubsection{Performance of Simulation and Reconstruction}
\begin{figure}[!htb]
    \centering
    \includegraphics[width=0.5\linewidth]{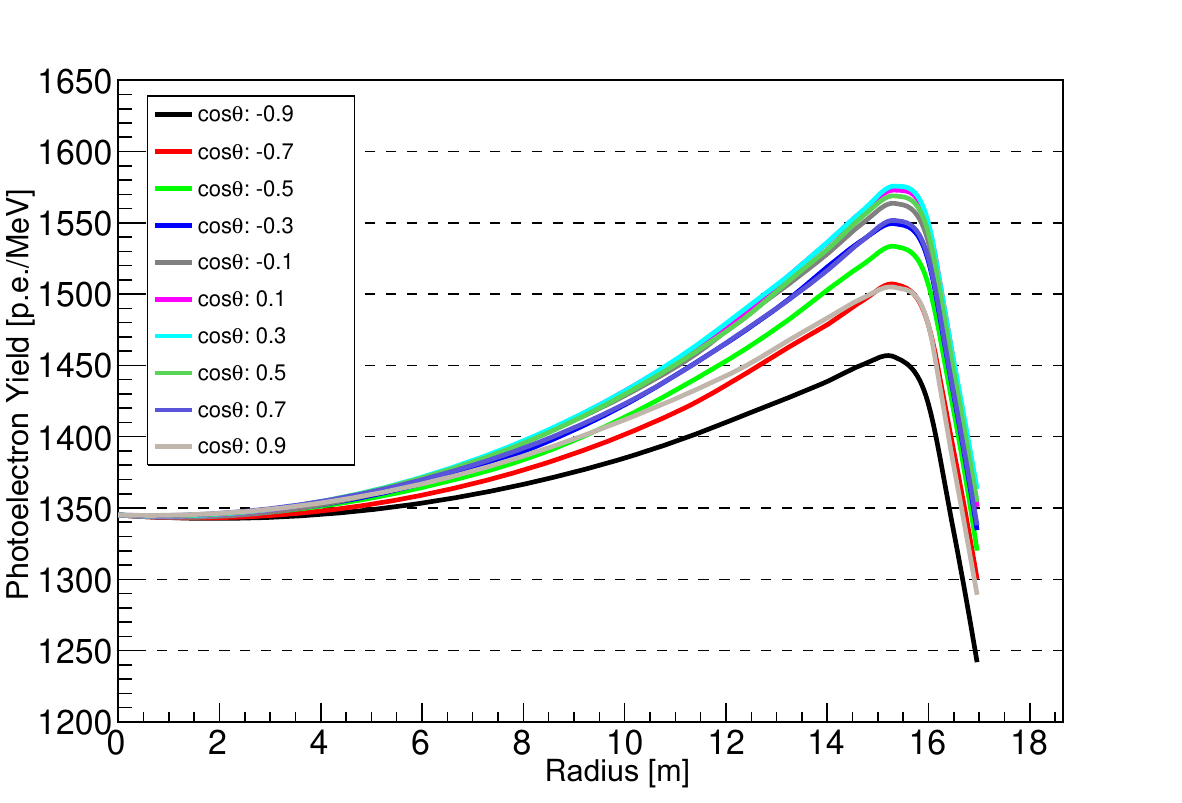}
    \caption{Mean number of photoelectrons per MeV for 2.22 MeV gammas as a function of radius at representative polar angles.
      \label{fig:PE_vs_R} }
\end{figure}

One of the most important parameters of the JUNO detector is the photoelectron yield, which is determined by optical parameters of materials in the Central Detector, such as the LS light yield, absorption lengths, Rayleigh scattering lengths, refractive indexes, and PMT detection efficiencies. All the critical parameters have been measured for JUNO and used in the simulation. Especially, the LS light yield is normalized to the Daya Bay data by comparing the Monte Carlo simulations with the Daya Bay and JUNO configurations.
Fig.~\ref{fig:PE_vs_R} shows the mean number of photoelectrons per MeV for 2.22 MeV gammas as a function of radius at several representative polar angles. The sharp decrease close to 15.5~m is due to the total reflection at the acrylic and water interface.

\begin{figure}[!htb]
    \centering
    \includegraphics[width=0.45\linewidth]{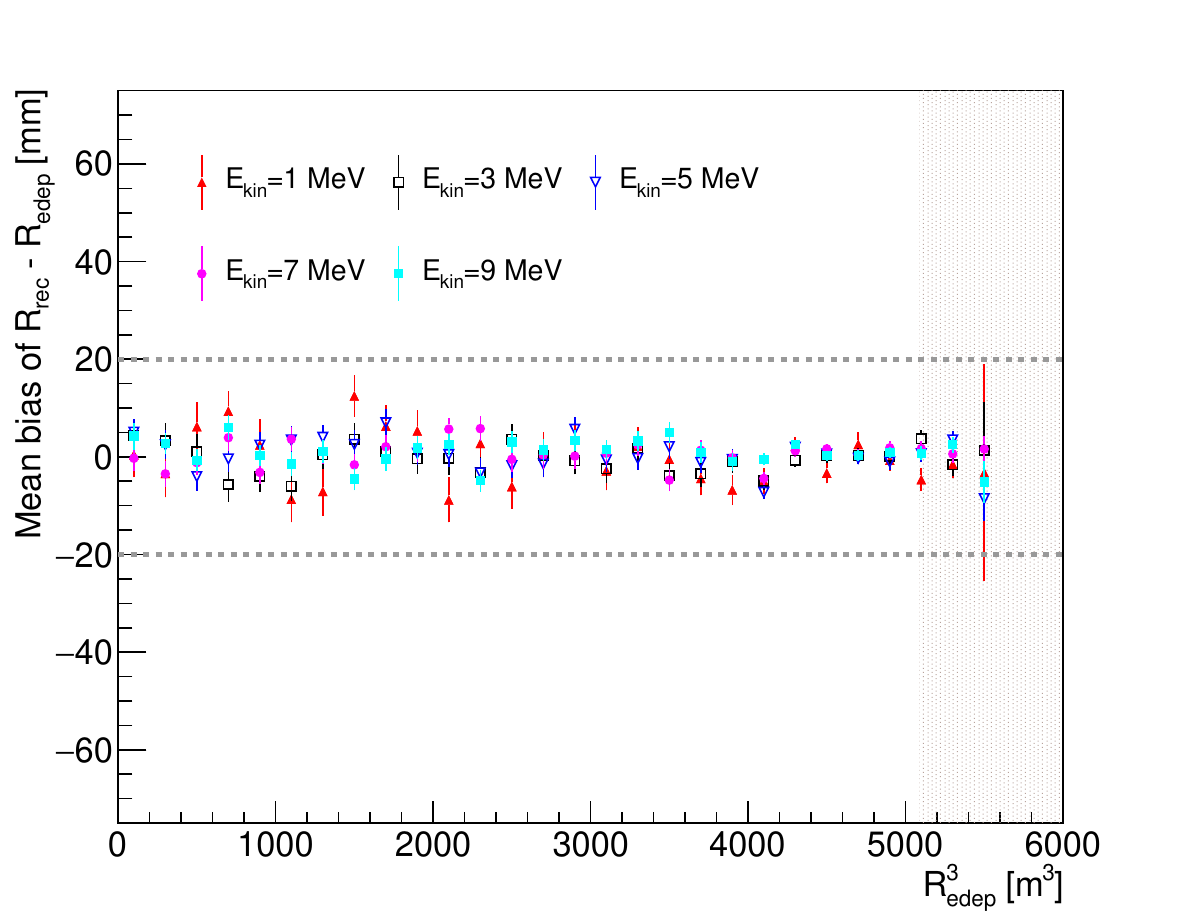}
    \includegraphics[width=0.45\linewidth]{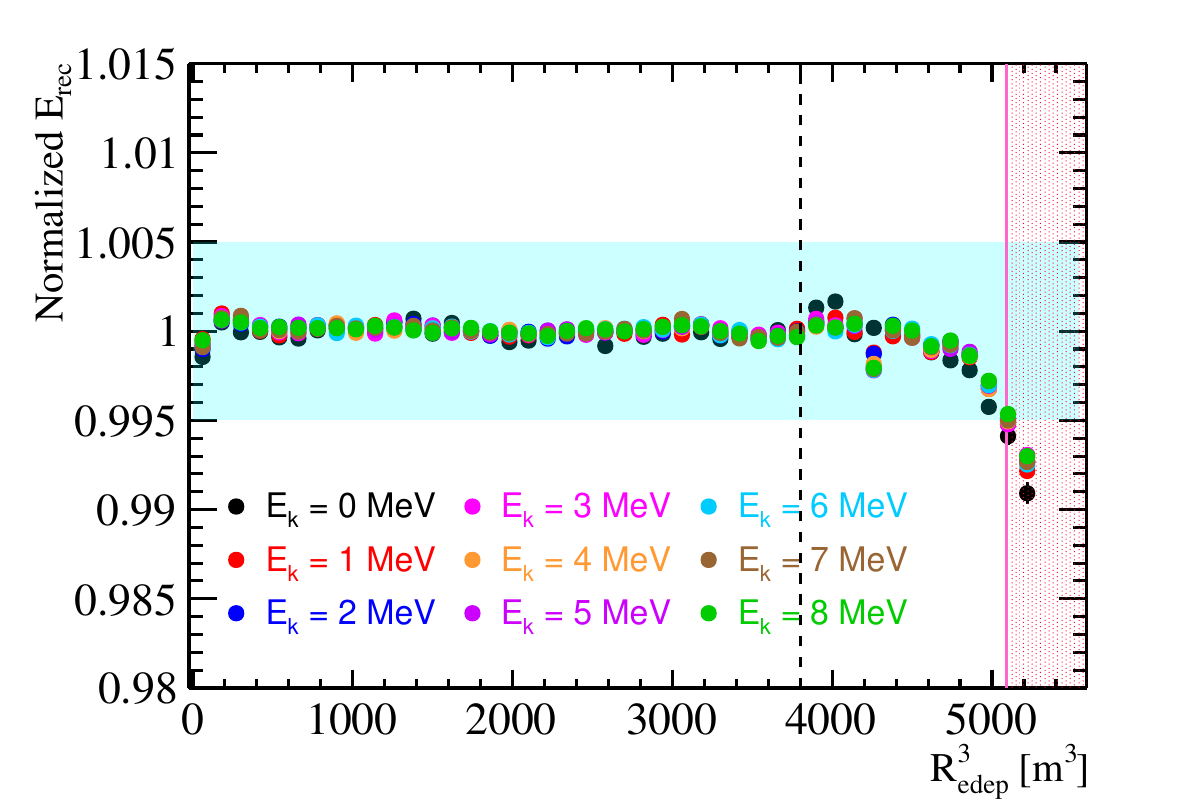}
    \caption{Left: Radial bias of the vertex reconstruction versus $R^3$ for different energies with deep learning. The two dashed lines are at y=$\pm$20~mm, and the shadow coresponds to r $>$ fiducial volumn. Right: Uniformity of the normalized $E_{rec}$ with respect to $R^3$ at various energies.
      \label{fig:ppnp_recPerformance} }
\end{figure}

Since the JUNO detector is nearly symmetric with respect to rotation, the radial bias of reconstructed vertex is of particular interest. Using the Machine Learning technique, the radial bias for positrons with different energies is controlled within 20~mm in the entire LS volume, as shown in the left panel of Fig.~\ref{fig:ppnp_recPerformance}. The vertex resolution, which depends on visible energy, can be described with an empirical formula $\sigma_{R} = a/\sqrt{E_{vis}} + b + c/E_{vis}$, where the parameters $(a, b, c)=(19.2, 22.7, 56.1)$~mm are obtained from the resolution curve using the ResNet-J model in Ref.~\cite{Qian:2021vnh}. The vertex reconstruction using the Machine Learning approach currently performs better than the conventional methods in Ref.~\cite{Liu:2018fpq,Li:2021oos}. However, for the latter there is still room for improvement, such as combining the time and charge PDFs near the detector boundary.

The non-uniform detector response in Fig.~\ref{fig:PE_vs_R} is eliminated to 0.3\% by using the multi-positional source deployment calibration strategy~\cite{Abusleme:2020lur}. The residual non-uniformity near total reflection zone and LS edge can be further mitigated by optimizing the choice of the calibration source and the calibration positions~\cite{Huang:2021baf}. The right panel of Fig.~\ref{fig:ppnp_recPerformance} shows the radial dependency of the normalized $E_{rec}$ for positrons with different energies, and the residual energy non-uniformity within the fiducial volume (across the whole detector) can be constrained to 0.17\% (0.23\%). The efforts to improve energy reconstruction and the energy resolution are still ongoing, such as reducing the random pile-up from the PMT dark noises and the $^{\rm 14}$C $\beta$-decays.

\subsubsection{Database and DQM}
Data on the detector condition including system configurations, calibration data and monitoring results are essential for event data processing and physics analysis. A JUNO Conditions Data Management System (JCDMS)~\cite{Huang:2020gpj} has been developed to homogenously collect and manage all these heterogeneous conditions data. The system provides several ways to manage and access status parameters via the web interface or the C++ Conditions Database (CondDB) service. Data caching based on Frontier and Squid~\cite{Dykstra:2011zz} are used to decrease resource burdens. The JCDMS is in use for JUNO Monte Carlo data production.

A JUNO Data Quality Monitoring (DQM) system has been developed that will provide real-time monitoring of DAQ data. Recent calibration constants and the full offline reconstruction algorithms will be used and a web interface will present processing results. In addition, the event display software~\cite{Zhu:2018vjj} will provide monitoring of recent events.

\subsubsection{Computing}
The raw data from the JUNO detector will be recorded at a rate of about 2~PB per year. A dedicated link of 1~Gbps bandwidth will be used to transfer raw data from the experimental site to the IHEP data center. The data will be stored on disk first and then archived in the tape library. Data sets will be replicated at European sites for easy access by the whole JUNO collaboration. The bandwidth of the international network between China and Europe, 10~Gbps, is adequate for normal data sharing. An application to join the Large Hadron Collider Open Network Environment (LHCONE)~\cite{Martelli:2015mkr} is in progress, which will enhance the performance of JUNO data transfers.

The JUNO distributed computing system has been established based on the DIRAC~\cite{Tsaregorodtsev:2008zz} interware in order to better organize the resources needed for data processing and analysis. The distributed computing system provides users with a common interface for storage and computing resources scattered across multiple data centers and provides a consistent infrastructure for the experiment. The system is structured in three layers or ``Tiers'' similar to the WLCG~\cite{Bonacorsi:2006zz}. Tier~0 owns the central Storage Element (SE) and is able to store a complete set of raw data as well as all derived data. Using its dedicated CPUs, Tier~0 is able to conduct prompt event reconstruction. Tier 1 holds a complete copy of the data and performs data reprocessing, simulation and analysis. The IHEP data center will act as both Tier~0 and Tier~1 and at least one large European data center will act as a Tier~1 site. Smaller sites with limited resources act as Tier~2 contributing to simulation and analysis. Other small sites can link caches with the SEs in order to support analysis.

Several rounds of data challenges producing Monte Carlo (MC) data have been performed successfully by the JUNO distributed computing system. Participants included data centers from JINR, CNAF, University of Padova, IN2P3 and IHEP. Software for the data challenges was distributed via CVMFS~\cite{Buncic:2010zz}, the Dirac File Catalogue (DFC) was used for data management and the workflow of the MC production was constructed using the DIRAC transformation system. Further data challenges are scheduled to improve the efficiency and reliability of the JUNO distributed computing system.

\subsection{Low background control}
\label{subsec:lowbkg}

The main expected background sources in JUNO are cosmic muons, fast neutrons, cosmogenic isotopes induced by muon spallation in the liquid scintillator, and natural radioactivity. Apart from the natural radioactivity, the impact of all these sources on JUNO measured event rate is determined by the choice of the experimental site and can only be diminished by efficient tagging, careful event reconstruction, and proficient veto strategies. The strict control of natural radioactivity is the only means to reduce the accidental count rate in the JUNO detector.

Natural radioactivity comes from all materials and the environment. Huge efforts on material screening and convenient arrangement of the experimental apparatus were the main driving actions during JUNO planning and designing. The Central Detector, i.e.\ the spherical acrylic vessel containing the 20\,kton of liquid scintillator, is immersed in a water pool that provides an effective shielding not only against the fast neutrons from muons, but also against the radioactivity from the rock. All materials comprising the JUNO detector, including the liquid scintillator itself, have been carefully selected according to their bulk radiopurity. Besides that, production, cleaning, storage, and construction protocols are carefully sorted out to avoid possible contamination of material surfaces, especially in the vicinity of the Central Detector. Finally, the radioactivity from external materials can also be efficiently removed by imposing an energy threshold and a fiducial volume cut. A simulation of the whole detector, based on the SNIPER framework and GEANT4 toolkit, is used to constantly check the impact of any modification or improvement on the JUNO background budget. A list of the target impurity concentrations in key materials of the JUNO detector is reported in Tab.~\ref{tab:bkgBudget}, together with their expected contribution to the count rate in the Region Of Interest (ROI).

Since the LS will be extensively purified during the production, and the purification efficiency is hard to predict accurately, we have assumed its $^{238}$U/$^{232}$Th content to be $1\times10^{-15}$~g/g in Tab.~\ref{tab:bkgBudget}, which is the minimum requirement for reactor neutrino studies. Similar requirement is put for $^{40}$K. The target  $^{238}$U/$^{232}$Th impurities of JUNO is $1\times10^{-17}$~g/g. The PMT glass radio-impurities are taken from sample measurements (dynode PMT) and constant monitoring of the glass radioactivities during production (MCP-PMT)~\cite{Zhang:2017ocm}. The $^{238}$U/$^{232}$Th contamination in acrylic can reach sub-ppt level according to the ICP-MS measurement of the pre-production samples~\cite{Cao:2020zyr}. The radiopurity during acrylic production will be monitored by both the ICP-MS measurement and neutrino activation analysis. Numbers for other materials are upper limits based on numerous measurements during R\&D.

\begin{table}[!htb]
  \begin{minipage}[c]{\textwidth}
  \resizebox{\textwidth}{!}{
\begin{tabular}{c|c|c|c|c|c|c|c|c} \hline
\hline
   \multirow{2}{*}{Material} & \multirow{2}{*}{Mass} & \multicolumn{5}{c|}{Target impurity concentration} & \multicolumn{2}{c}{Singles in ROI}  \\
   \cline{3-9}
   & & $^{238}$U & $^{232}$Th & $^{40}$K & $^{210}$Pb/$^{222}$Rn & $^{60}$Co & ALL & FV  \\
   & [t] & [ppb] & [ppb] & [ppb] &   & [mBq/kg] & [Hz] & [Hz]  \\ \hline
   LS & 20 k & 10$^{-6}$ &  10$^{-6}$ & 10$^{-7}$ &  10$^{-13}$ ppb & & 2.5 & 2.2  \\ \hline
   Acrylic &  610 & 10$^{-3}$ & 10$^{-3}$ & 10$^{-3}$ & & & 8.4 & 0.4  \\ \hline
   SS truss and nodes & 1 k & 0.2 & 0.6 &  0.02 && 1.5  & 15.8  & 1.1 \\ \hline
    dynode-LPMT glass & 33.5 & 400  & 400 &  40 && &\multirow{3}{*}{26.2} & \multirow{3}{*}{2.8} \\
   MCP-LPMT glass & 100.5 & 200  & 120 &  4 && &&\\
   dynode-SPMT glass & 2.6 & 400  & 400 &  200 && &&\\ \hline
   Water & 35 k & &&&10 mBq/m$^3$&  & 1.0 & 0.06  \\ \hline
   Other && &&&& & 5 & 0.6 \\ \hline
   \hline
   \multicolumn{7}{c|}{Sum} & 59 &  7.2  \\ \hline
  \hline
\end{tabular}}
  \end{minipage}
\caption{Background budget for key materials used in the JUNO detector. The ``LPMT" and ``SPMT" refer to 20-inch PMTs and 3-inch PMTs, respectively. The ``Other" components include all materials that have relatively smaller contribution to the background, such as the calibration parts, PMT cover, PMT potting, electronics and cables and the rock, etc. The expected rates are given both in the full detector volume (ALL, i.e., $r_{LS}=\rm{17.7\,m}$) and within the the default fiducial volume ($r_{LS}=\rm{17.2\,m}$). \label{tab:bkgBudget}}
\end{table}

Besides detector design and careful material control, particular care is devoted to materials that are in direct contact with the liquid scintillator.
For the panels of the acrylic vessel, a nylon film of $\sim$1~$\mu$m thickness is used to protect the acrylic surface against dust and radon diffusion until the Central Detector is filled with the scintillator.
Concerning the cleanliness of the tanks or equipment that will be in direct contact with the liquid scintillator during the purification process, the baseline limit for the residual dust on surfaces is $\sim$0.1\,mg/m$^2$, assuming that the radioactivity of the dust is similar to that of the rock. This should guarantee the U/Th contribution from surface pollution lower than 10$^{-16}$~g/g in 20\,kton of scintillator.

Even more importantly, the Radon concentration in underground air can reach more than 100~Bq/m$^3$. Radon can be dissolved in the liquid scintillator, placing important requirements for the gas tightness of all purification systems. Leakage tests will be performed based on vacuum force technology. Assuming the pressure difference is 1 bar and radon concentration in the underground air is 100 Bq/m$^3$, the leakage sensitivity for the single component has to reach 4$\times$10$^{-6}$ mbar$\cdot$l/s with the vacuum-air test to reach the goal for $^{210}$Pb/$^{222}$Rn.

Finally, the cleanliness of the environment during detector installation and liquid scintillator filling is also very important, since the acrylic will be exposed to the air after nylon removal and cleaning and before liquid scintillator filling. A class 100,000 environment is needed for the whole underground laboratory, and special requirements for the cleanliness of the air in the acrylic vessel should reach class 1000 -- 10,000 cleanliness level.

\subsection{TAO}
\label{subsec:tao}

The Taishan Antineutrino Observatory (TAO)~\cite{junotaocdr} is designed for a high precision measurement of the reactor antineutrino spectrum. With a ton-level liquid scintillator (LS) detector at $\sim30$ meters from a reactor core of the Taishan Nuclear Power Plant, TAO will provide a reference spectrum for the determination of neutrino mass ordering in JUNO, as well as a benchmark measurement to test nuclear databases~\cite{INDC-NDS-0786}. By monitoring one core for several fuel cycles, TAO will not only measure the total spectrum of this core with high energy resolution, but also be able to extract the isotopic spectra of $^{235}$U and $^{239}$Pu (or the combo of $^{239}$Pu and $^{241}$Pu) using the same technique in Ref.~\cite{Adey:2019ywk}. With the fission fractions provided by core simulations, all cores in the Taishan and Yangjiang power plants can be predicted to a good precision. The experiment is expected to start data taking at a similar time as JUNO.

With a 1-ton fiducial volume, the IBD event rate of TAO will be more than 30 times that of JUNO, providing enough statistics. Although $3\%/\sqrt{E}$ energy resolution ($E$ is visible energy in MeV) is enough for
TAO to serve as a reference detector of JUNO, the energy resolution should be as high as possible
to study the possible fine structure of the reactor antineutrino spectrum and create a highly resolved
benchmark to test nuclear databases. Therefore, Silicon Photomultipliers (SiPMs) of $>50$\% photon detection efficiency will be used to view the LS with close to full geometrical coverage. The detector has to operate at -50$^\circ$C to lower the dark noise of the SiPMs to an acceptable level. Simulations show that about 4,500 photoelectrons per MeV will be observed, corresponding to $1.5\%/\sqrt{E}$ in photon statistics. Taking into account systematic effects, the energy resolution can still reach $<2\%/\sqrt{E}$.

\begin{figure}[hbt]
    \centering
    \includegraphics[width=0.7\columnwidth]{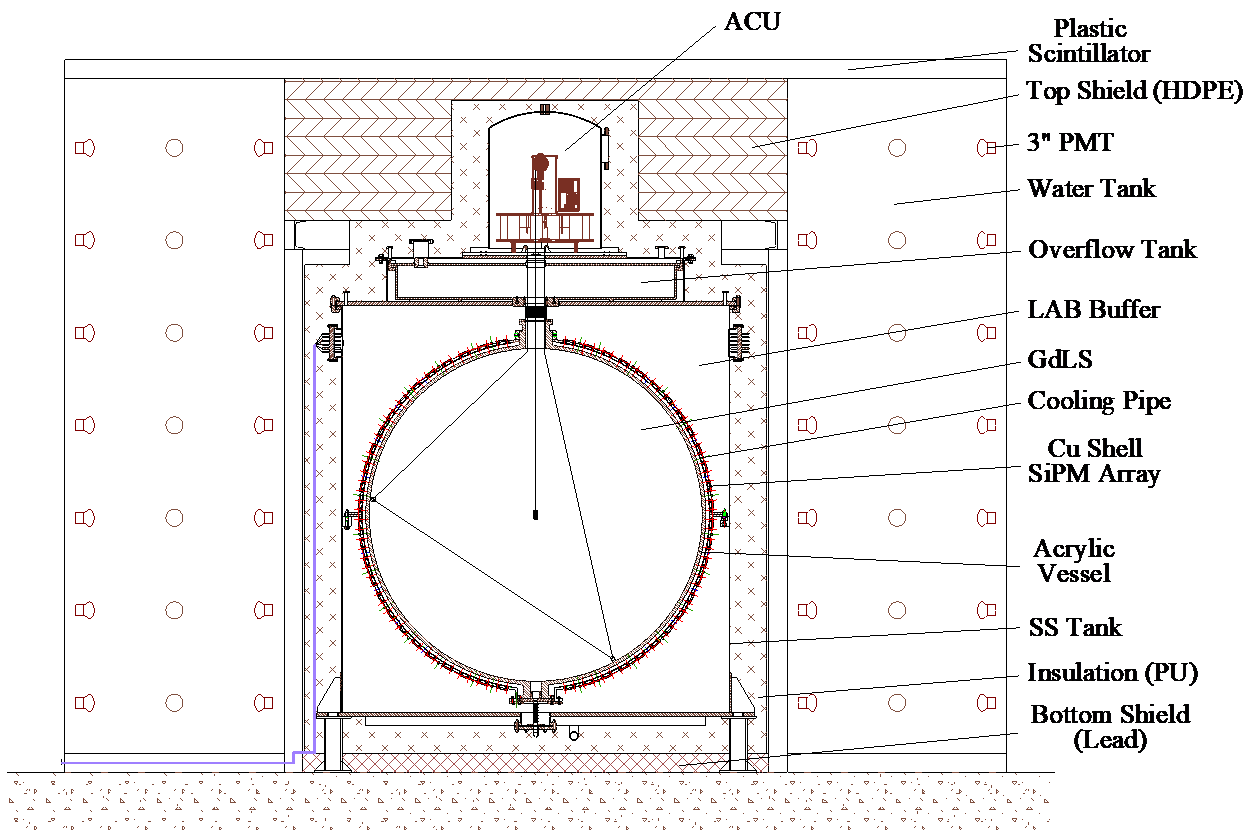}
    \caption{Schematic view of the TAO detector, which consists of a Central Detector (CD) and an outer shielding and veto system. The CD consists of 2.8~ton gadolinium-doped LS filled in a spherical acrylic vessel and viewed by 10~m$^2$ SiPMs, a spherical copper shell that supports the SiPMs, 3.45~ton buffer liquid, and a cylindrical stainless steel tank insulated with 20~cm thick Polyurethane. The outer shielding includes 1.2~m thick water in the surrounding tanks, 1~m High-Density Polyethylene on the top, and 10~cm lead at the bottom. The water tanks, instrumented with Photomultipliers, and the plastic scintillator on the top comprise the active muon veto system. The triangle and central axis inside the detector show the reach of the calibration.
    \label{fig:taodetector}}
\end{figure}

A schematic drawing of the TAO detector is shown in Fig.~\ref{fig:taodetector}. The Central Detector (CD) detects reactor antineutrinos with 2.8~ton Gadolinium-doped LS (GdLS) contained in a spherical acrylic vessel of 1.8~m in inner diameter. To contain the energy deposition of gammas from positron annihilation, a 25-cm selection cut from the acrylic vessel will be applied for the positron vertex, resulting in 1~ton fiducial mass. The IBD event rate in the fiducial volume will be about 2,000 (4,000) events per day with (without) the detection efficiency included. A spherical copper shell of 1.884~m in inner diameter and 12~mm in thickness wraps the acrylic vessel, and provides mechanical support and thermal stability for the SiPM tiles. The gap between the SiPM surface and the acrylic vessel is about 2~cm. The copper shell is installed in a cylindric stainless steel tank with an outer diameter of 2.1~m and a height of 2.2~m. The stainless steel tank is filled with Linear Alkylbenzene (LAB) as buffer liquid to shield the radioactivity of the outer tank and to couple optically the acrylic and the SiPM surface. The stainless steel tank is insulated with 20~cm thick Polyurethane (PU) to operate at -50$^\circ$C, at which the dark noise of SiPMs is expected to be reduced by 3 orders to $\sim$~100~Hz/mm$^2$.

An Automated Calibration Unit (ACU) from Daya Bay~\cite{Liu:2013ava} will be re-used to calibrate the detector along the central vertical axis with multiple radioactive sources and an LED. To calibrate the off-axis points, the ACU will be modified to include a Cable Loop System (CLS). The source moves along the side line of a triangle in the detector volume, providing a suitable calibration sample to study the detector's non-uniformity.

The CD is surrounded by 1.2~m thick water tanks on the sides, 1~m High-Density Polyethylene (HDPE) on the top, and 10~cm lead at the bottom to shield the ambient radioactivity and cosmogenic neutrons. Cosmic muons will be detected by the water tanks instrumented with PMTs and by Plastic Scintillator (PS) on the top.

The Taishan Nuclear Power Plant is located in Chixi town of Taishan city in Guangdong province, 53~km from the JUNO experiment. It has two cores currently in operation. Both are European Pressurised Reactors (EPR), each with 4.6~GW thermal power. At $\sim30$~meter baseline, the far core contributes about 1.5\% to the total reactor antineutrino rate in the TAO detector. The Taishan Neutrino Laboratory for the TAO detector is in a basement 9.6~m below the ground level, outside of the concrete containment shell of the reactor core. Muon rate and cosmogenic neutron rate are measured to be 1/3 of those on the surface. Simulations show that the background level due to cosmogenic fast neutrons, accidental coincidences, and cosmogenic $^8$He/$^9$Li can be controlled to $<10$\% of the signal with the designed shielding and muon veto detector.
The expected rates of IBD signal and the residual backgrounds passing the IBD selection cuts are summarized in Tab.~\ref{tab:TAOSB}.
\begin{table}[htb]
\setlength{\belowcaptionskip}{5pt}
\begin{center}
\caption{Summary of the IBD signal and background rates at TAO.\label{tab:TAOSB}}
\begin{tabular}{r l}
  \hline\hline
  IBD signal (after selections) & $\sim$  2000~events/day \\
  Muon rate & $\sim$  70~Hz/m$^2$ \\
  Singles from radioactivity & $< 100$~Hz \\
  Fast neutron background after veto & $< 200$~events/day \\
  Accidental background rate & $< 190$~events/day \\
  $^{8}$He/$^{9}$Li background rate & $\sim$  54~events/day \\
  \hline
\end{tabular}
\end{center}
\end{table}

\section{Summary}
\label{sec:summary}

In this article, we have reviewed the physics potential and final design of the JUNO detector.

With 20-kton liquid scintillator as the neutrino target and a $3\%/\sqrt{E({\rm MeV})}$ effective energy resolution, the neutrino mass ordering can be resolved at 3$\sigma$ with about 6 years of data by detecting the reactor antineutrinos from two nuclear power plants at equal distances of 53 km. Taking into account the precision at that the neutrino mass splitting will be measured by accelerator experiments by then, the significance could increase significantly. With reactor antineutrinos, the neutrino oscillation parameters $\sin^2 2\theta_{12}$, $\Delta m^2_{21}$, and $|\Delta m^2_{32}|$ can be precisely measured to a precision of better than 0.6\% with 6 years of data.

For a typical supernova burst at 10 kpc, JUNO will register $\sim5000$ IBD events, $\sim300$ $e$ES events and $\sim2000$ $p$ES events of all-flavor neutrinos, providing deep insights into the supernova burst mechanism. With 10 years of data, JUNO is expected to provide a 3$\sigma$ evidence of the DSNB signal, provide new low energy data of a possible spectrum distortion of the $^8$B solar neutrino with about 60,000 signal and 30,000 background, measure the geoneutrino flux to a 5\% precision, and constrain the proton lifetime in the $p \to \bar{\nu} K^+$ channel to a lower limit of $8.34 \times 10^{33}$ years (90\% C.L.). The spectrum of atmospheric neutrinos can be investigated with high energy resolution. Many interesting exotic searches may yield unforeseen surprises.

These exciting physics goals require challenging detector technologies. We present the final design of the detector and the key R\&D achievements. The 20-kton liquid scintillator will be purified with Alumina filtration, distillation, water extraction, and gas stripping, and tested by the 20-ton pre-detector OSIRIS to reach a $>20$-m attenuation length and $\sim 10^{-17}$ g/g U/Th/K backgrounds. The liquid scintillator is contained in a huge acrylic vessel with a diameter of 35.4~m and a thickness of  12~cm. The acrylic has a low background level of sub-ppt and $>96$\% transparency. Scintillation light is detected with 17,612 20-inch PMTs of an average detection efficiency of 29.1\% and 25,600 3-inch PMTs of an average detection efficiency of $\sim24\%$, resulting in a photocathode coverage of 77.9\% and a yield of about 1345 photoelectrons per MeV. PMTs are protected against implosion by well-tested covers made of acrylic and stainless steel. The 20-inch PMTs will be read out with 12-bit and 1-GS/s customized FADC and the 3-inch PMTs will be read out with CATIROC ASIC. The electronics are contained in underwater boxes with a required loss rate of $<0.5$\% in 6 years. The detector will be calibrated with multiple sources and laser, with an automated calibration unit, cable loop and guide tube systems, and a Remotely Operated Vehicle, to reach an energy resolution of 3\% at 1 MeV and an energy non-linearity of 1\%. Radioactivity backgrounds are expected to be well controlled to 59~Hz in the whole detector and 7~Hz in the fiducial volume with a 50-cm vertex cut from the acrylic vessel.

Liquid scintillator and PMTs are submerged in 35 kton of pure water instrumented with 2,400 20-inch PMTs. Muon detection efficiency of the water Cherenkov detector is expected to be 99.5\%. Radon in water will be controlled to be $\sim 10$~mBq/m$^3$ with a multi-stage degassing membrane and a micro-bubble system. Muons are also detected with a 3-layer Top Tracker made of plastic scintillator, with a detection efficiency of 93\% and an angular resolution of 0.2$^\circ$.

To eliminate the model dependence due to the possible fine structure in the reactor antineutrino spectrum, a satellite detector TAO will be installed at 30--35 meters from a core of the Taishan nuclear power plant. TAO consists of 2.8 ton gadolinium-loaded liquid scintillator and 10~m$^2$ SiPMs operated at -50$^\circ$C. About 4,500 photoelectrons per MeV and $<2$\% energy resolution is expected.

JUNO completed the excavation of the tunnel and the underground experimental hall at the end of 2020. The detector design has been finalized and all challenges regarding the detector technologies have been solved. The detector component production and facility and detector installation is underway. Both JUNO and TAO are expected to complete the detector construction at the end of 2022.

\section*{Acknowledgement}

We are grateful for the ongoing cooperation from the China General Nuclear Power Group.
This work was supported by
the Chinese Academy of Sciences,
the National Key R\&D Program of China,
the CAS Center for Excellence in Particle Physics,
Wuyi University,
and the Tsung-Dao Lee Institute of Shanghai Jiao Tong University in China,
the Institut National de Physique Nucl\'eaire et de Physique de Particules (IN2P3) in France,
the Istituto Nazionale di Fisica Nucleare (INFN) in Italy,
the Italian-Chinese collaborative research program MAECI-NSFC,
the Fond de la Recherche Scientifique (F.R.S-FNRS) and FWO under the ``Excellence of Science – EOS” in Belgium,
the Conselho Nacional de Desenvolvimento Cient\'ifico e Tecnol\`ogico in Brazil,
the Agencia Nacional de Investigacion y Desarrollo in Chile,
the Charles University Research Centre and the Ministry of Education, Youth, and Sports in Czech Republic,
the Deutsche Forschungsgemeinschaft (DFG), the Helmholtz Association, and the Cluster of Excellence PRISMA+ in Germany,
the Joint Institute of Nuclear Research (JINR) and Lomonosov Moscow State University in Russia,
the joint Russian Science Foundation (RSF) and National Natural Science Foundation of China (NSFC) research program,
the MOST and MOE in Taiwan,
the Chulalongkorn University and Suranaree University of Technology in Thailand,
and the University of California at Irvine in USA. 

\bibliographystyle{elsarticle-num}
\bibliography{juno_ppnp}
\end{document}